\documentclass{aa}  
\usepackage{graphicx}
\usepackage{txfonts}
\usepackage{xspace}
\usepackage[T1]{fontenc}

\usepackage{xcolor}

\usepackage{soul}

\newcommand{\nnhp}[0]{N$_2$H$^+$\xspace}
\newcommand{\hcop}[0]{HCO$^+$\xspace}
\newcommand{\tco}[0]{$^{13}$CO\xspace}
\newcommand{\CO}[0]{$^{12}$CO\xspace}

\newcommand{\SSFR}[0]{$\Sigma_\mathrm{SFR}$\xspace}
\newcommand{\Sstar}[0]{$\Sigma_\ast$\xspace}
\newcommand{\Smol}[0]{$\Sigma_\mathrm{mol}$\xspace}
\newcommand{\Pde}[0]{$P_\mathrm{DE}$\xspace}
\newcommand{\vdisp}[0]{$\sigma_\mathrm{CO}$\xspace}

\usepackage{hyperref}
\begin{document}

    \title{The SWAN view of dense gas in the Whirlpool}
    \subtitle{A cloud-scale comparison of \nnhp, \hcop, HNC and HCN emission in M51}

\author{Sophia K. Stuber\inst{\ref{MPIA}\ref{UniHD}}
        \and Eva Schinnerer\inst{\ref{MPIA}}
        \and Antonio Usero\inst{\ref{OAN}} 
        \and Frank Bigiel\inst{\ref{AlfA}}
        \and Jakob den Brok\inst{\ref{CfA}}
        \and Jerome Pety\inst{\ref{IRAM},\ref{LUX}}
        \and Lukas Neumann\inst{\ref{ESO}}
        \and Mar\'ia~J.~Jim\'enez-Donaire\inst{\ref{STScI},\ref{OAN}}
        \and Jiayi Sun\inst{\ref{UniPr}\thanks{NASA Hubble Fellow}}
        \and Miguel Querejeta\inst{\ref{OAN}}
        \and Ashley.~T.~Barnes\inst{\ref{ESO}}
        \and Ivana Be\v{s}li\'{c}\inst{\ref{LUX}}
        \and Yixian Cao\inst{\ref{MPE}}
        \and Daniel~A.~Dale\inst{\ref{UWyoming}}
        \and Cosima Eibensteiner \inst{\ref{NRAO}}\thanks{Jansky Fellow of the National Radio Astronomy Observatory}
        \and Damian Gleis\inst{\ref{MPIA}}
        \and Simon C.~O.~Glover\inst{\ref{ITA}}
        \and Kathryn Grasha\inst{\ref{ANU}}
        \and Ralf~S.~Klessen\inst{\ref{ITA},\ref{IWR},\ref{CfA},\ref{Rad}}
        \and Daizhong Liu\inst{\ref{PMO}} 
        \and Sharon Meidt\inst{\ref{UGent}}
        \and Hsi-An Pan \inst{\ref{TKU}}
        \and Toshiki Saito\inst{\ref{Shizuoka}}
        \and Mallory Thorp\inst{\ref{AlfA}}
        \and Thomas~G.~Williams\inst{\ref{Ox}}
        }

\institute{
    Max-Planck-Institut für Astronomie, Königstuhl 17, 69117 Heidelberg Germany\label{MPIA}
    \and
    Fakultät für Physik und Astronomie, Universität Heidelberg, Im Neuenheimer Feld 226, 69120 Heidelberg, Germany\label{UniHD}
    \and 
    Observatorio Astron\'omico Nacional (IGN), C/ Alfonso XII, 3, E-28014 Madrid, Spain\label{OAN}
    \and 
    Argelander-Institut für Astronomie, Universität Bonn, Auf dem Hügel 71, 53121 Bonn, Germany\label{AlfA}
    \and 
    Center for Astrophysics $\mid$ Harvard \& Smithsonian, 60 Garden St., 02138 Cambridge, MA, USA\label{CfA}
    \and 
    IRAM, 300 rue de la Piscine, F-38406 Saint Martin d’H\`eres, France\label{IRAM}
    \and 
    LUX, Observatoire de Paris,
    PSL Research University, CNRS, Sorbonne Universités, 75014 Paris, France \label{LUX}
    \and 
    European Southern Observatory, Karl-Schwarzschild 2, 85748 Garching bei Muenchen, Germany\label{ESO}
    \and 
    AURA for ESA, Space Telescope Science Institute, 3700 San Martin Drive, Baltimore, MD 21218, USA\label{STScI}
    \and
    Department of Astrophysical Sciences, Princeton University, 4 Ivy Lane, Princeton, NJ 08544, USA\label{UniPr}
    \and 
    Max-Planck-Institut f\"{u}r extraterrestrische Physik, Giessenbachstra{\ss}e 1, D-85748 Garching, Germany\label{MPE}
    \and 
    Department of Physics \& Astronomy, University of Wyoming, Laramie, WY 82071\label{UWyoming}
    \and
    National Radio Astronomy Observatory, 520 Edgemont Road, Charlottesville, VA 22903, USA\label{NRAO}
    \and 
    Universit\"{a}t Heidelberg, Zentrum f\"{u}r Astronomie, Institut f\"{u}r Theoretische Astrophysik, Albert-Ueberle-Str.\ 2, 69120 Heidelberg, Germany\label{ITA}
    \and 
    Research School of Astronomy and Astrophysics, Australian National University, Canberra, ACT 2611, Australia\label{ANU}
    \and 
    Universit\"{a}t Heidelberg, Interdisziplin\"{a}res Zentrum f\"{u}r Wissenschaftliches Rechnen, Im Neuenheimer Feld 225, 69120 Heidelberg, Germany\label{IWR}
    \and 
    Radcliffe Institute for Advanced Study, Harvard University,  10 Garden St, 02138 Cambridge, MA, USA \label{Rad}
    \and 
    Purple Mountain Observatory, Chinese Academy of Sciences, 10 Yuanhua Road, Nanjing 210023, China\label{PMO}
    \and
    Sterrenkundig Observatorium, Universiteit Gent, Krijgslaan 281 S9, B-9000 Gent, Belgium\label{UGent}
    \and 
    Department of Physics, Tamkang University, No.151, Yingzhuan Road, Tamsui District, New Taipei City 251301, Taiwan \label{TKU}
    \and 
    Faculty of Global Interdisciplinary Science and Innovation, Shizuoka University, 836 Ohya, Suruga-ku, Shizuoka 422-8529, Japan\label{Shizuoka}
    \and 
    Sub-department of Astrophysics, Department of Physics, University of Oxford, Keble Road, Oxford OX1 3RH, UK\label{Ox}
}
   \date{July 18, 2025}

 
  \abstract{
  Tracing dense molecular gas, the fuel for star formation, is essential for the understanding of the evolution of molecular clouds and star formation processes. 
  We compare the emission of HCN~(1-0), HNC~(1-0) and \hcop(1-0) with the emission of \nnhp(1-0) at cloud-scales ($125\,$pc) across the central $5\times7\,$kpc of the Whirlpool galaxy, M51a, from "Surveying the Whirlpool galaxy at Arcseconds with NOEMA" (SWAN). 
  We find that the integrated intensities of HCN, HNC and \hcop are more steeply correlated with \nnhp emission compared to the bulk molecular gas tracer CO, and we find variations in this relation across the center, molecular ring, northern and southern disk of M51.  
  Compared to HCN and HNC emission, the \hcop emission follows the \nnhp emission more similarly across the environments and physical conditions such as surface densities of molecular gas, stellar mass, star-formation rate, dynamical equilibrium pressure and radius. 
  Under the assumption that \nnhp is a fair tracer of dense gas at these scales, this makes \hcop a more favorable dense gas tracer than HCN within the inner disk of M51.  
  In all environments within our field of view, even when removing the central 2\,kpc, HCN/CO, commonly used to trace average cloud density, is only weakly depending on molecular gas mass surface density. 
  While ratios of other dense gas lines to CO show a steeper dependency on the surface density of molecular gas, it is still shallow in comparison to other nearby star-forming disk galaxies. 
  The reasons might be  physical conditions in M51 that are different from other normal star-forming galaxies.
  Increased ionization rates, increased dynamical equilibrium pressure in the central few kpc and the impact of the dwarf companion galaxy NGC~5195 are proposed mechanisms that might enhance HCN and HNC emission over \hcop and \nnhp emission at larger-scale environments and cloud scales. 
    }

   \maketitle
%
\section{Introduction}
\label{sec:Introduction}

Star formation is one of the most fundamental processes in the Universe \citep[see reviews by][]{Krumholz_2014,Klessen_Glover_2015, schinnerer_molecular_2024}. 
As the birth of a star is ultimately linked to collapsing clouds of dense gas, the molecular gas phase is the key objective to study. 
Modern extragalactic observations \citep[e.g., surveys like PHANGS-ALMA;][]{leroy_phangsalma_2021, leroy_phangsalma_2021-1}, show that molecular clouds are linked sensitively to their galactic environment. As an example, a cloud’s velocity dispersion, surface density and virial state all vary depending on whether the cloud is located in the main disk or near the center, in a spiral arm or interarm region, or in a stellar bar \citep{querejeta_dense_2019,beslic_dense_2021,Neumann_2024A&A}.
This implies that the host galaxy impacts the initial conditions for star formation. However, how exactly those changes in cloud properties translate into the global pattern of star formation is still unsettled \citep[see recent review by ][]{schinnerer_molecular_2024}.
As stars must form out of the densest gas in molecular clouds \citep{gao_star_2004,wu_connecting_2005, lada_star_2012, evans_star_2014}, two of the key open questions in astronomy are a) how to reliably access those densest molecular regions observationally and b) how the properties of dense gas and therefore the star formation is influenced by larger-scale environmental processes.

Line emission of molecules such as CO~(1-0) is well suited for tracing the bulk molecular gas distribution in regions of high metallicity, such as local galaxy disks \citep[e.g.,][]{leroy_phangsalma_2021-1}.
In contrast to CO, the higher-dipole moment molecular species such as HCN, HNC and \hcop preferentially emit at higher physical densities. 
HCN has long been used to trace dense gas by the extragalactic community due to its relatively bright transition lines making it accessible in extragalactic targets \citep[e.g.,][]{helfer_dense_1993,aalto_variation_1997,aalto_cn_2002, gao_star_2004, meier_spatially_2005, aalto_detection_2012, bigiel_empire_2016, jimenez-donaire_empire_2019,querejeta_dense_2019, Bemis2019, krieger_molecular_2020,beslic_dense_2021,eibensteiner_23_2022, imanishi_dense_2023, neumann_almond_2023}.

Still, at sub-cloud scales, studies from the Milky Way remain inconclusive about whether HCN reliably traces only the actual star-forming gas phase \citep[e.g.,][]{pety_anatomy_2017, kauffmann_molecular_2017,  mills_origins_2017, barnes_lego_2020, tafalla_characterizing_2021, tafalla_characterizing_2023},  
as sub-thermal emission from low density regions can in some of these cases dominate the emission output.
Further, HCN emission can be excited by electrons in addition to collisional excitation by H$_2$ molecules \citep{goldsmith_electron_2017} and, e.g., \citet{santa-maria_2023} showed that low visual extinction gas can amount to about 30\% of the total HCN luminosity in Orion B. 
Recent numerical simulations of star-forming clouds in conditions similar to the local ISM \citep[e.g., ][]{Jones_2023,priestley_neath_2023, Priestley_2024} also suggest that a significant fraction of the HCN emission emanates from regions with densities of a few thousand cm$^{-3}$ that are unlikely to form stars.

An improved approach to gauging the gas density is therefore to contrast the emission of HCN with emission of a bulk molecular gas tracer that emits preferentially at lower densities than HCN. 
This is a promising probe of the average physical cloud density as shown by simulations \citep{leroy_cloud-scale_2017, neumann_almond_2023} and observations at $\gtrsim 300\,$pc resolution in nearby galaxies \citep[e.g.,][]{neumann_almond_2023, Neumann_2024A&A,schinnerer_molecular_2024, Neumann_JimenezDonaire2025AA}. 

An alternative to these lines is the molecular ion diazenylium (\nnhp), a tracer of dense gas due to chemical reasons. 
Within the Milky Way \nnhp(1-0) seems to be exclusively detected in dense and cold regions 
\citep[H$_2$ column densities above 10$^{22}$ cm$^{-2}$;][]{pety_anatomy_2017, kauffmann_molecular_2017, tafalla_characterizing_2021}. 
At these column densities, CO molecules start to freeze to dust grains in the coldest and densest parts of clouds, allowing \nnhp molecules thrive as reactions of \nnhp and CO and H$_3^+$ and CO are inhibited, which is the main destruction mechanism of \nnhp and the main mechanism limiting the formation of \nnhp out of H$_3^+$. 
After the onset of star formation, stellar feedback heats the dust and  evaporates the CO molecules, which increases the destruction of \nnhp. 
This makes \nnhp molecules selective only for certain cold and high density regimes \citep[i.e.,][]{tafalla_characterizing_2023, barnes_lego_2020}. 
Hydrodynamical simulations show, in contrast to HCN, HNC and \hcop, \nnhp exist mostly in regions dense enough that they irreversibly undergo gravitational collapse \citep{priestley_neath_2023}. 

To understand the interplay between larger-scale dynamical features in galaxies and star formation, it is crucial to observe and study the properties of dense gas not just in individual clouds but across entire ensembles of clouds within different environments, making local galaxies the ideal testbeds to study dense gas. 
Since \nnhp(1-0) emission is several order of magnitude fainter than \CO(1-0) \citep{schinnerer_molecular_2024}, and several times fainter than HCN~(1-0) emission \citep{jimenez-donaire_constant_2023}, extragalactic observations of \nnhp have so far been limited to either larger-scale lower-resolution studies \citep[i.e., kpc-scales, ][]{sage_toward_1995,den_brok_co_2022, jimenez-donaire_constant_2023}, or higher-resolution studies of individual regions such as starburst galaxy centers \citep[e.g., the centers of NGC~253, IC342 and NGC~6946;][]{martin_alchemi_2021, meier_spatially_2005, eibensteiner_23_2022}.

Surveying the Whirlpool galaxy at Arcseconds with NOEMA (SWAN), a IRAM-NOEMA+30m Large Program, therefore represents the next logical step in understanding the dense molecular gas: 
SWAN \citep{Stuber_2025} is a high-resolution (125\,pc), high-sensitivity survey, mapping the emission of several 3\,mm lines, including the J=1-0 transition of so-called dense gas tracers HCN, HNC, \hcop and \nnhp across the central $\sim5\times7\,$kpc$^2$ of the Whirlpool galaxy (NGC~5194 or M51a). 
As the fraction of dense gas (as traced by HCN) is found to vary drastically with galaxy environment 
at both kpc and cloud-scale resolutions \citep{usero_variations_2015, gallagher_dense_2018, jimenez-donaire_empire_2019, querejeta_dense_2019, Bemis2019,beslic_dense_2021, neumann_almond_2023}, it is vital to reliably study the denser molecular gas phase across different environments. 
With the SWAN survey, we can for the first time access not only the common dense gas tracer HCN, but also the chemical tracer \nnhp at unprecedented resolution in M51 across the inner spiral arms and interarm regions, molecular ring, nuclear bar and center with known AGN and outflow \citep{ho_search_1997, dumas_local_2011, querejeta_agn_2016}. 
M51a, also known as the Whirlpool galaxy or NGC~5194, (M51 hereafter) is a massive star-forming galaxy in the northern hemisphere, known for its iconic grand-design spiral arm structure, high surface brightness, low inclination and proximity. 
Many molecular line studies have targeted M51 in the past \citep{Koda_2011,schinnerer_pdbi_2013, watanabe_spectral_2014,chen_dense_2017,querejeta_dense_2019,den_brok_co_2022} and its complex dynamics, partially triggered by interactions with the dwarf galaxy M51b, are well studied  \citep{meidt_gas_2013, colombo_pdbi_2014, querejeta_agn_2016}.
We list the main properties of M51 in Table~\ref{tab:M51Overview}.

In \citet[][S23 hereafter]{stuber_surveying_2023} we presented the first high-resolution map of \nnhp across a larger-scale field of view (FoV) (5$\times$7\,kpc$^2$) in an extragalactic target, M51. 
As the formation of stars is also closely connected with dense molecular clouds \citep{schinnerer_molecular_2024}, we further probe the impact of surface densities of SFR (\SSFR) and stellar mass (\Sstar) on the dense gas emission. In addition, we test hypothesized links between dense gas mass, CO velocity dispersion (\vdisp), galactocentric radius and dynamical equilibrium pressure (\Pde). In a hydrostatic equilibrium, the midplane pressure \Pde determines the ability of the ISM to form molecular gas, and is expected to rise with a cloud's mean density \citep{jimenez-donaire_empire_2019}. 
At $125\,$pc scales, we find a super-linear correlation between \nnhp(1-0) and HCN(1-0) when excluding the AGN-affected center. Here, we expand this work to include emission of additional high-dipole molecules HNC and \hcop(1-0), and to study what sets the density distribution and cloud-scale chemistry across M51's environments. We will compare those emission lines to tracers of the surface density of molecular gas, as a proxy for the average gas density and other physical properties of the gas disk related to star formation.

Section~\ref{sec:Data} describes our observations and data reduction, followed by our methodology (Section~\ref{sec:Methods}) and a direct comparison of the  molecular emission lines with each other and how the emission compares to the bulk molecular gas tracer \CO (Section~\ref{sec:Results}). 
We then probe the dependency of dense gas tracing molecular emission with  various surface densities (Section~\ref{sec:ResultsEnv}).
We discuss the chemical composition and environmental dependency of the quiescent and star-forming dense gas in Section~\ref{sec:Discussion} and summarize our results in Section~\ref{sec:Summary}.

\begin{table}
\caption{Overview of the main properties of M51 (NGC~5194).}\label{tab:M51Overview}
\centering
\begin{tabular}{lll}
\hline
\hline
\noalign{\smallskip}
Property & Value & Source \\
\noalign{\smallskip}
\hline
\noalign{\smallskip}
 RA  &  13h29m52.7s & 1\\
 DEC &  $47^\circ11\arcmin 43\arcsec$ & 1\\
 PA &  173$\pm$3\,$^{\circ}$ & 2 \\
 $i$ &  22$\pm$5\,$^{\circ}$ & 2\\
 $D$ &  8.58  \,Mpc & 3\\
 v$_\mathrm{sys}$ & 471.7 $\pm$ 0.3 \,km/s & 4\\
 log$_{10}\, \mathrm{M}_\ast / \mathrm{M}_\odot$ & 10.5 & 5 \\
$\langle \Sigma_\mathrm{SFR} \rangle$ &  $20 \times 10^{-3}$ M$_\odot$\,yr$^{-1}$\,kpc$^{-2}$ & 5\\
 morphology & SA(s)bc & 6 \\
\noalign{\smallskip}
\hline
\end{tabular}
\tablefoot{Main parameters of M51a. 1: galaxy coordinates from NED, 2: position angle and galaxy inclination from \citet{colombo_pdbi_2014}, 3: distance from \citet{mcquinn_distance_2016}, 4: systemic velocity \citep{shetty_kinematics_2007}, 5: average SFR surface density and integrated stellar mass (within $0.75\times R_{25}$) derived from 3.6\,\textmu m \citep{den_brok_co_2022} 6: galaxy morphology \citep{deVaucouleurs_1991rc3}}
\end{table}

\section{Observations and Data}
\label{sec:Data}
\begin{figure*}
    \centering
    \includegraphics[width = 0.99\textwidth]{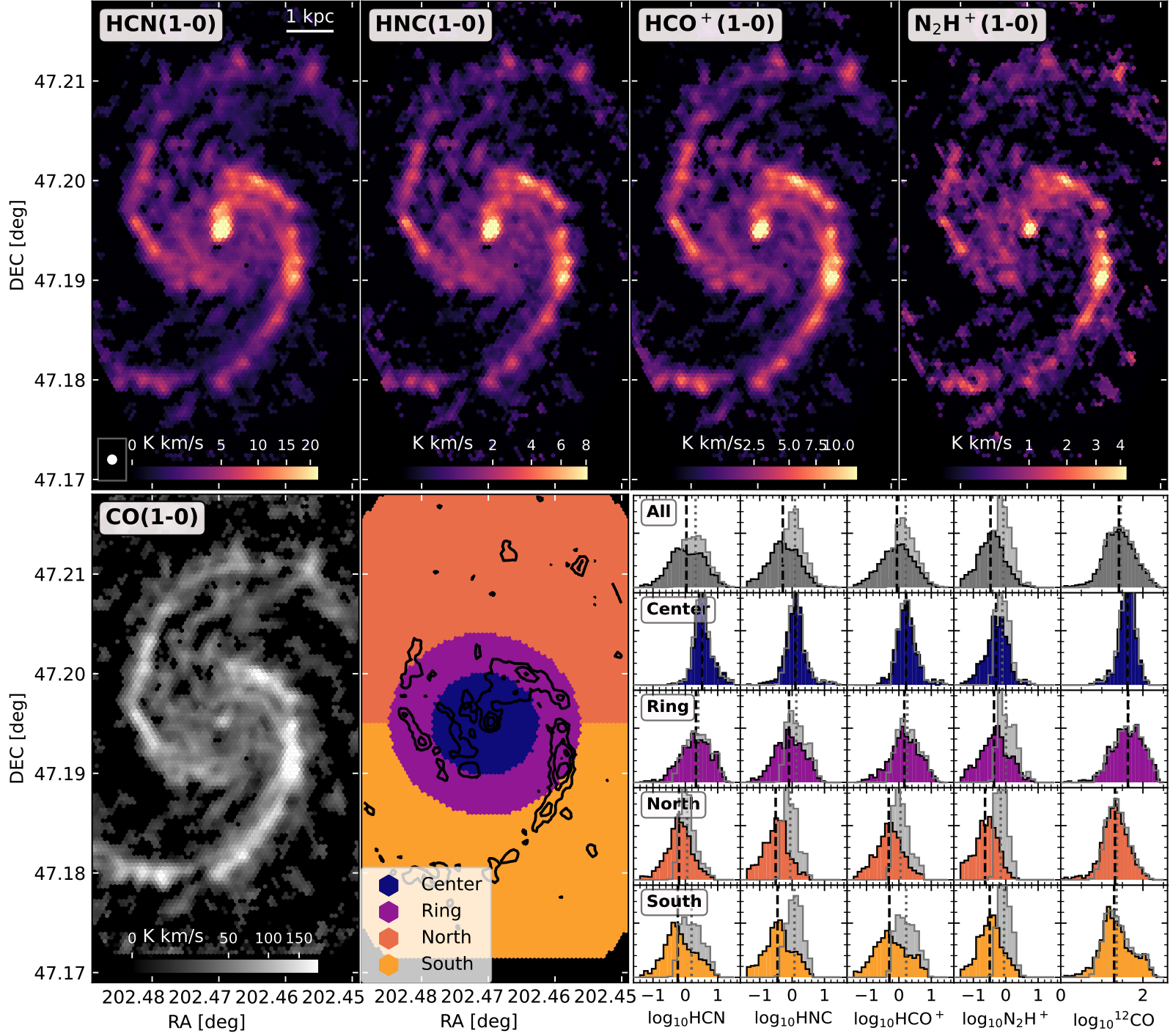}
    \caption{Integrated intensity maps of dense gas tracers HCN, HNC, HCO+ and \nnhp from SWAN at a common resolution of $3\arcsec$ (top row from left to right), as well as from \CO (PAWS, bottom left). We divide the disk into a center and ring environment \citep{colombo_pdbi_2014}, and the outer disk into northern and southern halves (bottom row, second panel from left). We add contours of integrated \nnhp emission of 0.75, 2 and 4\,K km/s to the environment map. 
 We show the pixel-based integrated intensity distribution (in K km/s) in various environments in the disk for all pixels in the FoV (colored shaded area), as well as for pixels where emission is detected (emission $>3\sigma$, light grey shaded area) in the bottom right panels. The area of each histogram is normalized to unity. 
 We indicate the median of all pixels (black dashed grey line) and median of masked pixels (dotted grey line) of each environment.  
 The median represents the median value of logarithmic emission (med$\left(\mathrm{log}_{10}\left( I \right) \right)$. Pixels with negative emission are excluded in the logarithmic scaling of the histograms.
 Since CO is detected across most of the FoV and is used as a prior in the creation of the moment-0 map (Section~\ref{sec:Data:SWANmoments}), its masked histogram distribution agrees well with the unmasked one.
 }
    \label{fig:Gallery}
\end{figure*}

To study the relation between galaxy environment and competing tracers of dense gas, we use the \nnhp, HCN, HNC and \hcop maps from the SWAN survey \citep[][;Section~\ref{sec:Data:SWAN}]{Stuber_2025} that cover the AGN and outflow, nuclear bar, molecular ring and inner spiral arms and interarm region in M51a at cloud-scales. 
We combine 3\,mm line emission observations with observations of bulk molecular gas tracer \CO(1-0) from the PdBI Arcsecond Whirlpool Survey \citep[PAWS,][]{schinnerer_pdbi_2013}. 
To test the physical conditions driving changes in the emission of dense gas tracing lines, we obtain surface density maps of SFR (\SSFR), molecular gas surface density (\Smol), stellar mass (\Sstar), as well as estimate the dynamical equilibrium pressure (\Pde) from ancillary data (Section~\ref{sec:Data:Ancillary}).
The utilized maps are presented in Figure~\ref{fig:Gallery}.

\subsection{Dense gas and CO observations}
\label{sec:Data:SWAN}

The SWAN survey combines observations from the IRAM large program LP003 (PIs: E. Schinnerer, F. Bigiel) that used $\sim214$ hours of observations with the Northern Extended Millimetre Array (NOEMA) and 69 hours of observations with the 30m single dish to map 3\,mm line emission from the central 5$\times$7\,kpc of the nearby galaxy M51. 
A more detailed description of this survey, the observations, data calibration, imaging and moment map creation can be found in a dedicated survey paper \citep{Stuber_2025}.
The NOEMA observations are performed by combining 17 pointings into a hexagonally spaced mosaic. 
The observations cover the $\sim85-110\,$GHz range allowing for simultaneous observations of 9 molecular lines: $^{13}$CO(J=1--0), HNCO(5--4), C$^{18}$O(1--0), \nnhp(1--0), HNC(1--0), \hcop(1--0), HCN(1--0), HNCO(4-3) as well as hyperfine transitions of C$_2$H(1--0). 
Calibration and joint deconvolution of NOEMA and 30m data were carried out using the IRAM standard calibration pipeline in GILDAS \citep{gildas_team_gildas_2013}.
The native resolution of the data varies from $\sim2.3\arcsec$ (for $^{13}$CO) to $3.0\arcsec$ (for HCN). 
The resulting data set has an rms of $\sim20\,$mK per 10\,km\,s$^{-1}$ channel for the brightest line ($^{13}$CO(1-0)).

For this study, we utilize the \nnhp, HCN, HNC and \hcop data cubes at 10\,km\,s$^{-1}$ spectral resolution per channel and native spatial resolution, which we convolve to a common spatial resolution of 3\arcsec ($\sim125\,$pc). 
For CO we directly use the PAWS data cube at 3\arcsec spatial resolution\footnote{Publicly available in the IRAM Data Management System \url{https://oms.iram.fr/oms/?dms=} or at the PAWS homepage \url{https://www2.mpia-hd.mpg.de/PAWS/PAWS/Data.html}}.

\subsection{Moment map creation}
\label{sec:Data:SWANmoments}

The creation of moment-0 maps is described in \citet{Stuber_2025}. 
We integrate the molecular data cubes along their spectral axis using the \textit{PyStructure} tool \citep{den_brok_co_2022,neumann_spectral_2023}. 
Creating moment maps is a common procedure to increase the signal-to-noise ratio (SNR), and to ease the comparison of the lines with each other and within the disk. 
These moment maps are created by using two selected priors, CO and HCN, to identify connected structures with emission above a selected SNR theshold of 2. This 3D signal mask is then used for integration. 
As described in \citet{Stuber_2025} this ensures that all emission is captured in the final maps, including emission from the center, where CO is comparably faint, in contrast to HCN. 
The emission of HCN, HNC, HCO$^+$, \nnhp and \CO is then integrated over these structures, ensuring that the
same pixels in the $ppv$ cube are used for integration for all lines. 
This approach differs slightly from the one used in S23, where only \CO was a prior for the mask instead of both CO and HCN. 
As described in \citet{Stuber_2025}, we interpolate and remove the area outside of the mosaic of pointings observed, as the noise increases towards the edges.   
The resulting maps have a resolution of 3\arcsec, and are regridded onto a hexagonal grid with four hexagonal spacings per beam (compare Figure~\ref{fig:Gallery}).

\subsection{Ancillary data}
\label{sec:Data:Ancillary}

\begin{figure*}
    \centering
    \includegraphics[width = 0.98\textwidth]{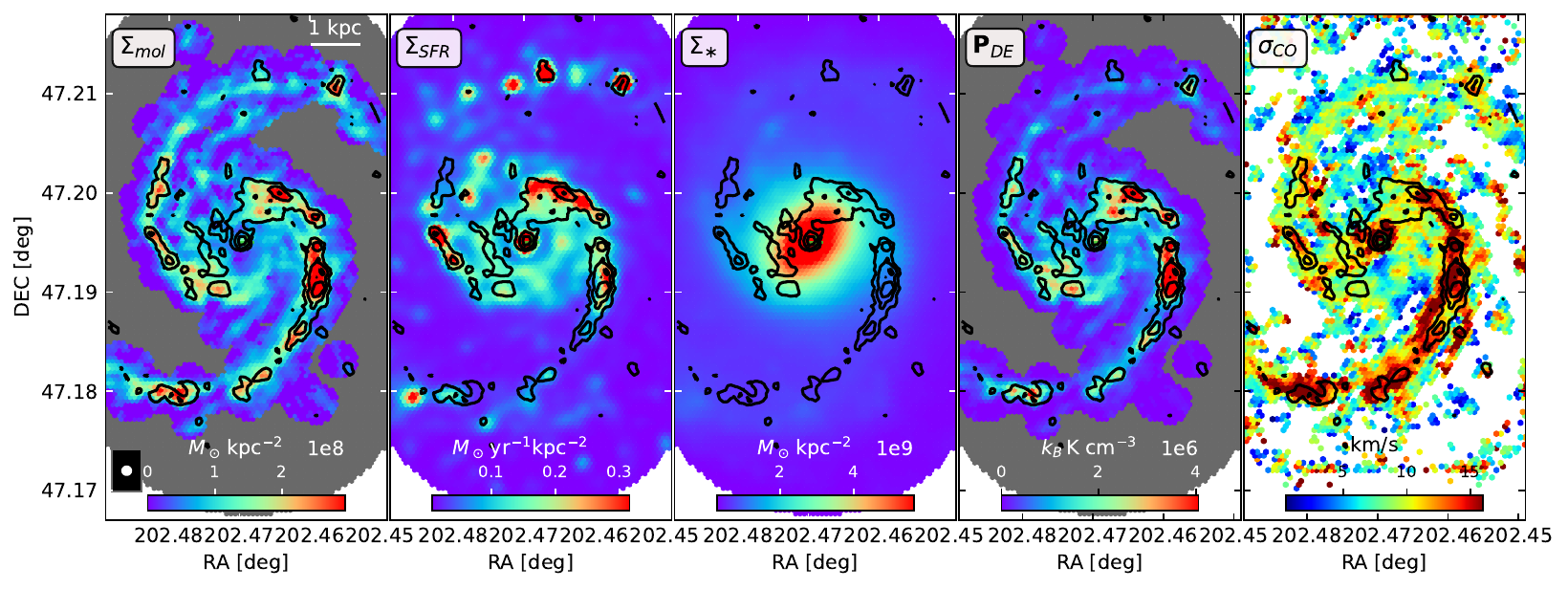}
    \caption{Molecular gas mass surface densities \Smol, star formation rate surface densities \SSFR,  stellar mass surface density $\Sigma_\ast$, dynamical equilibrium pressure P$_{DE}$ as well as CO velocity dispersion $\sigma_\mathrm{CO}$ at 3\arcsec\ resolution. We show contours of integrated \nnhp emission (0.75, 2,4 K km/s) on top.  }
    \label{fig:Gallery-Sigmas}
\end{figure*}

We use ancillary data to generate maps of SFR surface density (\SSFR), molecular gas mass surface density (\Smol), stellar mass surface density (\Sstar) and dynamical equilibrium pressure (\Pde) at a resolution of 3$\arcsec$. 
Those maps are incorporated into the PyStructure table and share the same hexagonal pixel grid as the molecular gas maps. 
All ancillary data are shown in Figure~\ref{fig:Gallery-Sigmas}.

\subsubsection{SFR tracers}
\label{sec:Data:SFR}

We combine the \textit{Spitzer} 24\,\textmu m map processed by \citet{dumas_local_2011}, tracing recent star formation obscured by dust with $3\arcsec$ resolution H$\alpha$ maps from \citet{kessler_pa_2020} tracing recent emission from HII regions powered by massive young stars.  
The 24\textmu m map is deconvolved with the so-called \textit{HiRes} algorithm \citep{Backus_2005ASPC} to achieve a resolution of $2-3\arcsec$ \citep{dumas_local_2011}. 
We obtain a map of SFR surface density ($\Sigma_\mathrm{SFR}$) via linear combination of the above mentioned SFR maps as described in equation 6 from \citet{leroy_molecular_2013} and correcting for inclination:

\begin{equation}
\begin{split}
    \Sigma_\mathrm{SFR}[M_\odot\, \mathrm{yr}^{-1}\,\mathrm{kpc}^{-2}] = [
    634\, I_{H\alpha} [\mathrm{erg}\,\mathrm{s}^{-1}\, \mathrm{sr}^{-1}\,\mathrm{cm}^{-2}] \\\ + 0.00325\,I_\text{24\textmu m} [\mathrm{MJy}\, \mathrm{sr}^{-1}] ] \times \mathrm{cos}(i)\;.
\end{split}
\end{equation}

\subsubsection{Molecular gas mass surface density and CO(1-0) velocity dispersion}
\label{sec:Data:CO}
We utilize observations of \CO(1-0) from the PAWS survey \citep{schinnerer_pdbi_2013}. The 1\arcsec\ resolution data are convolved to 3\arcsec\ spatial and 10\,km/s spectral resolution to match our observations.
Applying a CO-to-H$_2$ conversion factor  $\alpha_\mathrm{CO}$  map to the integrated \CO intensity ($I_\mathrm{CO}$) and correcting for inclination $i$ yields the molecular gas mass surface density map via: 

\begin{equation}\label{eq:sigma_mol}
    \Sigma_\mathrm{mol} = I_\mathrm{CO} \times \alpha_\mathrm{CO} \times \mathrm{cos}\left(i\right)
\end{equation}
The spatially varying $\alpha_\mathrm{CO}$ is estimated based on modeling the observed \CO(1-0), (2-1) and \tco (1-0), (2-1) lines from SWAN and SMA imaging \citep{den_Brok2025}. 
The measured values at $4\arcsec$ resolution are extrapolated with a Gaussian process regression towards neighboring pixels to cover a larger area  (compare area where \Smol is shown in Figure~\ref{fig:Gallery-Sigmas}).
We test the impact of using different $\alpha_\mathrm{CO}$ prescriptions on our results in Appendix~\ref{App:alphaCO}, including \Smol calculated with a constant $\alpha_\mathrm{CO}$ and with a metallicity and stellar mass based $\alpha_\mathrm{CO}$. 

We further estimate the CO velocity dispersion following the prescription of \citet{heyer_equilibrium_2001}, applied in, e.g.,\citet{beslic_dense_2021,neumann_almond_2023}.
This method estimates the 'effective width' of a line via the integrated intensity $I_\mathrm{CO}$ (a.k.a. moment-0) and the peak intensity ($T_\mathrm{peak}$) at each pixel.  
Assuming a Gaussian line profile for these spectra with peak $T_\mathrm{peak}$, we can estimate the rms velocity dispersion of the line $\sigma_\mathrm{measured}$ via: 
\begin{equation}
    \sigma_\mathrm{measured} = \frac{I_\mathrm{CO}}{\sqrt{2\pi} T_\mathrm{peak}}\;.
\end{equation}

To account for line broadening caused by the instrument, we subtract the instrumental contribution $\sigma_\mathrm{instrument}$ in quadrature \citep{rosolowsky_biasfree_2006, sun_cloud-scale_2018}. 
For M51, we use $\sigma_\mathrm{instrument}$ estimates for CO from \citet{sun_cloud-scale_2018} at 120\,pc resolution (see their equations 16, 17 and Table 9).

\subsubsection{Stellar mass map}
\label{sec:Data:Sigma_star}
We obtain stellar mass surface density maps ($\Sigma_\ast$) based on \textit{Spitzer} 3.6\,\textmu m maps from the \textit{Spitzer} Survey of Stellar Structure in galaxies 
\citep[S$^4$G;][]{sheth_spitzer_2010} that are further corrected for dust emission \citep{querejeta_spitzer_2015}. A more detailed description of this process and the obtained stellar mass map for M51 can be found in \citet{querejeta_dense_2019}.  
We convolve this map from 2.4\arcsec\ resolution to our common resolution of 3\arcsec, and apply an inclination correction to obtain surface densities.

\subsubsection{Dynamical Equilibrium Pressure}
\label{sec:Data:PdeVdisp}
The dynamical equilibrium pressure (\Pde) describes the ambient midplane pressure acting on the molecular gas disk, keeping it in vertical equilibrium and is set by the weight of the ISM in a galaxy's gravitational potential  
\citep{Elmegreen_1989,Ostriker_2010,field_2011, sun_dynamical_2020}. 
We estimate the ISM pressure in each individual, cloud-scale region following the prescription from \citet[][equation 16]{sun_dynamical_2020}. This ``cloud-scale dynamical equilibrium pressure'' formalism accounts for the weight of the gas from not only molecular cloud self-gravity but also external gravity by stars and gas across the galactic disk, as expressed below: 
\begin{equation}\label{eq:PDE}
    P_{\rm DE} = \frac{3 \pi G}{8} \Sigma_\mathrm{mol,\theta}^2 + \frac{\pi G}{2} \Sigma_\mathrm{mol,\theta} \Sigma_\mathrm{mol,\,kpc} + \frac{3 \pi G}{4} \rho_\mathrm{\ast,\,kpc} \Sigma_\mathrm{mol,\theta} D_\mathrm{cloud} \;.
\end{equation}
Here, $\Sigma_\mathrm{mol,\theta}$ is the molecular gas surface density at 3\arcsec\ resolution (as derived in Section~\ref{sec:Data:CO}); $\Sigma_\mathrm{mol,\,kpc}$ is the corresponding kpc-scale surface density, derived by convolving CO data from 3\arcsec\ to 1\,kpc resolution. $D_\mathrm{cloud}$ is the adapted cloud diameter, for which we use the beam size of 125\,pc assuming a single cloud fills each beam.
$\rho_\mathrm{\ast,kpc}$ is the stellar mass volume density near the disk mid-plane, which we estimate from the kpc-scale stellar mass surface density $\Sigma_\mathrm{\ast,\,kpc}$ 
and stellar disk scale height $H_\ast$:

\begin{equation}
    \rho_\ast = \frac{\Sigma_\mathrm{\ast,\,kpc}}{4 \times H_\ast}\;.
\end{equation}

Based on \citet{Kregel2002} and \citet{sun_dynamical_2020}, we estimate $H_\ast$ from the stellar disk radial scale length $R_\ast = 7.3 \times H_\ast$. For M51 we adopt $R_\ast= 4.0 \pm 0.2$\,kpc from \citet{dumas_local_2011}, adapted to our distance  of $D =8.6$\,Mpc, while \citet{dumas_local_2011} used $D=8.2$\,Mpc.

\section{Methodology and simulated data tests}
\label{sec:Methods}

We describe our methodology to obtain average intensities, average line ratios, as well as deriving relations between line intensities and line ratios and other galaxy parameters such as surface densities. 
To ensure that our results are not impacted by the different SNR of our dense gas tracing lines, we test all of the methods on a simple mock-data set created based on the \CO data, which we describe in more detail in Appendix~\ref{sec:Mockdata}. 
In short, we scale the \CO intensity in the CO data cube at 3\arcsec\ resolution to lower values. We use four different scaling factors, chosen to be plausible estimates of our faintest observed line (\nnhp(1-0)) from literature results. Those scaling factors further match our average HNC-to-\CO, \hcop-to-\CO and \nnhp-to-\CO ratios (Table~\ref{tab:Averagelineratios}) plus an even smaller scaling factor to test the effect of even lower SNR. 

For each scaled CO intensity we adjust the noise to simulate the corresponding SNR. 
The mock-data cubes are all then treated in the exact same way that we are treating our science data (Section~\ref{sec:Data:SWANmoments}), including the integration into the PyStructure to obtain moment maps that we can use for comparison (for more details see Appendix~\ref{sec:Mockdata:methods}). 

To investigate potential environmental variations, we divide the disk of M51 into four regimes: A kinematically determined center and molecular gas-rich ring based on \citet{colombo_pdbi_2014-1} and sub-dividing the remaining disk into northern (north of the galaxy center) and southern  (south of the galaxy center, see Figure~\ref{fig:Gallery}) halves. 
To highlight the importance of separating the environments, we note that the \CO distribution reveals a brightness asymmetry in gas emission between the fainter northern and brighter southern spiral arm which is speculated to be caused M51b \citep[e.g.,][]{Egusa_2017}. The center, which hosts both nuclear bar, AGN and low-inclined radio jet \citep{matsushita_resolving_2015, querejeta_agn_2016} also provides physical conditions very different from the molecular ring and spiral arms further out.  
In addition to the full FoV, we will apply the methods mentioned in this Section to the individual environments throughout this work.

\subsection{Determination of average line ratios}
\label{sec:Methods:Lineratios}

Average global and regional line ratios between the emission of two lines are an effective way to measure the underlying physical conditions of the observed region, while reducing the impact of noise or peculiar outliers. 
We calculate average line ratios between different lines ($I_\mathrm{line1}$, $I_\mathrm{line2}$)
by using the ratio of the integrated intensities in the full FoV.  We integrate the intensity of all pixels in the moment-0 map within the FoV for the first line (line1) and divide by the integrated intensity of all pixels in the moment-0 map within the FoV for the second line (line2): 

\begin{equation}\label{eq:averagelr}
    R^\mathrm{line1}_\mathrm{line2} = \mathrm{Sum}\left( I_\mathrm{line1} \right) / \mathrm{Sum}\left( I_\mathrm{line2} \right)\;.
\end{equation}

Statistical uncertainties are then calculated following standard Gaussian error propagation. 
Our tests with the mock data (Appendix~\ref{sec:Mockdata:intensities}) confirm that these methods robustly recover the expected line ratios. 
Masking will bias specifically data sets with lower SNR compared to those with higher SNR, and so do the other tested methods (median and mean line ratios, Appendix~\ref{sec:Mockdata:intensities}).

\subsection{Binning the data}
\label{sec:Methods:Binning}

Averaging data in increments of galactic parameters (e.g., surface density) is a useful technique (referred to as ``binning'' hereafter) to reduce the noise (i.e., negative and positive noise will cancel on average) and recover possible physical relations between intensity and those galactic parameters. 
To bin our data, we select bins of property $A$ ranging from three times its average uncertainty ($\Delta A$) up to the maximum value of $A$.  
For each bin, we first select the pixels that fall within the bin range and average the intensity of a line corresponding to those pixels, including non-detections ($<3\sigma$).
We calculate average line ratios $R$ (both CO-line ratios and line ratios of dense gas lines)
analogue to Equation~\ref{eq:averagelr}. 
As shown in Appendix~\ref{sec:Mockdata}, excluding pixels will introduce biases \citep[see also ][]{neumann_spectral_2023}. 

In Figure~\ref{fig:Mockdata-binningtests} we show that with the method described above and by including all pixels we can recover the expected relation for this simple mock-data model. 
To ensure that the binned intensity measurements are not predominantly noise, we remove measurements where the binned average is below 5 times the corresponding statistical uncertainty. The statistical uncertainty is much lower than the average noise per bin (see Appendix~\ref{sec:Mockdata}). We do not mask any data prior to the binning.

For binned intensities we provide the statistical uncertainty from error propagation. 
For binned line ratios $R = M_1/M_2$, the uncertainty is calculated as follows: 
\begin{equation}
    \Delta R = | R | \cdot \sqrt{\left(\frac{\Delta M_1}{M_1}\right)^2 + \left(\frac{\Delta M_2}{M_2}\right)^2 } \;.
\end{equation}

With $\Delta M_i$ being the relative error of binned intensity of line $i=1,2$, respectively. It is calculated following standard Gaussian error propagation $\Delta M_i = 1/N  \sqrt{ \Sigma_{a=1}^N \Delta s_a^2 }$, for pixels $a=1...N$ and the error of the intensity at each pixel being $s_a$. 
For ratios between fainter dense gas tracing lines, we further provide the 25th and 75th percentile range as visual guidance for each relation in addition to the statistical uncertainty presented above. 

In our analysis, we want to find simple correlations between our observed line emission or line ratios and physical quantities to identify possible driving mechanisms. 
To quantify the slope and scatter of some of the observed trends, we fit a linear function of shape $a \times x + b$ (in log space) to the binned data with \texttt{scipy-optimize}, including the bin uncertainties. 
The scatter is then measured as the median absolute deviation of the residual intensity (or line ratio) within the fitted range after subtracting the best-fit relation. 
This scatter depends on the average SNR of a line and is used to compare how the scatter in each line's intensity varies with different physical galactic parameters, rather than between different lines.
As this scatter is measured with respect to a linear relation, it measures both a non-linear behavior indicative of secondary dependencies, as well as the intrinsic uncertainty.

\section{Line emission of dense gas tracers in comparison}
\label{sec:Results}

\begin{table*}
\begin{small}
\caption{Average line ratios of dense gas tracers}\label{tab:Averagelineratios}
\centering
\begin{tabular}{lccccccc}
\hline\hline
\noalign{\smallskip}
line ratio & all & center & ring & north & south\\
\noalign{\smallskip}
\hline 
\noalign{\smallskip}
HCN/HNC & $2.729 \pm 0.013 $ & $3.007 \pm 0.014 $ & $2.583 \pm 0.013 $ & $2.612 \pm 0.056 $ & $2.665 \pm 0.048 $ \\
HCN/\hcop & $1.530 \pm 0.005 $ & $2.213 \pm 0.009 $ & $1.391 \pm 0.005 $ & $1.363 \pm 0.019 $ & $1.133 \pm 0.012 $ \\
HNC/\hcop & $0.560 \pm 0.003 $ & $0.736 \pm 0.004 $ & $0.538 \pm 0.003 $ & $0.522 \pm 0.012 $ & $0.425 \pm 0.008 $ \\
\noalign{\smallskip}
\hline 
\noalign{\smallskip}
HCN/\nnhp & $5.991 \pm 0.053 $ & $8.512 \pm 0.097 $ & $5.955 \pm 0.061 $ & $6.36 \pm 0.282 $ & $3.437 \pm 0.074 $ \\
HNC/\nnhp & $2.195 \pm 0.021 $ & $2.83 \pm 0.034 $ & $2.306 \pm 0.025 $ & $2.435 \pm 0.116 $ & $1.290 \pm 0.034 $ \\
\hcop/\nnhp & $3.917 \pm 0.036 $ & $3.846 \pm 0.045 $ & $4.283 \pm 0.044 $ & $4.667 \pm 0.21 $ & $3.033 \pm 0.066 $ \\
\noalign{\smallskip}
\hline 
\noalign{\smallskip}
HCN/\CO & $0.0504 \pm 0.0001 $ & $0.1153 \pm 0.0003 $ & $0.0558 \pm 0.0001 $ & $0.0273 \pm 0.0002 $ & $0.0274 \pm 0.0002 $ \\
HNC/\CO & $0.0185 \pm 0.0001 $ & $0.0384 \pm 0.0002 $ & $0.0216 \pm 0.0001 $ & $0.0104 \pm 0.0002 $ & $0.0103 \pm 0.0002 $ \\
\hcop/\CO & $0.0330 \pm 0.0001 $ & $0.0521 \pm 0.0002 $ & $0.0401 \pm 0.0001 $ & $0.0200 \pm 0.0002 $ & $0.0242 \pm 0.0002 $ \\
\nnhp/\CO & $0.0084 \pm 0.0001 $ & $0.0136 \pm 0.0002 $ & $0.0094 \pm 0.0001 $ & $0.0043 \pm 0.0002 $ & $0.0080 \pm 0.0002 $ \\
\hline 
\end{tabular}
\tablefoot{Average line ratios for all combinations of dense gas tracers (J=1--0 transitions) and \CO(1-0) for the full FoV, and for various galactic environments (center, ring, northern and southern disk, see Figure~\ref{fig:Gallery}). Averages are calculated 
as the ratio of the summed intensities with standard Gaussian error propagation as statistical uncertainty for line ratios between combinations of dense gas lines (Section~\ref{sec:Methods:Lineratios}).}
\end{small}
\end{table*}

HCN, HNC and \hcop are the favored tracers of dense gas in the extra-galactic community, as they are much brighter in emission than, for example, \nnhp and have higher critical densities than, e.g., CO \citep[e.g., see table~4 of][]{schinnerer_molecular_2024}. 
Therefore, this theoretically allows them to trace regions of denser gas. They all correlate well with star formation rate surface densities on galactic scales \citep{gao_star_2004}. 
To study the effect that the environment within M51 might have on these molecular lines, we compare their emission at  cloud-scales (Section~\ref{sec:Lineint_perpix}) and within larger scale galactic environments (northern and southern disk, ring, center,  Figure~\ref{fig:Gallery}, Section~\ref{sec:Lineint_perEnv}).
Comparing the emission of those dense gas tracing lines with emission of the Milky Way preferred dense gas tracer \nnhp will help understand their ability to trace dense regions. 
Lastly, we compare ratios of all dense gas tracers with \CO, which are found to be density-sensitive at $\gtrsim100\,$pc scales \citep{leroy_cloud-scale_2017,jimenez-donaire_empire_2019,usero_variations_2015}.

\subsection{Distribution of HCN, HNC, \hcop, \nnhp and \CO across environments}
\label{sec:Lineint_perEnv}

We show the emission of HCN, HNC, \hcop, \nnhp and \CO in the central disk of M51 in Figure~\ref{fig:Gallery}. 
In our FoV, HCN, HNC, \hcop and \nnhp are on average $\sim29, 71, 38, 142$ times fainter than \CO emission (Table~\ref{tab:Averagelineratios}) and their emission is well correlated  with each other (Spearman correlation coefficients all $\rho_\mathrm{Sp}>0.54$), while their intensity probes about 2 orders of magnitude. 

Figure~\ref{fig:Gallery} illustrates that all dense gas tracing lines are bright in the central few $\sim100$\,pc, and along the molecular ring (Figure~\ref{fig:Gallery}, bottom right). 
A peculiar bright region, where \nnhp is brightest, is located at the south-western brink of the molecular ring (compare also S23). 
\CO is distributed in the FoV in a similar fashion to the dense gas tracers, yet its emission is less enhanced in the galaxy center compared to the disk than is seen for the dense gas tracing lines. 
Average line ratios among dense gas tracing lines and \CO (Table~\ref{tab:Averagelineratios}) are enhanced in the central environment compared to other environments (factor of $\sim1.6-2.3$ from \nnhp to HCN between center and full FoV).
Further, all line ratios among the dense gas tracers  (HCN/HNC, HCN/\hcop, HNC/\hcop, HCN/\nnhp, HNC/\nnhp, \hcop/\nnhp) vary significantly (>$3\sigma$) between the central environment and other environments.  All line ratios are increased in the center compared to other regions, except the \hcop/\nnhp line ratio, which is decreased. 

While we see an asymmetry between a brighter southern and fainter northern arm in all lines, the overall intensity distribution of all molecules in the entire northern and southern disk match well (Figure~\ref{fig:Gallery}, bottom right panel). 
Still, average dense gas line ratios only agree between ring and northern disk environment, while line ratios in the southern disk are decreased  (except the HCN/HNC ratio) by up to a factor of $\sim1.9$ compared to both ring and northern disk (Table~\ref{tab:Averagelineratios}). 
Line ratios with CO are increased  in the southern disk compared to northern disk for \nnhp (factor $\sim2$) and \hcop (factor $\sim1/2$).

\subsection{Cloud-scale comparison of HCN, HNC, \hcop, \nnhp and \CO}
\label{sec:Lineint_perpix}

Larger-scale environments within galaxies are found to impact the physical conditions \citep{sun_molecular_2022, schinnerer_molecular_2024} and thus emission pattern of molecules.  Still, the exact mechanisms driving these large-scale changes, which we also see within M51 (Section~\ref{sec:Lineint_perEnv}) as well as variations within those environments are not well understood. 
In this Section, we analyze the cloud-scale variations within M51's environments by comparing emission from the extragalactic dense gas tracers HCN, HNC and \hcop with each other (Section~\ref{sec:hcnhnchcop}) and with the much fainter dense gas tracer \nnhp (Section~\ref{sec:linearityn2hp}). 
Cloud-scale variations of density sensitive line ratios with \CO are analyzed in Section~\ref{sec:linearityCO}.

\subsubsection{Correlation between HCN, \hcop, and HNC}
\label{sec:hcnhnchcop}

\begin{figure*}[t]
    \centering
    \includegraphics[width = 0.99\textwidth]{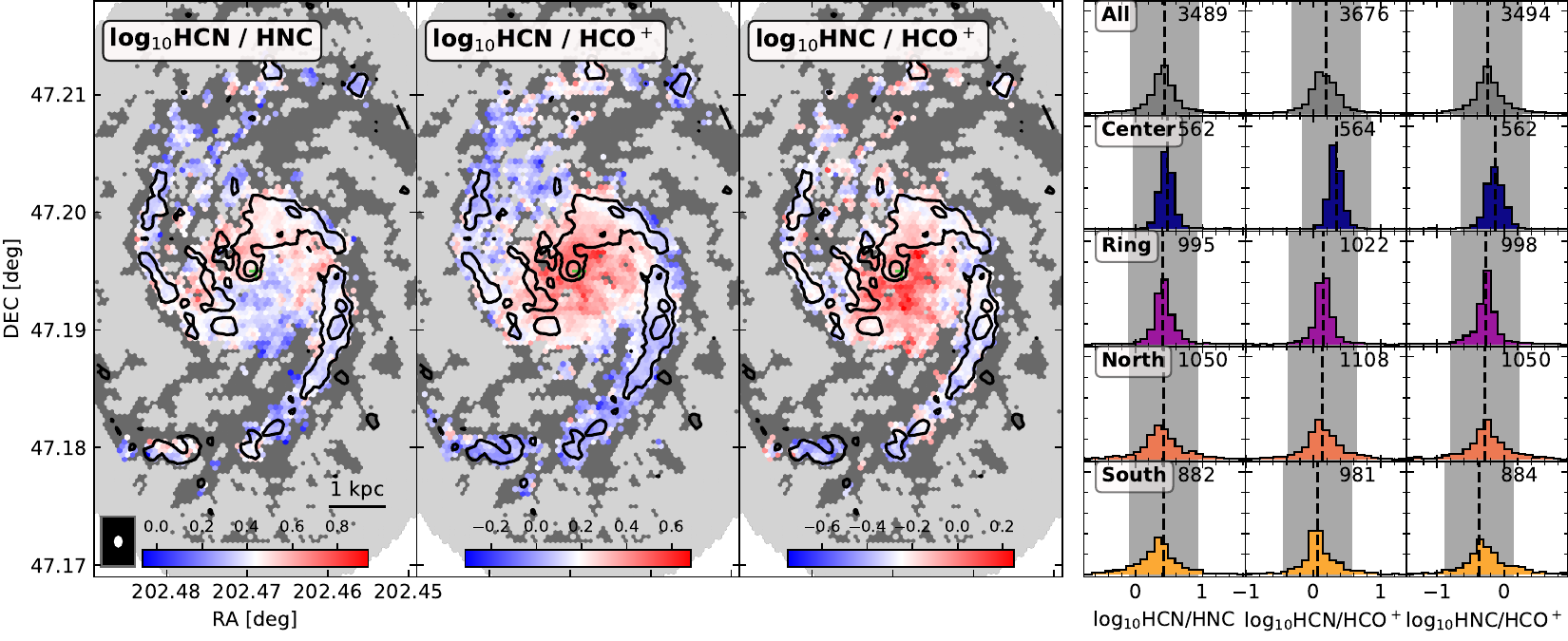}
    \caption{Left panel: Line ratios of integrated line emission from dense gas tracers HCN, HNC and \hcop. For visual purpose, we only show line ratios for significant detected pixels ($>3\sigma$), but include non-detections in all calculations. We mark pixels in which CO is detected (grey points) and the center of the galaxy (green plus). The intensity scale is centered logarithmically on the average line ratios (log$_{10} R$, with $R$ from Table~\ref{tab:Averagelineratios}) determined for all pixels in the FoV, including non-detections.  The average line ratio (log$_{10} R$) and the total range of 1\,dex covered by the color bar are indicated by the black dashed line and greyshaded area in the right panel. We show contours of integrated \nnhp emission (0.75, 3 K km/s) on top. 
    Right panel: Histogram of line ratios per environment analogous to Figure~\ref{fig:Gallery} but for HCN/HNC, HCN/\hcop and HNC/\hcop (colored histograms). 
    We indicate the amount of pixels shown in the histogram (top right corner), which varies slightly, as values with negative noise can not be shown in the logarithmic scale. 
    }
    \label{fig:Gallery_linerationon2hp}
\end{figure*}

In Figure~\ref{fig:Gallery_linerationon2hp} we display the cloud-scale distribution of line ratios HCN/HNC, HCN/\hcop and HNC/\hcop 
where for visual purpose only line ratios for detected sightlines ($>3\sigma$) are shown in the maps.
Of all line combinations, the HCN/HNC line ratio has the smallest variations within the disk ($\pm 0.2$\,dex to the global average). It is slightly increased in the northern disk and molecular ring compared to the southern half and spiral arms. 

The line ratios with \hcop exhibit larger variations ($\pm0.2-0.4$\,dex to the global average). HCN/\hcop increases azimuthally symmetrically in the center out to a radius of $\sim1\,$kpc. The line ratio in the southern arm is smaller than the global average, but shows little variations, while there are $\sim$ beam-sized variations in the line ratios in the northern arm. 
Similarly to the HCN/\hcop ratio, the HNC/\hcop ratio is increased in the center relative to the global average and spiral arms, with a slightly stronger azimuthally asymmetric increase towards the south-western half of center and molecular ring up to a radius of $\sim1.5\,$kpc, in accordance with the asymmetric HCN/HNC distribution. Beam-sized variations in the HNC/\hcop ratio are seen across the entire disk. 

In summary, both HCN and HNC are enhanced in the central $\sim1-1.5$\,kpc compared to \hcop and compared to the spiral arms and global average.

\subsubsection{Comparing line intensities to \nnhp}
\label{sec:linearityn2hp}

\begin{figure*}
    \centering
    \includegraphics[width = 0.99\textwidth]{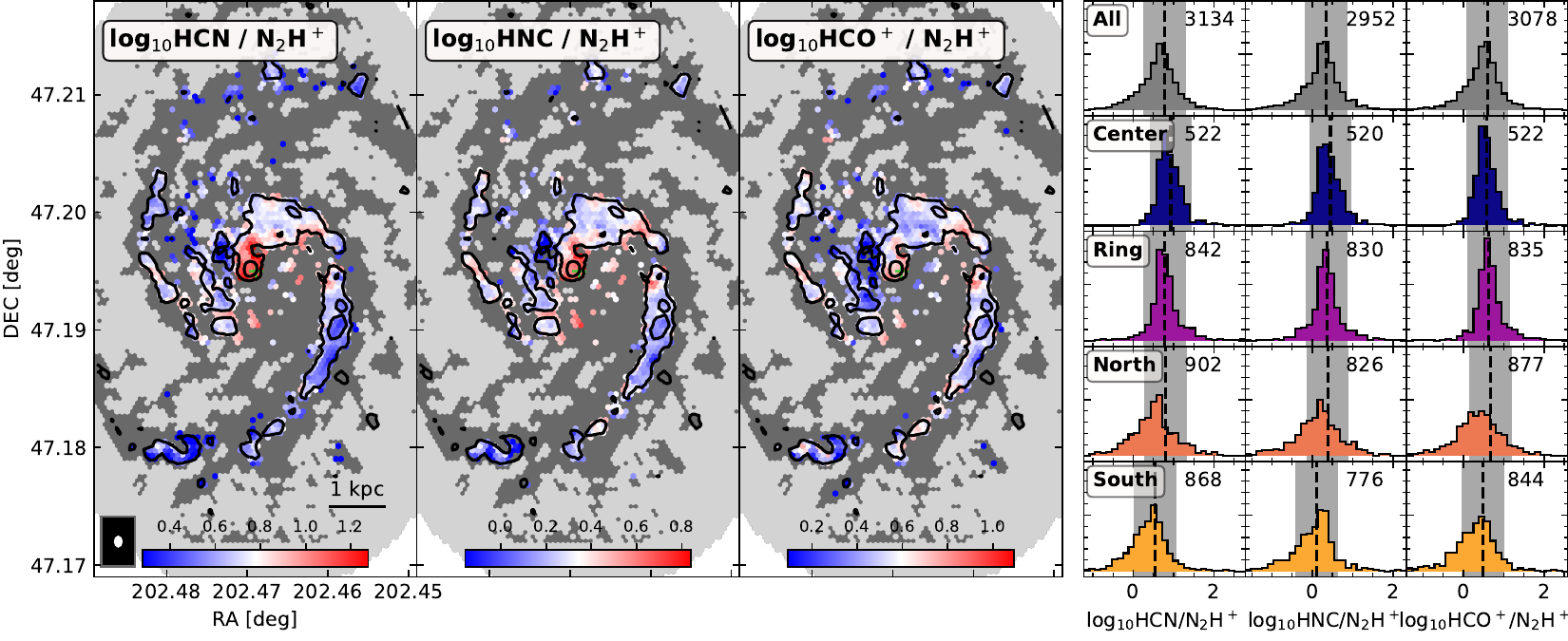}
    \caption{Same as Figure~\ref{fig:Gallery_linerationon2hp} but for the HCN-to-\nnhp, HNC-to-\nnhp and \hcop-to-\nnhp line ratios. 
    The colorbar spans the same 1\,dex range as in Figure~\ref{fig:Gallery_linerationon2hp}, and is centered logarithmically on the average line ratios (log$_{10} R$, with $R$ from Table~\ref{tab:Averagelineratios}).
    }
    \label{fig:Gallery_lineration2hp}
\end{figure*}

Unlike HCN, HNC and \hcop, emission from the molecular ion \nnhp has been the favored tracer of dense regions in Milky Way clouds \citep[e.g.,][]{kauffmann_molecular_2017, pety_anatomy_2017}.
We show the spatial distribution of line ratios of HCN, HNC and \hcop with \nnhp emission in Figure~\ref{fig:Gallery_lineration2hp}. 
The intensity scale depicts the same range (in dex) around the average line ratio as in Figure~\ref{fig:Gallery_linerationon2hp}. 
The variations seen in the \nnhp-line ratios are larger ($\gtrsim 0.5\,$dex around the global average) than the variations seen among the brighter dense gas lines (e.g., HCN/HNC).

All \nnhp line ratios increase compared to the global average in a region north-east of the center of a few $\sim100$\,pc in size, and decrease at even larger radii east of the center. The increase in line ratios is strongest for HCN/\nnhp ($\sim0.5$dex) and weakest for \hcop/\nnhp ($\sim0.1$dex) compared to the global average. 
\nnhp is not detected in the south-western half of the center and molecular ring, where the HNC/\hcop line ratios are increased. 

\nnhp line ratios are decreased in the spiral arms in pixels where \nnhp is significantly detected  compared to the global average. Again, the strongest variations in line ratios are seen for HCN/\nnhp ($\pm0.5$\,dex across many pixels), the smallest for \hcop/\nnhp (mostly $\pm0.2$\,dex with only few pixels with larger line ratios).

\subsubsection{Proposed average gas density tracer: Line ratios with CO}
\label{sec:linearityCO}

\begin{figure*}
    \centering
    \includegraphics[width = \textwidth]{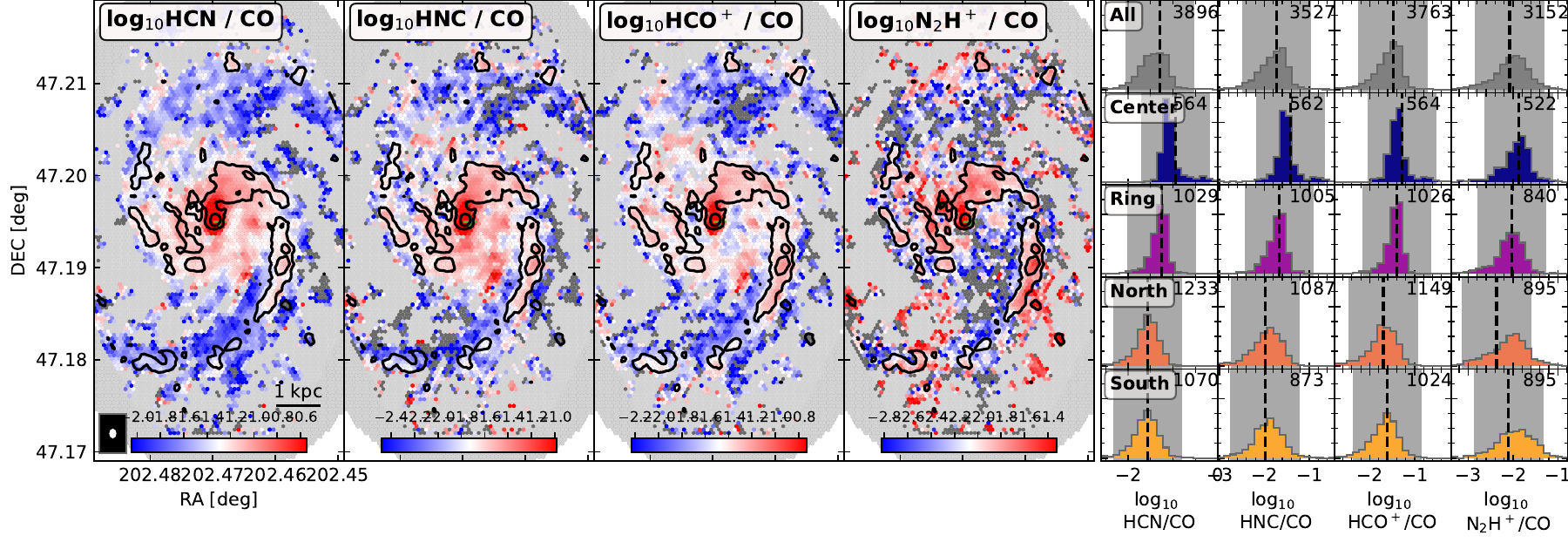}
    \caption{Same as Fig.~\ref{fig:Gallery_linerationon2hp} and ~\ref{fig:Gallery_lineration2hp} but for line ratios with CO. In contrast to the previous Figures, the colorbar spans a larger range of 1.5\,dex, centered on the average line ratios (Table~\ref{tab:Averagelineratios}) and we add line ratios in pixels with non-detections.   
    Since we are showing the logarithmic line ratio, negative values arising due to negative noise can not be shown in either the spatial map or the histograms. We mark pixels where CO is significantly detected, but the line ratio can not be shown in logarithmic scaling in dark grey. 
    The contours depict integrated \nnhp emission at 0.75 and 3 K km/s. 
    }
    \label{fig:Gallery_fdense}
\end{figure*}

While dense gas tracers such as HCN, HNC and \hcop are efficiently emitting at densities above their critical densities, less efficient emission from sub-critical density regions can still significantly contribute to the total integrated intensity of those lines \citep{kauffmann_molecular_2017,Leroy2017_MMwave}. 
In spite of those effects, line ratios with the bulk molecular tracer CO are found to be sensitive to the average gas density  \citep[e.g., review by ][]{neumann_almond_2023,schinnerer_molecular_2024, Neumann_JimenezDonaire2025AA}. 
We show the spatial distribution of \CO line ratios of HCN, HNC, \hcop and \nnhp in Figure~\ref{fig:Gallery_fdense}. We refer to these line ratios as $R_{\rm CO}^{\rm Line}$.

We find that for all lines, $R_{\rm CO}^{\rm Line}$ increases in the center and molecular ring compared to the global average, with a particular increase of up to $\gtrsim0.8\,$dex in an elongated feature that expands towards north-east of the center, which corresponds to the same region where \nnhp line ratios are increased. 
Visually, the \hcop/CO ratio is less enhanced in the central $\sim2$~kpc than the HCN/CO and HNC/CO ratio. 
All $R_{\rm CO}^{\rm Line}$ are decreased in the spiral arms compared to the global average, and exhibit a gradient from low to high line ratios across the arms from trailing to leading side, which is more prominent in the southern arm.

\subsection{Average line intensities and CO line ratios compared to \nnhp emission}
\label{sec:Fdense_vs_n2hp}

\begin{figure}
    \centering
    \includegraphics[width=0.5\textwidth]{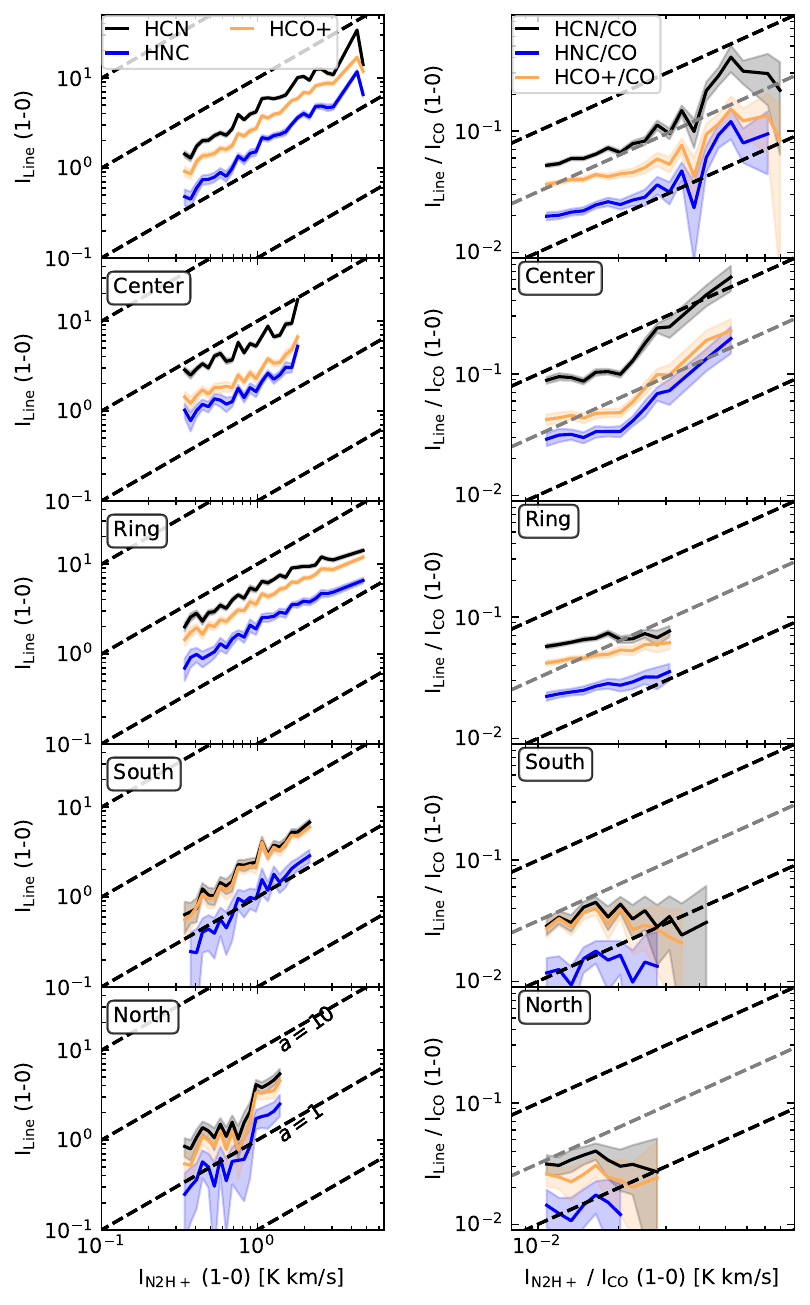}
    \caption{Average intensities of HCN, HNC and \hcop as function of \nnhp intensity (left) as well as average \CO line ratios of HCN, HNC and \hcop (right) as function of the \nnhp/CO line ratio emission. We do not apply any masking, but utilize all pixels in the FoV. 
    We show the binned intensities/line ratios for the total FoV (top) as well as for individual environments (Figure~\ref{fig:Gallery}). 
    We provide lines of constant slopes in log space for visual guidance (black dashed lines). 
    The bins cover a range between 3 times the average \nnhp (\nnhp/CO) uncertainty up to the maximum \nnhp intensity (\nnhp/CO line ratio). 
    }
    \label{fig:binning_vs_n2hp}
\end{figure}

\begin{table*}
\centering
\caption{Fitting average line intensities and average CO-line ratios as function of \nnhp intensity or \nnhp/CO ratio per environment}\label{tab:Fitparams_binnedbyN2HP}
\begin{tabular}{cccc|cccc}
\hline 
\hline 
\noalign{\smallskip}
 & \multicolumn{3}{c}{Full FoV} & Center & Ring & North & South \\
\noalign{\smallskip}
line & slope & offset [dex] & scatter [dex] & slope & slope & slope & slope  \\
\noalign{\smallskip}
\hline 
\noalign{\smallskip}
\multicolumn{6}{c}{Binned line intensities per environments vs \nnhp intensity}\\
\noalign{\smallskip}
\hline 
\noalign{\smallskip}
CO & 0.71$\pm$0.02 & 1.9$\pm$<0.1 & 0.24 & 0.46$\pm$0.06 & 0.61$\pm$0.02 & 0.89$\pm$0.09 & 0.77$\pm$0.07 \\
HCN & 1.13$\pm$0.06 & 0.7$\pm$<0.1 & 0.30 & 0.98$\pm$0.10 & 0.75$\pm$0.03 & 1.39$\pm$0.11 & 1.19$\pm$0.05 \\
HNC & 1.14$\pm$0.05 & 0.2$\pm$<0.1 & 0.28 & 0.91$\pm$0.08 & 0.82$\pm$0.03 & 1.57$\pm$0.13 & 1.27$\pm$0.06 \\
\hcop & 1.07$\pm$0.03 & 0.5$\pm$<0.1 & 0.26 & 0.90$\pm$0.07 & 0.80$\pm$0.02 & 1.48$\pm$0.13 & 1.18$\pm$0.06 \\
\noalign{\smallskip}
\hline 
\noalign{\smallskip}
\multicolumn{6}{c}{Binned CO line ratios per environments vs the \nnhp/CO line ratio}\\
\noalign{\smallskip}
\hline 
\noalign{\smallskip}
HCN/CO & 0.80$\pm$0.08 & 0.3$\pm$0.2 & 0.39 & 0.90$\pm$0.18 & 0.26$\pm$0.05 & 0.22$\pm$0.19 & 0.29$\pm$0.19 \\
HNC/CO & 0.77$\pm$0.09 & -0.2$\pm$0.2 & 0.34 & 1.00$\pm$0.16 & 0.42$\pm$0.02 & 0.35$\pm$0.32 & 0.44$\pm$0.26 \\
\hcop/CO & 0.54$\pm$0.07 & -0.4$\pm$0.1 & 0.30 & 0.91$\pm$0.15 & 0.37$\pm$0.02 & 0.05$\pm$0.19 & 0.18$\pm$0.20 \\
\noalign{\smallskip}
\hline 
\noalign{\smallskip}
\multicolumn{6}{c}{Binned CO-line ratios per environments vs \nnhp intensity}\\
\noalign{\smallskip}
\hline 
\noalign{\smallskip}
HCN/CO & 0.37$\pm$0.06 & -1.2$\pm$<0.1 & 0.25 & 0.39$\pm$0.11 & 0.14$\pm$0.02 & 0.48$\pm$0.06 & 0.51$\pm$0.04 \\
HNC/CO & 0.47$\pm$0.06 & -1.7$\pm$<0.1 & 0.24 & 0.42$\pm$0.11 & 0.23$\pm$0.02 & 0.68$\pm$0.07 & 0.62$\pm$0.06 \\
\hcop/CO & 0.39$\pm$0.04 & -1.4$\pm$<0.1 & 0.17 & 0.40$\pm$0.10 & 0.20$\pm$0.01 & 0.58$\pm$0.05 & 0.51$\pm$0.04 \\
\noalign{\smallskip}
\hline
\end{tabular}
\tablefoot{Fit parameters when fitting a linear relation to binned line emission as function of \nnhp emission in log space, binned \CO line ratios as function of \nnhp/CO and binned \CO line ratios as function of \nnhp intensity. We provide slopes and offsets of linear trends applied to the logarithmic binned values for the full FoV, and provide slopes when fitting binned values obtained in individual environments. For the full FoV we provide the average scatter after subtracting the best-fit from the data (Section~\ref{sec:Methods:Binning}).}
\end{table*}

Combining the results from Section~\ref{sec:linearityn2hp} and ~\ref{sec:linearityCO}, we see clear variations in line ratios both per galactic environment and from cloud-to-cloud. 
To quantify the impact of these variations on the average \nnhp emission, we probe the binned intensity of HCN, HNC, \hcop as well as their binned \CO line ratios as a function of \nnhp emission in Figure~\ref{fig:binning_vs_n2hp} for the total FoV and the individual environments.  
Fit parameters obtained by fitting the binned intensities and binned CO line ratios as function of \nnhp emission and \nnhp/CO (Section~\ref{sec:Methods:Binning}) are provided in Table~\ref{tab:Fitparams_binnedbyN2HP}. 

Across the total FoV, the average HCN, HNC and \hcop emission agrees well with the \nnhp emission (slopes are within $3\sigma$ of unity), while the average \CO emission is significantly sub-linearly related with \nnhp emission across the full FoV (Table~\ref{tab:Fitparams_binnedbyN2HP}) and our defined environments. 
This is in agreement with the super-linear inverse \nnhp-to-\CO and \nnhp-to-HCN relations,  found by S23 using the preliminary SWAN data. The scatter in the full FoV for the line vs \nnhp relation is highest for HCN, followed by HNC, \hcop and \CO as function of \nnhp intensity (Table~\ref{tab:Fitparams_binnedbyN2HP}).
In contrast to \CO, the slopes of HCN, HNC and \hcop as function of \nnhp intensity agree well with each other across all environments. Still, their slopes are sub-linear in the center and ring, and super-linear in northern and southern disk. 

While it is unclear wether $R_{\rm CO}^{\rm N2H+}$  is sensitive to the average cloud density similar to what is proposed for $R_{\rm CO}^{\rm HCN}$, we test the $R_{\rm CO}^{\rm Line}$ vs $R_{\rm CO}^{\rm N2H+}$ relation (Figure~\ref{fig:binning_vs_n2hp}) and list scatter and slopes in Table~\ref{tab:Fitparams_binnedbyN2HP}. 
As the \nnhp molecule is chemically bound in existence to very dense and cold regions where CO, its main reactant, is frozen to dust grains, its emission is found to depend non-linearly on column density in clouds in the Milky Way \citep{tafalla_characterizing_2023}. 
On our $\sim100\,$pc scales we find $R_{\rm CO}^{\rm Line}$  to depend mostly sub-linearly on $R_{\rm CO}^{\rm N2H+}$ across the environments. The slopes range between very low values $0.05-0.44$ in ring, northern and southern disk, up to slopes of  $\sim1.0$ in the center.
The scatter of the $R_{\rm CO}^{\rm Line}$ vs $R_{\rm CO}^{\rm N2H+}$ relation is the largest among all relations tested.

The \CO line ratios of HCN, HNC and \hcop as function of \nnhp intensity reveal significantly sub-linear slopes across all environments, with slopes as low as 0.14 for $R_{\rm CO}^{\rm HCN}$ vs \nnhp in the ring and as large as 0.68 for $R_{\rm CO}^{\rm HNC}$ in the northern disk. 
Despite the low slopes, the average scatter of  $R_{\rm CO}^{\rm Line}$ vs \nnhp are the lowest among all relations tested, and is particularly low for the  $R_{\rm CO}^{\rm HCO+}$ vs \nnhp relation.

\section{Environmental impact - physical parameters driving the line emission}
\label{sec:ResultsEnv}

While all dense gas tracers are similarly distributed in the disk of M51, systematic variations as function of both larger-scale and cloud-scale environment are seen (see Section~\ref{sec:Results}). 
Here we aim to identify the physical conditions that best describe the intensity distribution of dense gas tracing molecules and variations between them by comparing the dense gas emission with different physical parameters, including \Smol, \Sstar, \SSFR, \Pde, \vdisp and galactocentric radius. 
All values are corrected for inclination and described in more detail in Section~\ref{sec:Data}. Figure~\ref{fig:Gallery-Sigmas} shows our estimates of \Smol, \SSFR, \Sstar, \Pde and \vdisp across our FoV. 

We test both the intensity of HCN, HNC, \hcop and \nnhp and their ratios with the \CO line as a function of this set of properties in Section~\ref{sec:ResultsEnv:Intensity} and ~\ref{sec:ResultsEnv:DGfraction}, respectively. 
In Section~\ref{sec:ResultsEnv:DGlineratios} we investigate how line ratios between dense gas lines depend on the environmental properties.

\subsection{Line intensity as a function of physical parameters}
\label{sec:ResultsEnv:Intensity}

\begin{figure}
    \centering
    \includegraphics[width = 0.5\textwidth]{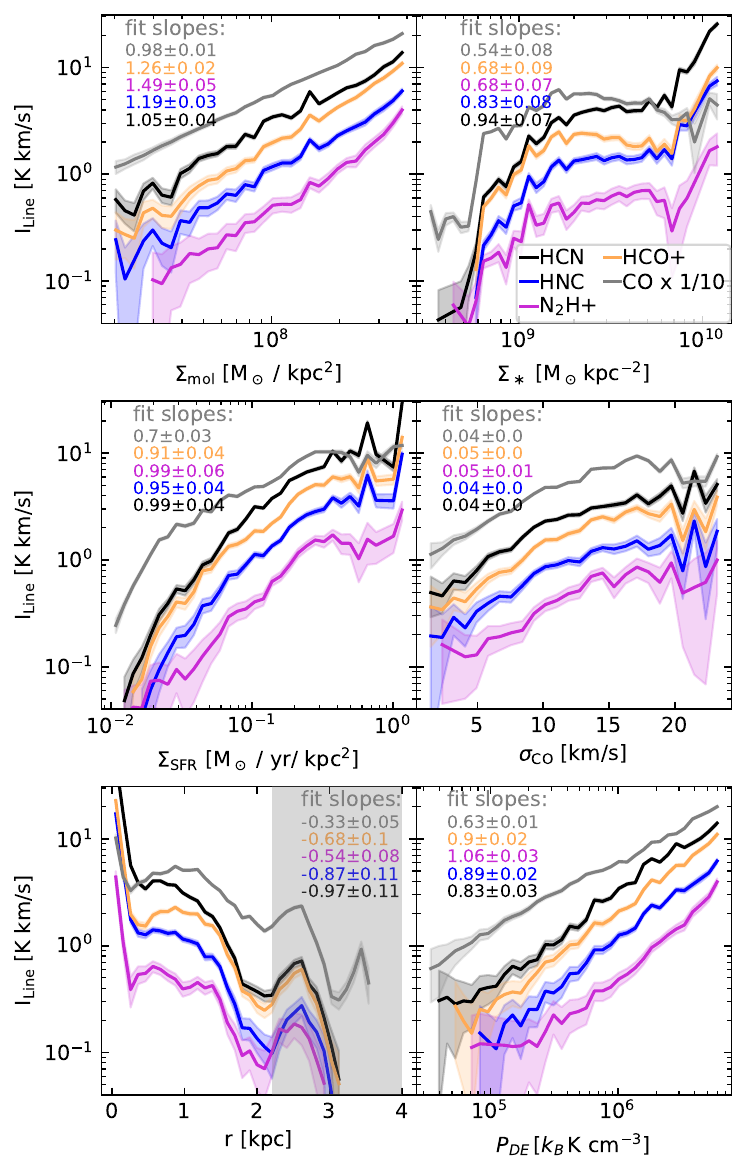}
    \caption{Average line intensity $I_\mathrm{Line}$ as a function of surface densities of molecular gas mass (\Smol), stellar mass (\Sstar) and star formation rate (\SSFR), as well as velocity dispersion (\vdisp), dynamical equilibrium pressure (\Pde) and galactocentric radius for all dense gas tracers. 
    Shaded areas mark the standard deviation per bin. The \CO intensity scaled by a factor of $1/10$ is added for comparison (grey line).  
    For $\rm r>2.2kpc$, the shape of our FoV leads to incomplete sampling of these radial bins (grey shaded area). The obtained slopes for a linear fit (in log space) to the binned averages are provided in each panel.}
    \label{fig:average_intensity_sigmas}
\end{figure}

\begin{figure}
    \centering
    \includegraphics[width=0.24\textwidth]{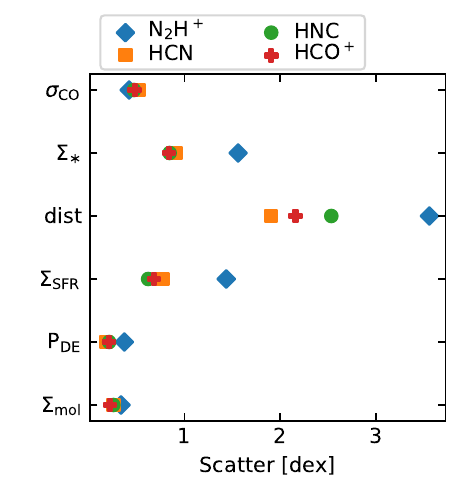}
    \includegraphics[width=0.24\textwidth]{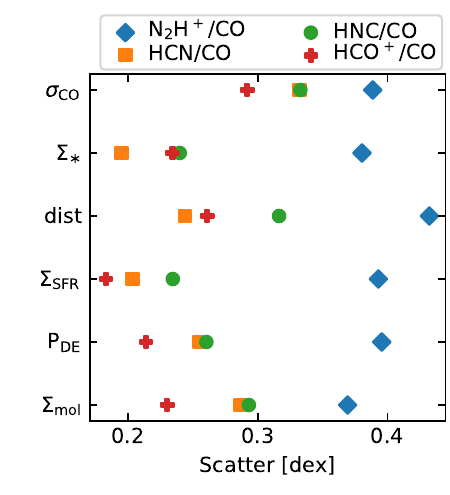}
    \caption{Average scatter of dense gas line intensity (left panel) and average dense gas to CO line ratios (right panel) as function of physical galactic properties with respect to a linear relation (methodology described in Section~\ref{sec:Methods:Binning}). The impact of SNR on this plot is shown in Appendix~\ref{sec:Mockdata:Scatter}. }
    \label{fig:scatter-lineintensities}
\end{figure}

\begin{figure}
    \centering
    \includegraphics[width=0.24\textwidth]{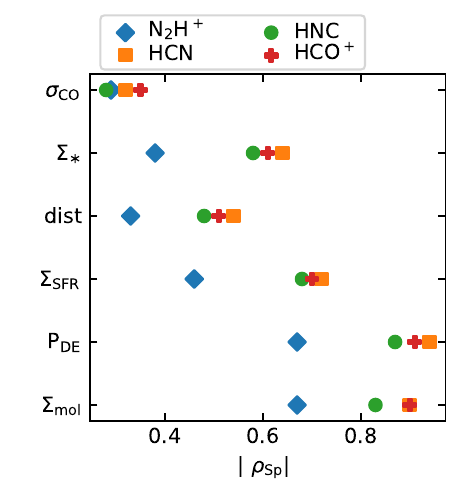}
    \includegraphics[width=0.24\textwidth]{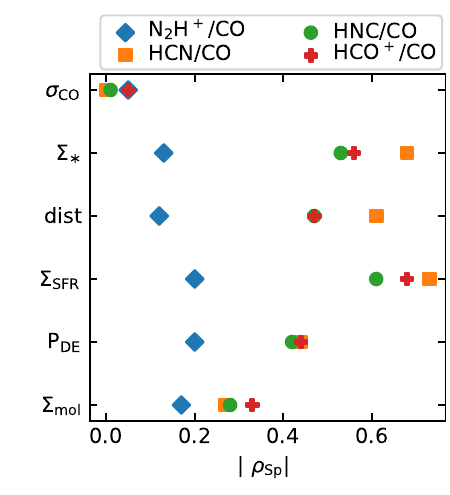}
    \caption{Spearman correlation coefficients of line intensity (left panel) and average dense gas to CO line ratios (right panel) as function of physical galactic properties (Section~\ref{sec:Data:Ancillary}). An increased SNR will directly result in a lower correlation coefficient. We therefore expect higher correlation coefficients for HCN, followed by \hcop, then HNC, then \nnhp as function of the same galactic properties. All p-values are below 5\% except p-values for the Line/CO ratios as function of \vdisp. }
    \label{fig:spearman-intensityvsproperties}
\end{figure}

We show the binned average intensities of HCN, HNC, \hcop, \nnhp and \CO as function of physical parameters in 
Figure~\ref{fig:average_intensity_sigmas}. 
The same analysis per spatially distinct environment (center, ring, north, south), as well as a pixel-by-pixel version of these plots is provided in Appendix~\ref{App:Environments} and \ref{app:pixelbased}, respectively.
We provide fit parameters from the full FoV in Figure~\ref{fig:average_intensity_sigmas} and Table~\ref{tab:Fitparams_Int_vs_all_env}, and for each environment in Appendix~\ref{App:Environments}. The scatter with respect to the full FoV fits is depicted in Figure~\ref{fig:scatter-lineintensities}. Further, we show Spearman correlation coefficients of the full FoV in Figure~\ref{fig:spearman-intensityvsproperties}. 

The average line intensity of all lines is monotonically positively correlated with \Smol and \Pde (deviations from a monotonic relation $\lesssim 0.2\,$dex) and non-monotonically with \Sstar, \SSFR, \vdisp, and galactocentric distance, which suggests secondary dependencies. 
Consequently, the scatter with respect to a linear relation between line intensity and physical properties is lowest  (Figure~\ref{fig:scatter-lineintensities}) and Spearman correlation coefficients are highest for \Pde and \Smol (Figure~\ref{fig:spearman-intensityvsproperties}).  

Fitted slopes of average intensity as function of \Smol range from unity (CO, HCN), to significantly super-linear (HNC, \hcop, \nnhp).  
These values confirm that \nnhp has a significantly steeper dependency on \Smol than the other lines, while HCN has a similar shallower dependency on \Smol than CO does. While slightly reduced in the center, the slopes obtained in the environments are consistent with this result (Appendix~\ref{App:Environments}). 
We note a turn-up in the \nnhp-to-\Smol relation at high values of \Smol, where the trend steepens, which corresponds to the \nnhp-brightest region located at the south-western brink of the molecular ring (Section~\ref{sec:Lineint_perEnv}). 

Slopes of average intensity as function of \Pde range from significantly sub-linear (CO, HCN, HNC, \hcop) to linear (\nnhp) and while they vary slightly with environment (Appendix~\ref{App:Environments}), the increased order or slopes from CO to \nnhp is mostly consistent with the full FoV environment. 
The dependency of CO and HCN on \Pde is significantly different. 

As the relations of line intensity with other physical galactic parameters are depicting larger variations deviating from a straight-line relation, we suspect a dependency on secondary properties. 
We find the following: 
Average line intensities all increase with increasing \Sstar, except for a plateau (\nnhp, HNC, HCN) or decline (CO, \hcop) at \Sstar $\sim 2-7\times 10^9$ M$_\odot$\,kpc$^{-2}$, which corresponds to radii of $r\sim 0.2-1\,$kpc.
This is consistent with the flatter trend seen in line ratio as function of galactocentric radius. 

All lines are similarly well correlated with \SSFR up to values of $4\times 10^{-1}$ M$_\odot$\,yr$^{-1}$\,kpc$^{-2}$, where the distribution flattens.  
We identify a shallower distribution in the northern disk compared to the other environments, which, when averaged, produces the flattening (Appendix~\ref{App:Environments}).
The same effect is driving the flattening seen with \vdisp.

\subsection{Line ratios with CO as function of physical parameters}
\label{sec:ResultsEnv:DGfraction}

\begin{figure}
    \centering
    \includegraphics[width = 0.5\textwidth]{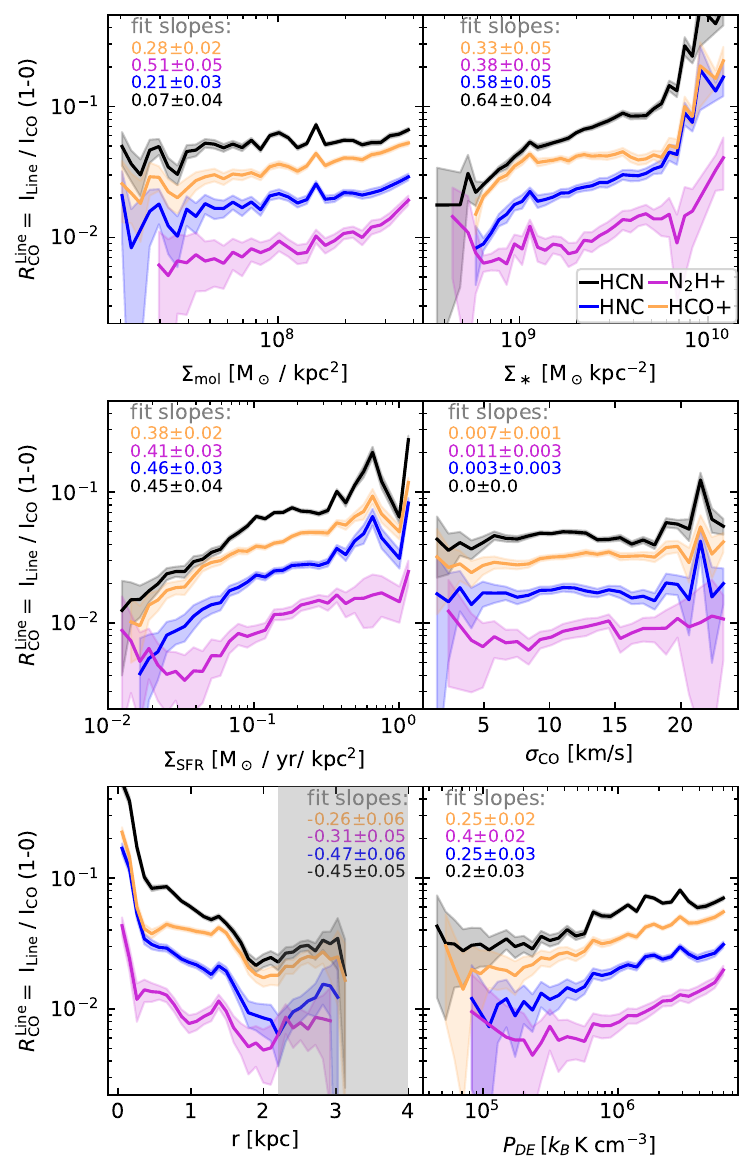}
    \caption{Same as Figure~\ref{fig:average_intensity_sigmas} but for line ratios with \CO. }
    \label{fig:average_DGfraction_sigmas}
\end{figure}

Ratios between dense gas tracers and bulk tracer CO are found to be density sensitive \citep[e.g., ][]{leroy_cloud-scale_2017,schinnerer_molecular_2024}. 
To explore the physical mechanisms driving these CO line ratios, we show $R_{\rm CO}^{\rm HCN}$, $R_{\rm CO}^{\rm HNC}$, $R_{\rm CO}^{\rm HCO+}$ and $R_{\rm CO}^{\rm N2H+}$  as function of physical properties \Smol, \Sstar, \SSFR, \vdisp, \Pde and galactocentric radius in Figure~\ref{fig:average_DGfraction_sigmas}. 
We provide fit slopes when fitting a linear to the binned data in Figure~\ref{fig:average_DGfraction_sigmas} and for the distinct environments and full FoV in Appendix~\ref{App:Environments}.

The average $R_{\rm CO}^{\rm Line}$ follows a nearly straight-line relation with \Smol and \Pde, albeit the $R_{\rm CO}^{\rm N2H+}$-to-\Smol relation steepens at high values of \Smol. 
Ratios related with other parameters depict stronger variations and deviations from a linear relation. 
Despite this linear dependency of CO line ratios and \Smol, fitted slopes are surprisingly low, with slopes for $R_{\rm CO}^{\rm HCN}$ consistent with being zero. Spearman correlation coefficients are lowest for $R_{\rm CO}^{\rm HCN, HNC, HCO+}$-to-\Smol ($\rho_\mathrm{Sp}\lesssim0.35$) compared to any other property (except \vdisp, Figure~\ref{fig:spearman-intensityvsproperties}). \\
This indicates that in the full FoV, $R_{\rm CO}^{\rm HCN}$ is nearly independent of \Smol at cloud-scales in M51.\\
We explore possible biases that affect these measurements, such as different estimates of \Smol (Appendix~\ref{App:alphaCO}), the impact of the AGN (Appendix~\ref{App:fdense_Sigmamol_radialtrend}), or the relation within individual environments (Appendix~\ref{App:Environments}). 
All tests agree that the $R_{\rm CO}^{\rm HCN}$ ratio has a very shallow and variable dependency on \Smol, that even turns negative depending on the prescription used to estimate \Smol and the environment studied. 
$R_{\rm CO}^{\rm HNC}$ and $R_{\rm CO}^{\rm HCO+}$ behave similarly, although they generally have steeper slopes as a function of \Smol. 
Within distinct environments, the slope of the $R_{\rm CO}^{\rm N2H+}$ to \Smol relation varies from $-0.04\pm0.1$ in the center up to  $0.8\pm0.05$ in the molecular ring (Appendix~\ref{App:Environments}).
Removing the central 1\,kpc data including the brightest pixels of HCN, HNC and \hcop, increases the slopes of all $R_{\rm CO}^{\rm Line}$ to \Smol relations, particularly for the $R_{\rm CO}^{\rm HCN}$ ratio (i.e., slope of $0.36\pm0.02$ for $R_{\rm CO}^{\rm HCN}$, and $0.68\pm0.06$ for $R_{\rm CO}^{\rm N2H+}$, Appendix~\ref{App:fdense_Sigmamol_radialtrend}). 
Even in the environments outside of the center, the  $R_{\rm CO}^{\rm HCN}$ to \Smol relation is shallower than common literature results (compare Section~\ref{sec:Discussion}).

The dependency of $R_{\rm CO}^{\rm Line}$ to \Pde is positive albeit shallow across all environments (Appendix~\ref{App:Environments}) with lowest slopes for $R_{\rm CO}^{\rm HCN}$ followed by $R_{\rm CO}^{\rm HNC}$, $R_{\rm CO}^{\rm HCO+}$ and $R_{\rm CO}^{\rm N2H+}$. Spearman correlation coefficients are higher than compared to \Smol, but lower than \Sstar, $r$ and \SSFR (Figure~\ref{fig:spearman-intensityvsproperties}. 
While the scatter relative to a linear correlation (Figure~\ref{fig:scatter-lineintensities}) is lower for $R_{\rm CO}^{\rm HCN}$, $R_{\rm CO}^{\rm HNC}$ and $R_{\rm CO}^{\rm HCO+}$ as function of \Pde compared to \Smol, the scatter is even lower for other physical parameters, i.e. \SSFR for  $R_{\rm CO}^{\rm HCO+}$  and \Sstar for $R_{\rm CO}^{\rm HCN}$.
Similarly, Spearman correlation coefficients (Figure~\ref{fig:spearman-intensityvsproperties}) are highest for $R_{\rm CO}^{\rm Line}$ for all lines except \nnhp as function of
\Sstar, $r$ and \SSFR, followed by \Pde, \Smol and are lowest for \vdisp. 

Due to the low SNR of the \nnhp emission, the variation seen in the scatter and Spearman coefficients of $R_{\rm CO}^{\rm N2H+}$ as function of physical parameters is consistent with noise (compare Appendix~\ref{sec:Mockdata:Scatter}).

Similar to Section~\ref{sec:ResultsEnv:Intensity}, the \CO line ratios as function of \Sstar and radius change slope at radii of $\sim 0.2-1.2\,$kpc.

\subsection{Dense gas line ratios as function of physical parameters}
\label{sec:ResultsEnv:DGlineratios}

\begin{figure*}
    \centering
    \includegraphics[width = 0.49\textwidth]{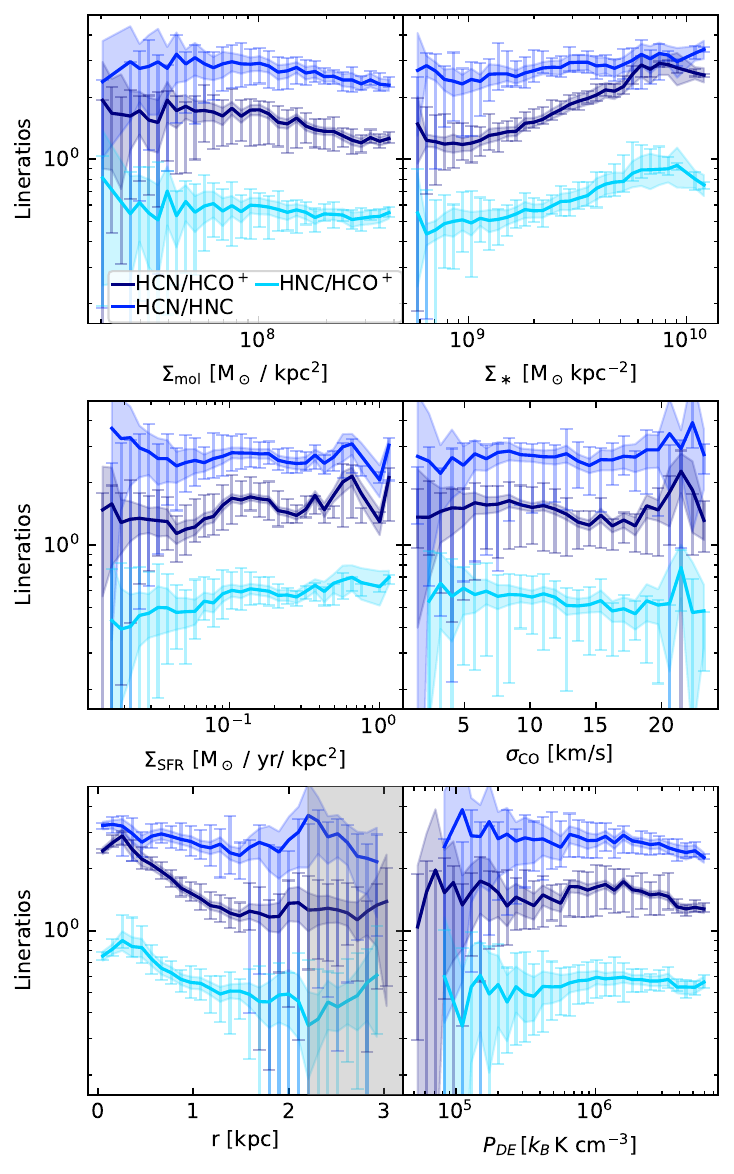}
    \includegraphics[width = 0.49\textwidth]{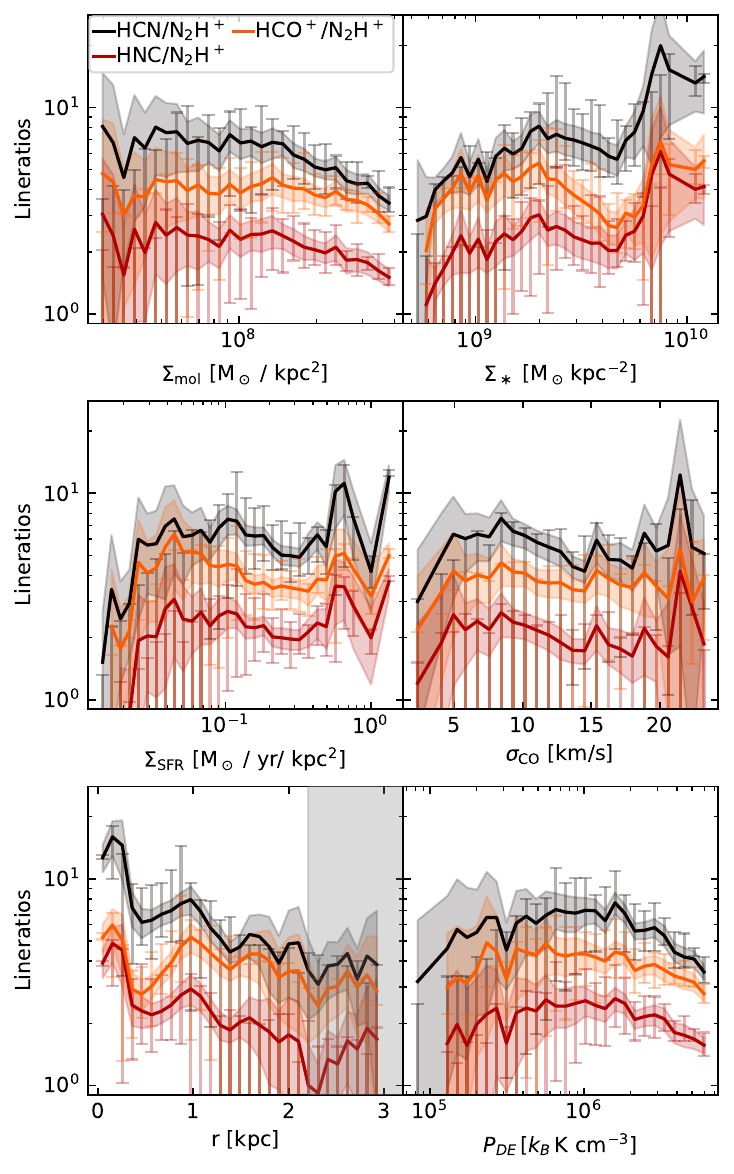}
    \caption{Average line ratios of HCN/HNC, HCN/\hcop and HNC/\hcop (left panels) as function of physical properties as well as for HCN/\nnhp, HNC/\nnhp, \hcop/\nnhp (right panels). Error bars depict the 25/75th percentiles.}
    \label{fig:average_lineratio_n2hp}
\end{figure*}

We show average line ratios between HCN, HNC and \hcop and line ratios between those lines and \nnhp as function of \Smol, \Sstar, \SSFR, \vdisp, galactocentric radius and \Pde in Figure~\ref{fig:average_lineratio_n2hp}, and separated into environments in Appendix~\ref{App:Environments}.  

Among the brighter dense gas tracers HCN, HNC and \hcop we find the largest dependency on physical properties of line ratios with \hcop, with a nearly linear dependency with \Smol and radius.   
Consistent with findings in Section~\ref{sec:hcnhnchcop}, the HCN/HNC ratio is basically independent of physical properties or dependencies are very shallow (slopes range from $\rm -0.14\,to\,0.11$, Table~\ref{tab:Slopes_Lineratios_vs_properties}).
HCN/\hcop is positively correlated with  
\SSFR and negatively correlated with radius and \Smol. 

Ratios of HCN, HNC and \hcop with \nnhp show no clear linear dependencies on any of the properties. 
All ratios monotonically decrease at high values of \Smol and \Pde and all except for the \hcop/\nnhp ratio decrease with galactocentric radius. 
We find the steepest slopes for the HCN/\nnhp-to-\Smol relation ($-0.42$) and shallowest for \hcop/\nnhp-to-\Smol ($-0.2$).  
A local minima in \nnhp line ratios as function of both \Sstar and radius can be associated with radii of $r\sim0.2-1\,$kpc.

\begin{figure}
    \centering
    \includegraphics[width=0.5\textwidth]{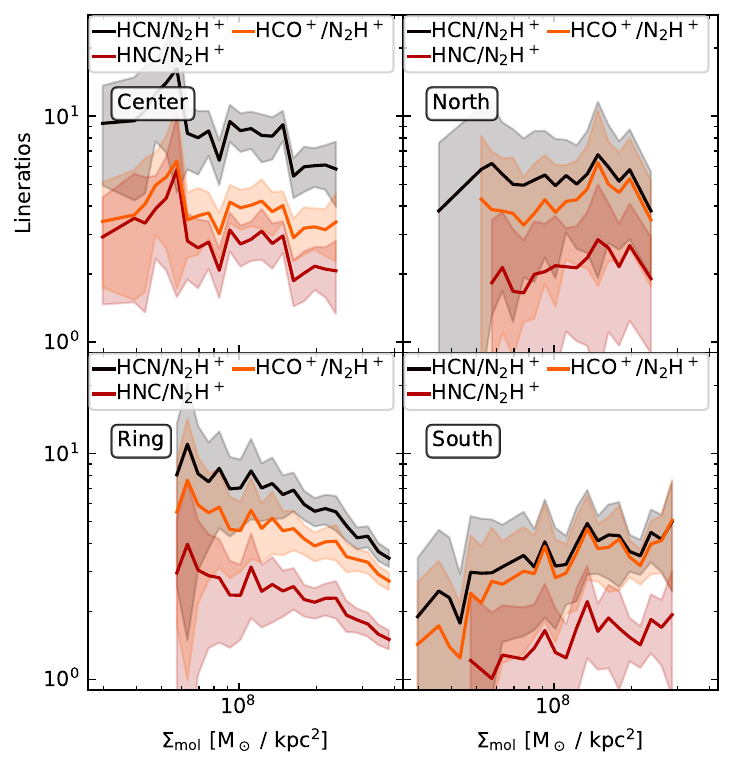}
    \caption{Same as right panels of Figure~\ref{fig:average_lineratio_n2hp}, but only for relations with \Smol and separated into the different environments. We note that fitting the binned averages as function of \Smol results in slopes that are within $5\sigma$ to a slope of zero for all line ratios as function of \Smol, except for the ring environment, where the slopes are significantly negative. The slopes for the HCN-, HNC- and \hcop-to-\nnhp line ratio as function of \Smol in the ring environment are $-0.63\pm0.04$, $-0.48\pm0.04$ and $-0.46\pm0.04$, respectively.}
    \label{fig:average_lineratio_vs_sigmamol_perenv}
\end{figure}

\begin{table*}
\caption{Fitted slopes to binned line ratios as function of \Smol, \SSFR, \Sstar, \Pde, \vdisp and radius. }\label{tab:Slopes_Lineratios_vs_properties}
\centering
\begin{tabular}{lcccccc}
\hline\hline
\noalign{\smallskip}
LR & \Smol & \SSFR & r & \Sstar & \vdisp & \Pde \\
HCN/\nnhp & -0.42$\pm$0.04 & 0.09$\pm$0.05$^\ast$ & -0.32$\pm$0.04 & 0.29$\pm$0.04 & -0.29$\pm$0.10$^\ast$ & -0.25$\pm$0.04 \\
\hcop/\nnhp & -0.20$\pm$0.03 & -0.01$\pm$0.04$^\ast$ & -0.08$\pm$0.04$^\ast$ & -0.04$\pm$0.05$^\ast$ & -0.10$\pm$0.06$^\ast$ & -0.16$\pm$0.03 \\
HNC/\nnhp & -0.27$\pm$0.03 & 0.07$\pm$0.04 $^\ast$& -0.24$\pm$0.04 & 0.20$\pm$0.04 $^\ast$& -0.27$\pm$0.08 $^\ast$& -0.18$\pm$0.03 \\
\noalign{\smallskip}
\hline 
\noalign{\smallskip}
HCN/HNC & -0.14$\pm$0.01 & 0.04$\pm$0.02$^\ast$ & -0.07$\pm$0.01 & 0.11$\pm$0.01 & -0.01$\pm$0.03$^\ast$ & -0.08$\pm$0.01 \\
HCN/\hcop & -0.21$\pm$0.02 & 0.13$\pm$0.03 & -0.22$\pm$0.02 & 0.34$\pm$0.02 & -0.18$\pm$0.05$^\ast$ & -0.08$\pm$0.02$^\ast$ \\
HNC/\hcop & -0.07$\pm$0.01 & 0.10$\pm$0.01 & -0.14$\pm$0.02 & 0.24$\pm$0.01 & -0.15$\pm$0.04$^\ast$ & -0.01$\pm$0.01$^\ast$ \\ \hline
\end{tabular}
\tablefoot{Fitted slopes of binned line ratios as function of \Smol, \SSFR, \Sstar, $r$, \Pde and \vdisp. Binned averages are calculated as described in Section~\ref{sec:Methods:Binning} for all pixels in the FoV. The fitting is performed as before (i.e, Table~\ref{tab:Fitparams_binnedbyN2HP}). 
We mark values where the slope agrees within 5$\sigma$ to a slope of zero with $^\ast$. }
\end{table*}

We highlight the impact environment has on the slopes of average \nnhp line ratios as function of \Smol  in Figure~\ref{fig:average_lineratio_vs_sigmamol_perenv}. We show the same relation for all physical properties in Appendix~\ref{App:Environments}, but provide here a version easier to compare. In contrast to the other environments, the slope measured in the ring environment is significantly negative with a steep slope of $-0.63$ (see Figure caption).

Our results suggest that the line most closely related to \nnhp is \hcop, as the \hcop/\nnhp line ratio shows the smallest slopes as function of galactic parameters compared to the HCN/\nnhp and HNC/\nnhp ratios. 
The most different to \nnhp is the molecule HCN, with the largest line-ratio to galactic property slope of HCN/\nnhp-to-\Smol of -0.42. The molecular line emission closest to the HCN(1-0) emission is HNC(1-0), as the HCN/HNC line ratio is smaller than HCN line ratios with \hcop or \nnhp.  
The largest differences between \nnhp ratios (if present) is revealed by \Smol, followed by galactocentric radius, \Sstar and \Pde. 
The largest difference between line ratios among HCN, HNC and \hcop is revealed by \Sstar, followed by galactocentric radius and \Smol. 
In summary, at our 125\,pc scales, there are no two molecules that have the exact same emission pattern across all environmental conditions.

\section{Discussion on dense gas in M51}
\label{sec:Discussion}

Our observations of HCN, HNC, \hcop and \nnhp in the disk of M51 cover different galactic environments, including spiral arms, a molecular ring and AGN affected center, and span a range of physical conditions.  
We discuss the impact of environment and physical conditions on the dense gas emission, and ultimately the ability of this emission to trace dense regions.

\subsection{Are dense gas tracers reliably tracing dense gas?}

\begin{figure}
    \centering
    \includegraphics[width=0.4\textwidth]{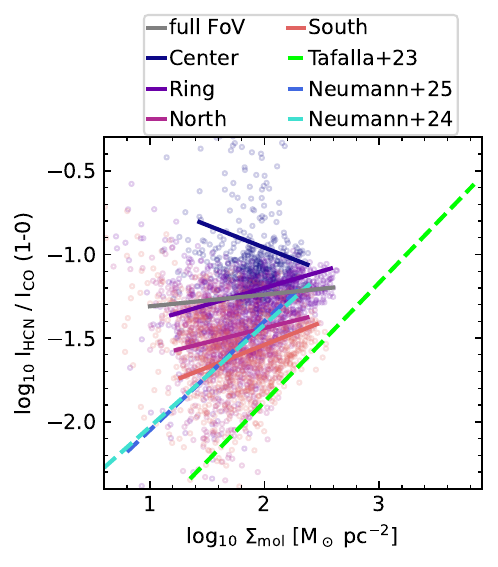}
    \caption{Literature comparison of fits best describing HCN/CO as function of \Smol. We include estimates from Milky Way clouds \citep[][green dashed line]{tafalla_characterizing_2023}, fits of individual environments and full FoV in M51 (this work, solid lines), fits based on HCN/CO at $260\,$pc resolution in NGC~4321 \citep[][]{Neumann_2024A&A} and fits based on 31 local spiral galaxies from the ALMOND \citep{neumann_almond_2023} and EMPIRE \citep{jimenez-donaire_empire_2019} surveys at $\sim$kpc resolution \citep{Neumann_JimenezDonaire2025AA}. We add our HCN/CO observations for visual comparison for each environment within M51 (colored open circles).}
    \label{fig:fdense_Smol_Literature}
\end{figure}

The emission of \nnhp, which originates from regions where \CO is frozen out on dust grains, is undisputedly an ideal indicator of cold and dense regions within Milky Way clouds, and likely also a good indicator of dense regions in M51. 
The emission of a ``dense gas tracer'' is thus expected to agree reasonably well with the emission of \nnhp. 
Our data show that, in contrast to the emission of the bulk tracer \CO, the emission of HCN, HNC and \hcop performs well at tracing \nnhp emission linearly in our FoV, but this correlation clearly varies with environment. 
When compared on cloud-scales and across the environments and physical conditions present in M51, the emission of \hcop is favoured over the emission of HCN and HNC, as it is more closely related to the \nnhp emission and has higher values of $\rho_\mathrm{Sp}$ across the environments when compared to \Smol despite being fainter than HCN on average.  

When comparing the emission of dense gas molecules with estimates of \Smol, often assumed to be indicative of the gas volume density, we find that the relation between HCN and \Smol is more similar (in slope) to the relation between CO and \Smol, while the relationship between \hcop and \Smol (in slope) is closer to the relationship between \nnhp and \Smol. 
At much smaller than our resolutions within three individual clouds in the Milky Way, \citep{tafalla_characterizing_2023-1} find strong and similar correlations between HCN, HNC and \hcop emission and column density and clear deviations from the CO relation. 
\citet{tafalla_characterizing_2023-1} further find that \nnhp emission rises more steeply with estimates of column density than any of the other lines, and emission only arises above column densities of $10^{22}\,$cm$^{-2}$ 
(corresponding to \Smol$\sim 2\times10^8\,$M$_\odot$/kpc$^2$).
This is in disagreement with our general similar trends found between \nnhp, \hcop, HCN and HNC, but at our resolution (and different measurements of \Smol).
However, we note that in our observations, \nnhp emission as function of \Smol changes slope at highest values of \Smol ($\gtrsim2\times10^{8}$\,M$_\odot$\,kpc$^{-2}$), which is not the case for the other lines. 
While this could point to a boosted production of \nnhp above a threshold of \Smol, in agreement with \citet{barnes_lego_2020,tafalla_characterizing_2023-1}, we note that at our resolution, we are likely averaging several clouds within one beam and do not expect to recover the same relation. 
Further, 
this steepening in the \nnhp-to-\Smol relation is mainly driven by the \nnhp-brightest region located at the south-western brink of the molecular ring. 
We discuss possible mechanisms responsible for the high \nnhp emission in Section~\ref{sec:Discussion:Ring}.

The \CO line ratios depend super- or sub-linearly on the \nnhp/CO ratio depending on the environment, suggesting that the molecules are not similarly sensitive to density. 
The scatter is minimized when comparing CO line ratios directly with \nnhp intensity, albeit the relations are significantly sub-linear with exception of the central environment. 
When comparing their emission with estimates of surface density, \Smol, a commonly used proxy of gas density,  we find that the HCN/\CO line ratio is not significantly correlated with \Smol across the FoV of our SWAN data, with Spearman correlation coefficients of $\rho_\mathrm{Sp} \sim 0.3$, which are smaller than when comparing HCN/\CO to any other physical parameter (except \vdisp). 
Typical HCN/CO line ratios from lower-resolution surveys follow a relation with \Smol of slope $\sim0.65$ \citep{Neumann_JimenezDonaire2025AA}, which is obtained by combining $\gtrsim$ kpc-scale observations of 31 local spiral galaxies in the EMPIRE survey \citep{jimenez-donaire_empire_2019} and ALMOND survey \citep{neumann_almond_2023}. 
Similarly, the 300\,pc resolution observations in NGC~4321 match these literature slopes well \citep[][Figure~\ref{fig:fdense_Smol_Literature}]{Neumann_2024A&A}
Even estimates from three Milky Way clouds find slopes in the HCN/CO-to-\Smol relation of $\sim0.71$ \citep{tafalla_characterizing_2023}. 
The relation found in our data, both from the total FoV as well as obtained from individual environments, is significantly shallower and weaker ($\rho_\mathrm{Sp} \lesssim0.4$) than any of these literature correlations or even negative, as we show in Figure~\ref{fig:fdense_Smol_Literature}. 

Our results are similar to 150\,pc resolution observations in the central 1\,kpc of NGC~6946 \citep{eibensteiner_23_2022}, that only find a weak and shallow correlation between HCN(1-0)/CO(2-1) and CO(2-1) intensity as proxy for \Smol (slopes of $\sim0.23$, $\rho_\mathrm{Sp}\sim0.39$), and instead find a stronger correlation for HNC/CO followed by \hcop/CO. 
In their observations, \hcop is best correlated with \SSFR. NGC~6946 does not exhibit signs to host an AGN that could explain an enhancement in HCN emission.

Our results show that in the inner disk of M51 at cloud-scales, the HCN/CO ratio does not trace \Smol well at all and also the \hcop/CO or HNC/CO ratios are shallow. 
Removing central apertures (Appendix~\ref{App:fdense_Sigmamol_radialtrend}) or using different prescriptions to calculate \Smol (Appendix~\ref{App:alphaCO}) do not resolve this issue. While we find steeper slopes for some of those treatments, the high variability of the slope with aperture and \Smol prescription makes the HCN/CO ratio a very questionable tracer of \Smol in M51. 
Spearman correlation coefficients do not change when removing central apertures, and are all significantly lower than when comparing $R_\mathrm{CO}^\mathrm{Line}$ to any other physical parameter except \vdisp. 
The dynamic range of \Smol is not reduced when removing the central 1\,kpc (Section~\ref{App:fdense_Sigmamol_radialtrend}). 
While the \nnhp/CO-to-\Smol relation reveals higher slopes, those slopes vary greatly across the different environments. 

Potentially, the Line/CO ratio is not solely dependent on gas density, but other conditions that are possibly more extreme in the interacting M51 than in the other galaxies mentioned above.  However, since this is the first time that observations at this resolution across such a large FoV have been taken, it is unclear whether the conditions in M51 are peculiar, or whether the resolution is causing the differences.

\subsection{Physical conditions driving gas emission}
\label{sec:Discussion:Drivingconditions}

\begin{figure}
    \centering
    \includegraphics[width=0.4\textwidth]{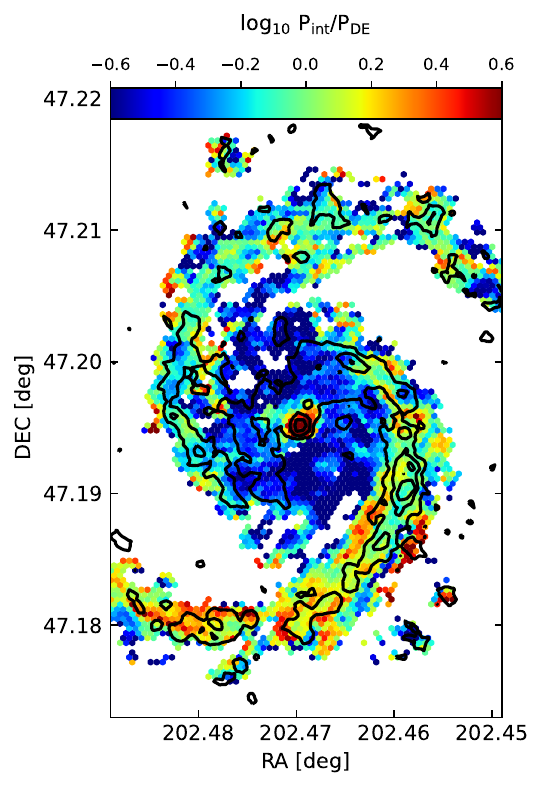}
    \caption{Logarithmic ratio of P$_\mathrm{int}$/\Pde. We calculate P$_\mathrm{int} = 3/2 \, \Sigma_\mathrm{mol}\times \sigma_\mathrm{CO}^2 /D$ according to equation 10 from \citet{sun_dynamical_2020} and with $D = 125$\,pc. We show \nnhp contours at 0.5, 2, 4\,K.
    }
    \label{fig:PintPDE}
\end{figure}

While molecular gas mass surface density is commonly used as a proxy for the volume density and thought to directly drive the emission of dense gas molecules, dynamical equilibrium pressure is suggested to directly determine the ability of gas to form stars \citep[e.g.,][]{usero_variations_2015,jimenez-donaire_empire_2019, Neumann_JimenezDonaire2025AA}. 
In comparison to \Smol, we find less variation in the slopes of  R$_{\rm CO}^{\rm Line}$ as a function of \Pde across environments (i.e., no negative slopes), but slopes are similarly shallow. 
In contrast to a linear correlation, \citet[][]{Neumann_2024A&A} suggests a dynamical equilibrium pressure threshold above which the HCN/CO to \Pde and HCN/SFR to \Pde relations change slope, falling at 
\Pde$\sim 4\times10^5 k_B$K cm$^{-3}$ for HCN/CO vs.\ \Pde and \Pde$\sim1\times10^6 k_B$K cm$^{-3}$ for HCN/SFR vs.\ \Pde.
Considering that this threshold of the measured \Pde ~may change with resolution \citep{sun_dynamical_2020}, this is in agreement with various theoretical works that suggest that clouds decouple from their environment above a pressure threshold and collapse in a universal fashion \citep[e.g.,][]{Ostriker_2010,Ostriker_2022}. 
Our data reveals no clear change in the HCN/CO vs \Pde relation, despite covering the same dynamic range, but all line ratios with \nnhp visually appear to turn from positive to negative slopes at \Pde$\sim1\times10^6 k_B$K cm$^{-3}$.  
At high pressures the fractions of immediately star forming gas is expected to increase so that the ratio of higher critical density \nnhp to HCN or \hcop is expected to change.  
As shown by sub-cloud-scale simulations by \citet{priestley_neath_2023}, \nnhp largely traces gas in such dense regions that irreversibly collapse, in contrast to HCN, HNC and \hcop which also trace a significant amount of gas that will not eventually form stars, in agreement with our observations.

While \Pde and \Smol are promising drivers of the gas emission, line ratios with CO have an increased scatter when compared to those physical properties compared to others. 
Despite there being secondary trends and local minima in relations with other physical quantities, the scatter measured with respect to a linear relation of the \hcop/CO ratio is minimized when compared to \Sstar, whereas the scatter of HCN/CO is smallest when compared to \Sstar and \SSFR. 
Both \Sstar and \SSFR however depict local minima and changes of slopes that might be connected with \Pde as follows. 
Figure~\ref{fig:PintPDE} shows the ratio of internal to dynamical equilibrium pressure ($P_{\rm int}$/ \Pde). 
The internal pressure $P_{\rm int}$ refers to the pressure inside a molecular cloud, likely dominated by turbulence (hence often known as $P_{\rm turb}$, which supports the cloud against its own weight and the weight of the ambient gas\citep[e.g.,][and references within]{sun_dynamical_2020}. 
A ratio of $P_{\rm int}$/ \Pde ~of unity might suggest that the cloud is pressure-confined and in approximate balance with its environments.  A ratio larger than unity suggests that molecular gas is overpressurized compared to its surrounding environment. 
This ratio is below unity in the central $\sim3\,$kpc, except for the very central few $\sim100$\,pc, suggesting that the 
clouds are not in equilibrium with their surrounding environment. 
The P$_{\rm int}$/\Pde ratio further reveals an asymmetry between the northern and southern spiral arms, which we discuss in more detail below. 
The decrease of P$_{\rm int}$/\Pde in the center coincides spatially with the shallower relation between line intensity, CO line ratios and \Sstar or radius found at radii of $\sim0.2-1.2\,$kpc. 
In comparison, the bulge of M51 is thought to be only $\sim 450\times 650\,$pc in size \citep[][]{Lamers2002}. 
In the central environment coinciding with this region, emission of HCN, HNC and \hcop is sublinearly correlated with \Smol, and CO line ratios are negatively correlated with \Smol while being most steeply correlated with both \Sstar and \vdisp compared to other environments. 

Further, the  HCN/\hcop (and HNC/\hcop) ratio is higher in the central $\sim1.5\,$kpc. 
 A possible mechanism to regulate the molecular line emission in the center could be provided by a high cosmic ray ionization rate (CRIR). A high CRIR
 can increase the formation of \hcop \citep[e.g.,][]{harada_molecular-cloud-scale_2019}, by increasing the abundance of H$_3^+$, which is a key molecule for \hcop formation. 
This occurs via the ionization of H$_2$ to H$_2^+$, which can form H$_3^+$ in a secondary step \citep[e.g.,][]{le_petit_2016}. 
However, in lower-density environments or at high cosmic ray ionization rates, the electrons produced via the ionization of H2 recombine with and destroy H$_3^+$ and \hcop \citep{le_petit_2016}, yielding an equilibrium \hcop abundance that is independent of the cosmic ray ionization rate.
An increased free electron fraction also enhances the HCN excitation rate  \\
\citep{Krolik1983, maloney_x-ray--irradiated_1996, Papadopoulos2007, goldsmith_electron_2017}. 
Combined, this might lead to an increased HCN to \hcop emission ratio.\\

While some studies find a correlation between cosmic ray ionization rates and X-ray emission produced by inverse Compton scattering of hot gas \citep{Schober2015MNRAS}, the contribution of this mechanism to the total X-ray flux is expected to be fairly small for local non-starbursting galaxies. 
\citet{Zhang2025} find a sharp decrease in the surface brightness profile of hot gas from X-ray observations with Chandra in M51 at $r\sim2$\,kpc as well as a steeper relation between kpc-scale HCN emission in the central $\sim2\,$kpc and hot gas luminosity compared to the disk. They further conclude a negligible impact of the AGN on the hot X-ray gas luminosities, and suggest that instead core-collapse supernovae (SN) dominate the energy source in M51's center.

Increased ionization rates can also be found in the CMZ of the Milky Way compared to its disk, where they also result in increased gas temperatures \citep{Oka_2019, indriolo_herschel_2015}.
Variations in kinetic gas temperature can increase the scatter of HCN intensity as function of gas column density  \citep{tafalla_characterizing_2023-1}. 
With the gas temperature in merging galaxy systems generally being increased \citep{schinnerer_molecular_2024}, this can increase the scatter of the HCN/CO ratio. 
However, kinetic gas temperature estimates by \citet{den_Brok2025} combining SWAN and SMA CO and CO isotopologue observations reveal average gas temperatures in our FoV of $\sim13\pm5$\,K with no significant increase towards M51's  center.
High HCN/\hcop ratios similar to our results (Figure~\ref{fig:Gallery_linerationon2hp}) are also common in centers in Ultra Luminous Infrared Galaxies \citep[ULIRGs, e.g.,][]{aalto_chemistry_2006,imanishi_dense_2023}, LIRGs \citep{Papadopoulos2007} as well as centers hosting AGNs of normal galaxies \citep[][]{usero_molecular_2004,nakajima_molecular_2023}, including M~51 \citep{kohno_prevalence_2005}.
However, the molecular gas in centers of ULIRGs is often warm \citep[i.e., $T\gtrsim100$\,K][]{imanishi_dense_2023}, which is -- again -- not the case for M51's center.

\begin{figure}
    \centering
    \includegraphics[width=0.3\textwidth]{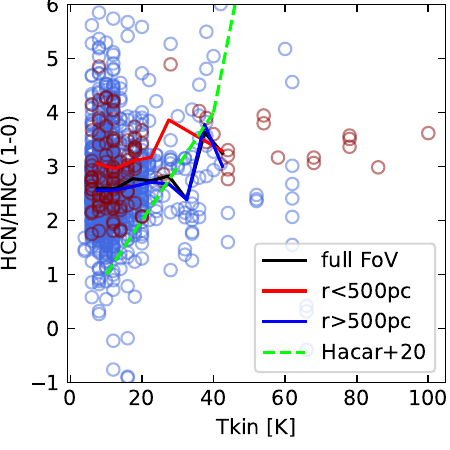}
    \caption{HCN/HNC line ratio from this work as function of kinetic gas temperature measurements at $4\arcsec$ resolution from \citet{den_Brok2025}. We separate pixels in the center ($r<500\,$pc, red points) and disk ($r>500\,$pc, blue points), and provide binned estimates for the full FoV (black line),  center and disk (red, blue, respectively) in steps of $5$K in the range $\sim10-50$K, suggested by \citet{hacar_hcn--hnc_2020}.  We add the fitted relation from \citet{hacar_hcn--hnc_2020} (green dashed line).}
    \label{fig:HCNHNCTKin}
\end{figure}

The HCN/HNC ratio has been suggested to track kinetic gas temperature in Milky Way clouds \citep{hacar_hcn--hnc_2020}. We show the HCN/HNC ratio as function of T$_{\rm kin}$ from \citet{den_Brok2025}  in Figure~\ref{fig:HCNHNCTKin}. 
Theoretically, at a temperature of T$>30\,$K, the equal formation of HCN and HNC is offset by the conversion of HNC into HCN. 
We find no clear correlation between our line ratios and kinetic gas temperature, in agreement with other extragalactic studies at kpc scales in the EMPIRE galaxies \citep{eibensteiner_23_2022} and $\sim100\,$pc in NGC~6946, M51, NGC3627 \citep{eibensteiner_23_2022} and M83 \citep{Harada_2024}. The latter find HCN/HNC to depend on UV illumination instead.

We note that the effects of AGN feedback on the molecular ISM and the extent of this impact 
is not well known in M\,51. 
Our data reveal both a clear enhancement in line intensities and line ratios in the very center (r$\lesssim500\,$pc, i.e., Figure~\ref{fig:Gallery_fdense}), which might agree with radio emission from a low-inclined radio jet  \citep{matsushita_resolving_2015,querejeta_agn_2016}. \citet{querejeta_dense_2019} find a clear enhancement of HCN/CO due to the AGN in M51's center at $4''$ resolution. 
In fact, observations of dense-gas tracing molecules such as HCN, \hcop and HNC in galactic starburst-driven \citep[NGC~253;][]{meier_alma_2015} or AGN-driven outflows \citep[Mrk231; NGC1068;NGC1377;][]{aalto_detection_2012, aalto_2015_Mrk, Garcia_Burillo_2014_1068,aalto_alma_2020} at resolutions down to a few pc suggest the survival, formation and/or emission enhancement of dense gas in outflows \citep{Jorgensen_2004_Protostellar, Tafalla_2010, aalto_2015_Mrk}. 
The nuclear outflow will be discussed in much greater detail in upcoming papers (Thorp et al. in prep., Usero et al. in prep).

\subsection{Anomalies in the molecular ring}
\label{sec:Discussion:Ring}

\begin{figure}
    \centering
    \includegraphics[width = 0.5\textwidth]{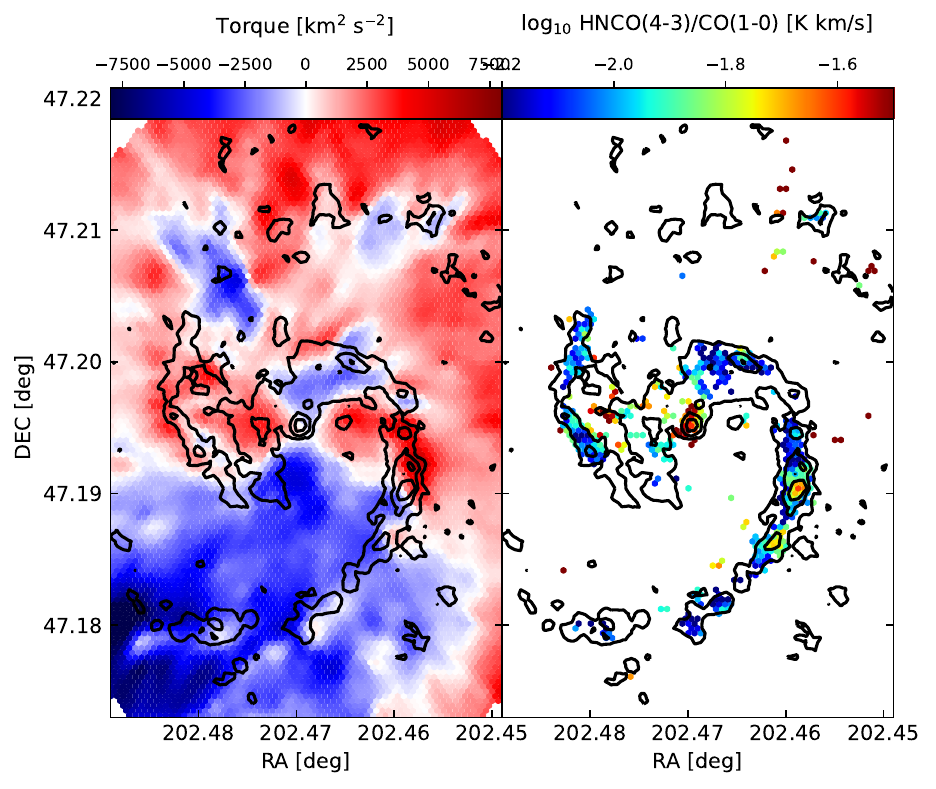}
    \caption{(left) Torques per unit mass from \citet{querejeta_inflow_2016} estimated from dust corrected (3.6$\,\textmu$m) Spitzer images. (right) HNCO(4-3)/CO(1-0) emission from SWAN \citep{Stuber_2025} for pixels where HNCO(4-3) is detected >$3\sigma$. \nnhp contours are plotted on top. }
    \label{fig:Dynamics}
\end{figure}

The molecular ring in M51 stands out as a peculiar environment: 
Average line intensities of HCN, \hcop and HNC follow \nnhp emission only sub-linearly (Table~\ref{tab:Fitparams_binnedbyN2HP}), and consequently, line ratios with \nnhp as function of \Smol exhibit significantly negative trends (Figure~\ref{fig:average_lineratio_n2hp}). 
The \nnhp/CO-to-\Smol relation in the ring has the largest slope ($0.8\pm0.05$, Appendix~\ref{App:Environments}) compared to all other environments and all other CO line ratios.  
The ring environment has the largest absolute number of pixels in which \nnhp is detected ($>3\sigma$) compared to the other environments. As it has the second highest fraction of detected pixels (28\%, compared to the center with 52\%),
effects of SNR and beam dilution are expected to play a much lesser role than for the northern and southern spiral arm. 
In addition, the brightest \nnhp emission is located in the ring.
This region was previously speculated to drive the super-linear relation between \nnhp and HCN emission (S23) and the steepening of \nnhp intensity function of \Smol and \Pde (Figure~\ref{fig:average_intensity_sigmas}) is likely driven by this region. 
While \Pde is generally high in this region (Figure~\ref{fig:Gallery-Sigmas}), the ratio of internal to dynamical equilibrium pressure (Figure~\ref{fig:PintPDE}) is not particularly enhanced in the \nnhp bright region compared to other regions along the southern arm. 

The redistribution of gas and gas flows are crucial in shaping cloud conditions and are not explicitly contained in the P$_\mathrm{int}$/\Pde ratio. 
Therefore, we show a map of torques in M51 from \citet{querejeta_inflow_2016} based on 3.6\,$\textmu$m Spitzer images that are corrected for dust emission in Figure~\ref{fig:Dynamics}.  
These estimates give insights into the gravitational torques exerted by the stellar potential on the gaseous disk. The torque map multiplied with the total gas mass per pixel - which can be inferred from the line intensity - describes the change rate of angular momentum over time. 
This \nnhp-brightest region is located at the transition between normal star-forming regions and the star-formation \textit{desert} where torques are high \citep{querejeta_inflow_2016}. 
We find that this \nnhp-bright region is located at a spot of minimum torque, between two regions of opposing torques (Figure~\ref{fig:Dynamics}). 
We speculate that the observed high fractions of dense gas might stem from
the action of torques in the immediate vicinity that preferentially bring material to that location from both smaller and larger galactocentric radii (where torques are positive and negative, respectively).
This is in agreement with a peak in HNCO(4-3)/CO(1-0) emission which we also display in Figure~\ref{fig:Dynamics}). 
HNCO emission is known to trace shocks induced by the spiral arms, and is comparably bright in the exact location of the \nnhp-bright spot. 
HNCO/CO being elevated perhaps supports this opposing flow/shock interpretation. 

While this is a speculative explanation for non-linear behavior of the \nnhp-bright spot, it does not explain the general offset of the ring environment compared to other environments.
We find no global trends between average line ratios and torques, nor with average line intensity and torques.

\subsection{Differences and similarities in northern vs southern spiral arm}
\label{sec:Discussion:spiralarmcontest}

M51 is known for an asymmetric behavior between northern and southern spiral arm likely induced by the dwarf companion galaxy M51b \citep[e.g., ][]{Henry_2003Asymmetry,schinnerer_pdbi_2013}. 
While the northern spiral arm shows prominent regions of high \SSFR, the southern arm is rich in molecular gas emission (CO, HCN) but lacks high values of \SSFR (see Figure~\ref{fig:Gallery}). 
Many literature studies have found the southern arm to be inefficient in forming stars \citep[e.g.,][]{querejeta_dense_2019}. 
\citet{querejeta_dense_2019} studied HCN at the same resolution in three pointings in M51 and when comparing peaks of HCN emission to the corresponding SFRs traced by 33\,GHz emission, they find lower dense gas SFEs in the southern arm in contrast to the northern arm. 

Our data, however, reveals that when averaging the northern disk and southern disk instead of focusing on bright spots only, the northern disk is overproducing stars, while the southern arm shows consistent behavior with the molecular ring and center. 
Per unit value of \SSFR, all average line intensities of HCN, HNC, \hcop and \nnhp are reduced in the northern disk compared to not only the southern disk, but also the center and ring environment (Appendix~\ref{App:Environments}). 
This overproduction seems independent of dense gas properties, as the dense gas line ratios as function of all physical properties agree between northern and southern disk (Figure~\ref{fig:Environments}). Thus while there are differences related to \SSFR in the two spiral arms, the molecular gas chemistry might be more or less unaffected. 

Potential explanations for these differences could include that the high SFR in the northern arm is destroying denser gas, thus reducing the average dense gas emission, whereas the southern arm, revealing high values of \vdisp, and high values of P$_\mathrm{int}$/\Pde (Figure~\ref{fig:PintPDE}) is in a complex dynamical state.

\section{Summary}
\label{sec:Summary}

We present observations of the  J=1-0 emission of HCN, HNC, \hcop and \nnhp tracing the dense gas in the  inner 5$\times$7 kpc of the Whirlpool galaxy (M51) at 125\,pc resolution from \textit{Surveying the Whirlpool at Arcseconds with NOEMA} (SWAN). 
These observations cover very diverse environments, including spiral arms, a molecular ring, and the AGN affected center. 
We compare the emission of these molecules with each other, as well as with surface densities of molecular gas mass (\Smol), SFR (\SSFR), stellar mass (\Sstar) and estimates of \Pde and \vdisp, and find the following:

\begin{enumerate}

    \item The average emission of HCN, HNC, \hcop and \nnhp  is generally correlated with each other. 
    Average HCN, HNC and \hcop intensities follow \nnhp intensities, and so do HCN/CO, HNC/CO and \hcop/CO ratios, but we find clear variations in slopes with galactic environments.  

    \item While average HCN, HNC, \hcop and \nnhp are well correlated with 
    \Smol, \SSFR, \Sstar, \Pde and \vdisp over several orders of magnitude, the slopes of these correlations vary from line to line and from environment to environment.  
    The average \nnhp emission depends significantly steeper on \Smol (slope of $1.48\pm0.05$) compared to HCN, HNC and \hcop ($1.05\pm0.04$, $1.26\pm0.02$, $1.19\pm0.03$, 
    respectively). 
    The slope of the average HCN-to-\Smol relation is in agreement with the slope of the average CO-to-\Smol relation. 
    Consequently, the HCN-, HNC- and \hcop-to-\nnhp ratios are anti-correlated with \Smol. 
    The different dependency of dense gas lines on \Smol could imply a different dependency on column densities as found in Milky Way studies, or it could  stem from beam dilution effects. 

    \item The average HCN/CO, HNC/CO, \hcop/CO and \nnhp/CO relation shows a surprisingly shallow relation with \Smol, with slopes of the average HCN/CO-to-\Smol relation as low as $0.07\pm0.03$ and consistently shallower (or negative) in all environments compared to typical literature trends. 
    While the \nnhp/CO-to-\Smol relation is steeper (slope of $0.50\pm0.05$), large variations in the slopes across the environments can be found.

    \item Dynamical equilibrium pressure \Pde is a promising parameter well related to both dense gas emission and line ratios with CO, with smaller variations in slope across the environments compared to \Smol.  We find indications of a \Pde threshold, above which \nnhp emission is increased compared to HCN, HNC and \hcop. 
    Still, these CO line ratios have increased scatter when correlated with \Pde compared to \Sstar and \SSFR. 

    \item Variations in line ratios can be seen from one environment to the other, and on cloud-scales within individual environments. 
    First, the nuclear outflow towards north-east of the center increases emission of HCN, HNC and \hcop compared to \nnhp, and of all those lines compared to \CO. 
    Secondly, just outside the very center, within the ring, where we find an increased HCN/\hcop ratio, the decreased ratio of internal to dynamical equilibrium pressure 
    suggests that the clouds are not in balance with their environment. 
    
    Thirdly, the molecular ring hosts the region with the brightest \nnhp emission, where gas flows might converge resulting in large-scale shocks. 
    Lastly, we find asymmetries between the spiral arms. 
    While the intensity pattern of all molecular lines as function of \SSFR in the northern arm deviates from those of all other environments, its the southern spiral arm in which we see enhanced velocity dispersion, enhanced $P_\mathrm{int}$/\Pde and generally high quantities of bulk and dense molecular gas.

\end{enumerate}

These cloud-scale relations of HCN, HNC, \hcop and \nnhp across these environments in the inner disk of M51 reveal, that while to first-order, the molecules agree in their emission, when considering different environments and different physical properties, no two molecules have exactly the same emission pattern at those scales. 
Only SWAN can reveal these variations that occur on both larger-scale environments and on cloud-scales within environments. 
In other galaxies,  where the faint \nnhp emission is difficult to obtain, and the conditions are similar to the conditions in the inner disk of M51, the emission of \hcop, which is nearly as bright as HCN, might be the favorable option to trace dense molecular gas.

\bibliographystyle{aa} 
\bibliography{M51paper2.bib} 

\newpage 
\begin{acknowledgements}
This work was carried out as part of the PHANGS collaboration and is based on data obtained by PIs E.\ Schinnerer and F.\ Bigiel with the IRAM-30 m telescope and NOEMA observatory under project ID M19AA.
IRAM is supported by INSU/CNRS (France), MPG (Germany) and IGN (Spain)
SKS thanks several PHANGS team members, specifically Adam Leroy for helpful discussions. 
SKS acknowledges financial support from the German Research Foundation (DFG) via Sino-German research grant SCHI 536/11-1.
AU, MJD, and MQ acknowledge support from the Spanish grant PID2022-138560NB-I00, funded by MCIN/AEI/10.13039/501100011033/FEDER, EU.
JPe acknowledges support by the French Agence Nationale de la Recherche through the DAOISM grant ANR-21-CE31-0010 and by the Thematic Action “Physique et Chimie du Milieu Interstellaire” (PCMI) of INSU Programme
National “Astro”, with contributions from CNRS Physique \& CNRS Chimie, CEA, and CNES.
JS acknowledges support by the National Aeronautics and Space Administration (NASA) through the NASA Hubble Fellowship grant HST-HF2-51544 awarded by the Space Telescope Science Institute (STScI), which is operated by the Association of Universities for Research in Astronomy, Inc., under contract NAS~5-26555.
HAP acknowledges support from the National Science and Technology Council of Taiwan under grant 113-2112-M-032 -014 -MY3.
 J.d.B. acknowledges support from the Smithsonian Institution as Submillimeter Array (SMA) Fellow.
SCOG and RSK acknowledge financial support from the European Research Council via the ERC Synergy Grant ``ECOGAL'' (project ID 855130),  from the German Excellence Strategy via the Heidelberg Cluster of Excellence (EXC 2181 - 390900948) ``STRUCTURES'', and from the German Ministry for Economic Affairs and Climate Action in project ``MAINN'' (funding ID 50OO2206). 
RSK also thanks the 2024/25 Class of Radcliffe Fellows for highly interesting and stimulating discussions. 
 \end{acknowledgements}

\begin{appendix}

\section{Mock data to test the effect of low SNR}
\label{sec:Mockdata}

Comparing the emission of molecular lines with different intrinsic intensities in data sets with similar noise levels is challenging, as fainter emission lines (such as \nnhp(1-0)) are much more strongly affected by noise, resulting in a lower fraction of significant detections than brighter lines. 
Including/excluding regions based on their SNR alone can thus bias any measurement of lines of different brightness differently. 
To address this issue, we simulate the effect of different brightness levels using the CO(1-0) data cube, which contains a lot of signal compared to its noise level with significant detections across most of our FoV. 
We present the methodology to generate data cubes representing fainter lines (Section~\ref{sec:Mockdata:methods}), and create moment-maps. 
We explore different commonly used approaches to determine average intensities and line ratios based on the mock-data in Section~\ref{sec:Mockdata:intensities}. 
Since the noise in observations averages to zero, binning is commonly used to recover  relations. We test the effect on binning on our mock-data sets in Section~\ref{sec:Mockdata:binning}.

\subsection{Creation of Mock-data sets}
\label{sec:Mockdata:methods}

We take the CO(1-0) data cube from PAWS matched to the SWAN data at 3$\arcsec$ resolution with a spectral resolution of 10\,km/s (see Section~\ref{sec:Data:CO}). 
First, we scale the intensity of the CO data cube by a constant line ratio $S$. This factor represents the true line ratio we want to recover in our analysis. We call this scaled data cube $A$.
Next, we create a data cube filled with only simulated noise. At each pixel and for each channel we randomly draw values from a Gaussian distribution with arbitrary amplitude. 
We then spatially convolve the Gaussian noise for each channel of the noise cube by our beam size of $3\arcsec$. This will correlate the noise of neighboring pixels as it is the case for our interferometric observations from SWAN. 
While technically the noise is not just correlated across a beam, but across larger scales as well, this is not included in this simple test. 
The amplitude of the correlated noise in the noise data cube is scaled to  match an RMS of 8.6\,mK which corresponds to the RMS of \nnhp from our SWAN observations \citep{Stuber_2025}.  
Lastly, we add the cube containing the correlated scaled noise to the scaled CO data cube $A$. 
We do not add the cube in quadrature, as the relative noise in the scaled CO cube is negligible compared to the noise added. 
This final data cube is our ''mock-data cube''. 

We perform these steps with line ratios of $S=0.024, 0.016, 0.007, 0.0036$.  
The noise cube is drawn anew for each $S$. 
Those are line ratios obtained from different methods to calculate the \nnhp-to-CO line ratio within the literature \citep[i.e., see][]{jimenez-donaire_constant_2023,schinnerer_molecular_2024}. 
Those values further roughly match our typical \hcop-to-\CO, HNC-to-\CO and \nnhp-to-\CO line ratios ($0.033, 0.018, 0.008$, Table~\ref{tab:Averagelineratios}) with an additional even smaller line ratio to account for individual regions (i.e. norhtern disk) with increased noise.

\begin{figure*}
    \centering
    \includegraphics[width=1\textwidth]{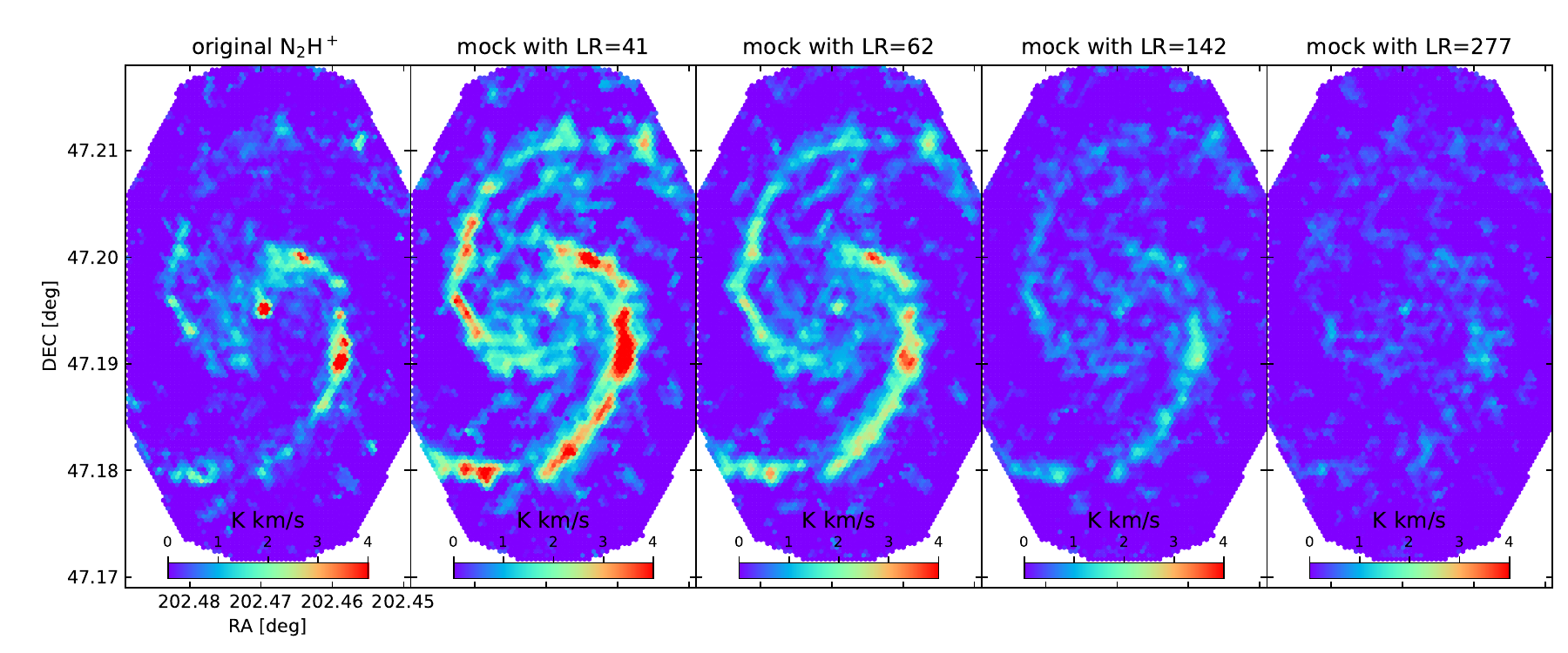}
    \caption{Integrated emission of \nnhp(1-0) and Mock data cubes based on the original CO(1-0) data cube, scaled down by a line ratio of (from left to right) LR=41,62,142,277 and adding noise to simulate lower SNR observations. }
    \label{fig:Mock-Gallery}
\end{figure*}

Our measurements of the observed SWAN line emission (Section~\ref{sec:Results} and following) is not performed on the data cubes directly, but on the moment maps (Section~\ref{sec:Data:SWANmoments}). 
Therefore, we create a PyStructure based on the exact same criteria as before (using HCN and CO as priors, Section~\ref{sec:Data:SWANmoments}), integrating the Mock data cubes. 
The resulting moment-0 maps are shown in Figure~\ref{fig:Mock-Gallery}.
We add the moment-0 map of \nnhp(1-0), the faintest of the dense gas tracing lines, for comparison. 

\subsection{Recovering average intensity ratios}
\label{sec:Mockdata:intensities}

\begin{figure}
    \centering
    \includegraphics[width=0.5\textwidth]{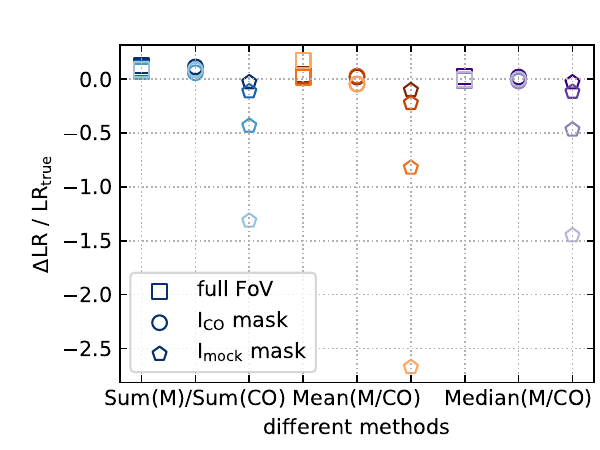}
    \caption{Testing different methods to calculate CO line ratios and the effect of masking. We show the relative difference between the true and measured line ratio $\Delta$LR/LR$_\mathrm{true}$ with $\Delta$LR=LR$_\mathrm{true}$ - LR$_\mathrm{measured}$. We estimate the measured line ratio LR$_\mathrm{measured}$ per three different methods (sum: sum(M)/sum(CO) in blue, mean: mean(M/CO) in orange, median: median(mock/CO) in purple) with M and CO indicating the intensity of the mock and CO data, respectively. We further test the impact of different masking on the moment maps: Utilizing all pixels in the FoV (squares), only pixels where CO is significantly detected ($>3\sigma$, circle), and only pixels where \nnhp is significantly detected ($>3\sigma$, pentagons). The color gradient indicates the SNR of the data cube (measured via LR$_\mathrm{true}$), with darker color being LR$_\mathrm{true}=0.024$ and the faintest color LR$_\mathrm{true} = 0.0036$. Points of the same color (albeit different gradient) belong to the same method.}
    \label{fig:Mockdata-averageCOlineratios}
\end{figure}

We compare different methods to recover average line ratios between our mock-data and the original \CO data in Figure~\ref{fig:Mockdata-averageCOlineratios}. 
These methods are the mean and median line ratio of all pixels, as well as the integrated intensity of the mock data divided by the integrated CO data (sum(M)/sum(CO) = $\Sigma i_\mathrm{mock}$ / $\Sigma i_\mathrm{CO}$). 
We apply these methods to a) the full FoV, excluding the edges where the noise is increased \citep[see][]{Stuber_2025}, b) a I$_\mathrm{CO}$-based mask and c) a I$_\mathrm{mock}$-based mask. 
The I$_\mathrm{CO}$ (I$_\mathrm{mock}$) mask exclude pixels in which the CO (mock) intensity is below $3\sigma$. 
In Figure~\ref{fig:Mockdata-averageCOlineratios} we show the relative difference between the measured line ratio and the true line ratio (LR$_\mathrm{true}$) used in the generation of the mock cubes. 

The deviations between measured and true line ratio between different methods are small compared to the deviations between different masking techniques.  
The largest deviations occur when applying the I$_\mathrm{mock}$-based mask. The smaller the true CO-line ratio (the smaller the SNR), the larger the deviation due to masking. 
The I$_\mathrm{mock}$-based mask is biased towards fewer and brighter pixels compared to the other masks. 
This will result in a bias towards more positive measurements. 

The difference between values derived using the full FoV and within the I$_\mathrm{CO}$-based mask is small as expected. As the moment-0 maps are created using CO and HCN emission as a prior, there are only few pixels within the resulting moment-0 map where the integrated \CO emission is less than 3 times its corresponding value.

\begin{figure}
    \centering
    \includegraphics[width=0.5\textwidth]{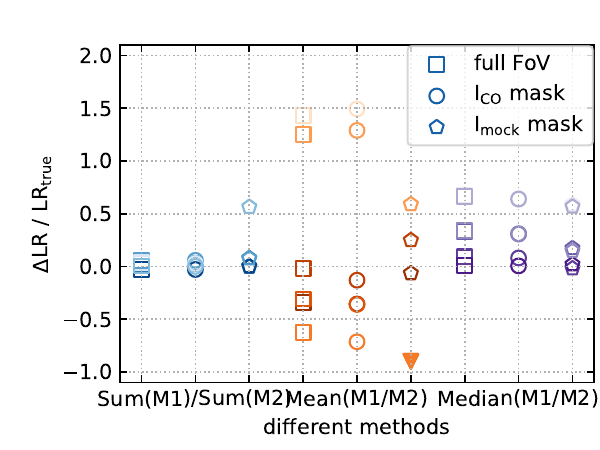}
    \caption{Testing the effect of masking on different methods to calculate average line ratios of mock-data. Same as Figure~\ref{fig:Mockdata-averageCOlineratios} but for line ratios between different mock data sets. 
    The color gradient indicates the expected line ratio (measured via LR$_\mathrm{true}$). The ratio of Mock41 and Mock227 results in the largest line ratio (and thus faintest data point). Two outliers with large $\Delta$LR/LR$_\mathrm{true}$ are only shown as lower limits (colored triangle). 
    }
    \label{fig:Mockdata-averagelineratios}
\end{figure}

We test the calculation of line ratios between the mock data, that are all much fainter than the CO line, in Figure~\ref{fig:Mockdata-averagelineratios}. 
Similar to before we test both different calculation methods, and different masking techniques. 
We find the sum-method (integrated intensity of mock1 divided by integrated intensity of mock2) results in the most reliable results, even for large true line ratios. 
This method is also least effected by masking.

\subsection{Binning Mock data intensities and ratios}
\label{sec:Mockdata:binning}

\begin{figure*}
    \centering
    \includegraphics[width=1\textwidth]{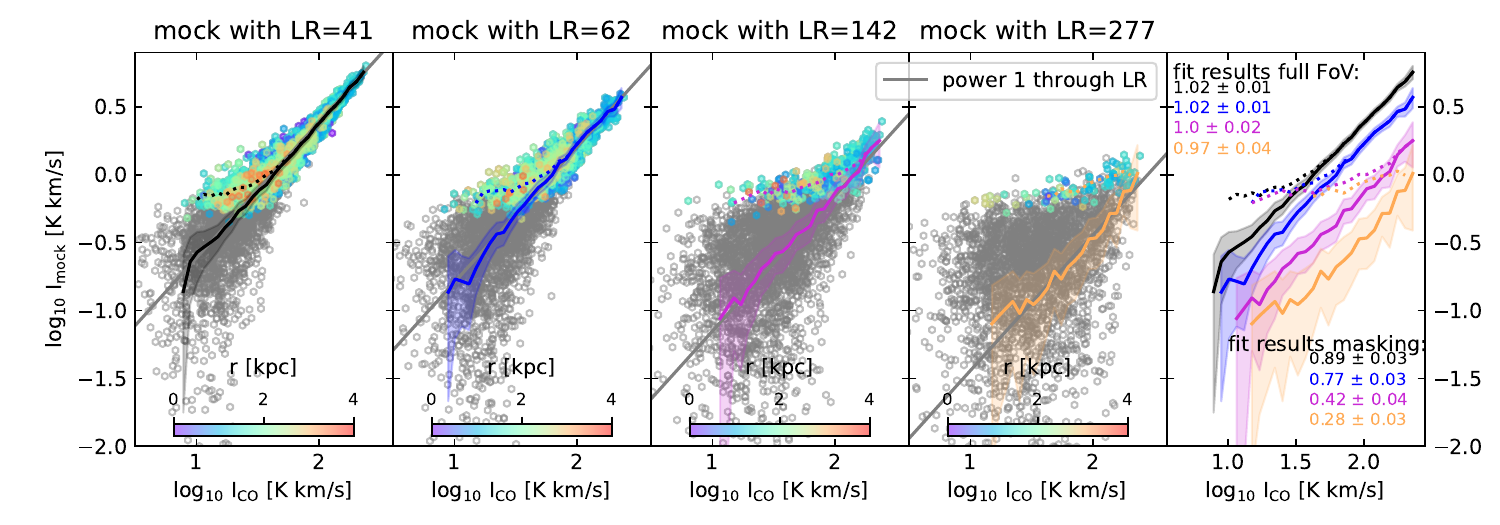}
    \caption{Testing the effect of binning to recover average line relations. We show the intensity of each pixel in the mock-moment-0 map (I$_\mathrm{mock}$) as function of the correponding intensity in the \CO moment-0 map (I$_\mathrm{CO}$. The different mock-data are simulated based on \CO data scaled by a line ratio of 0.024, 0.016, 0.007, 0.0036 (aka LR: 41, 62, 142, 277). 
    The expected relation (black dashed line) corresponds to line ratio used to create the mock data (power of 1, offset corresponds to the line ratio).  
    We highlight pixels where I$_\mathrm{mock}$ is significantly detected (>$3\sigma$, colored data points) and indicate their galactocentric distance by color. 
    We bin I$_\mathrm{mock}$ in logarithmic steps of I$_\mathrm{CO}$ for all pixels in the FoV (solid line), as well as when only including pixels with significant detections (dotted line). We add the slopes of linear fits fitted to the binned data for the full FoV and the masked data. }
    \label{fig:Mockdata-binningtests}
\end{figure*}

When measuring relations between molecular line emission and other parameters such as, e.g., galactocentric radius or molecular gas surface density, binning the emission (a.k.a. averaging the emission in increments of, e.g., galactocentric radius) is a commonly used technique to recover the physical relation that is obscured by noise. 
Ideally, the noise (including positive and negative noise values) in each bin will average to zero and the actual signal can be recovered. 
We test how well binning recovers the known relation between our mock data and the \CO data in Figure~\ref{fig:Mockdata-binningtests}. The binning method used is described in Section~\ref{sec:Methods:Binning}.
The true relation is very well recovered when all data points are used. Masking pixels ($>3\sigma$)  results in a flattening of the average relation for low CO intensities. 

Fitting  those masked and unmasked binned line ratios (Section~\ref{sec:Methods:Binning}) confirms this bias: 
When using all data, we can recover the true relation (slope of 1) in all of the mock data sets, even in the data with lowest SNR (LR=277). 
When restricting the data to $>3\sigma$, slopes are artificially decreased with decreasing average SNR of the mock data and significantly deviate from the true slope of unity.

\begin{figure}
    \centering
    \includegraphics[width=0.5\textwidth]{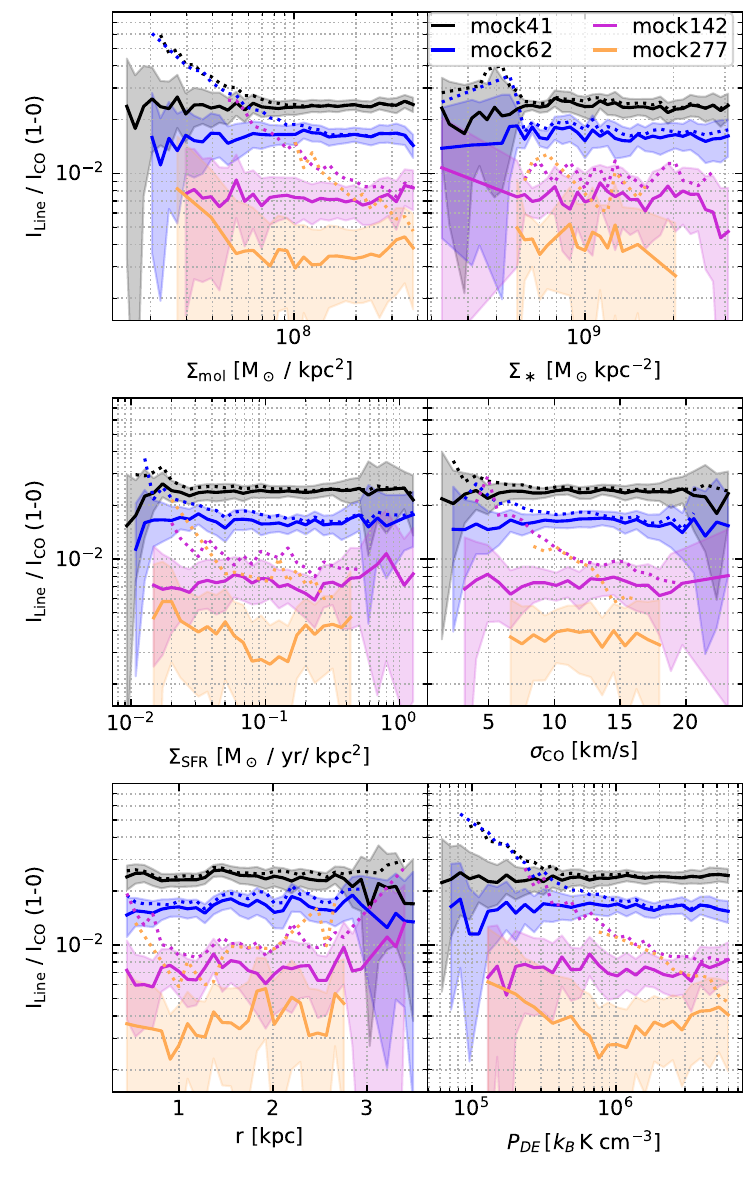}
    \caption{Binned mock-to-CO line ratios as function of physical parameters such as molecular gas mass, stellar mass and SFR surface densities, velocity dispersion, galactocentric radius and dynamical equilibrium pressure. We show the binned line ratios using all pixels (solid line), as well as restricting to detections (intensity >$3\sigma$, dotted lines). }
    \label{fig:Mockdata-binningvsparameters}
\end{figure}

In addition to binning line intensities, we test if we are able to recover the relation between line ratios and other more independent parameters, such as molecular gas, star-formation rate and stellar mass surface densities, galactocentric radius, velocity dispersion and dynamical equilibrium pressure used in our analysis (Section~\ref{sec:ResultsEnv}).
Figure~\ref{fig:Mockdata-binningvsparameters} shows the binned mock-to-CO line ratios as function of those parameters. We expect a flat relation (slope 0) with different offsets along the y-axis depending on the mock data used.  

Most mock-to-CO line ratios are flat as expected (Figure~\ref{fig:Mockdata-binningvsparameters}). 
Again, masking pixels $<3\sigma$ introduces significant biases and results in artificial slopes that are largest for the mock-data with the lowest SNR. 
This bias is strongest for relations of line ratios with \Smol, $\sigma_\mathrm{CO}$ and PDE, with a $>5\sigma$ deviation of slopes obtained from the lowest SNR mock data (LR=277) to the true slope of zero. 
When including all data, we can recover the expected slope of zero for all mock data sets as function of all galactic parameters.

This simple mock-data set is not a perfect representation of the actual noise, as it does not consider noise correlated over scales larger than a beam-size. 
Still, it shows the impact of excluding parts of the data on the ability to retrieve the correct result. 
The actual correlations shown might suffer from additional noise sources not considered in this mock-data set.

\subsection{Estimating the scatter of mock data relations}
\label{sec:Mockdata:Scatter}

\begin{figure}
    \centering
    \includegraphics[width=0.24\textwidth]{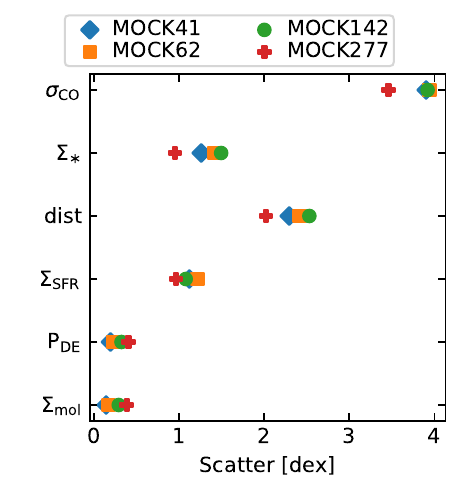}
    \includegraphics[width=0.24\textwidth]{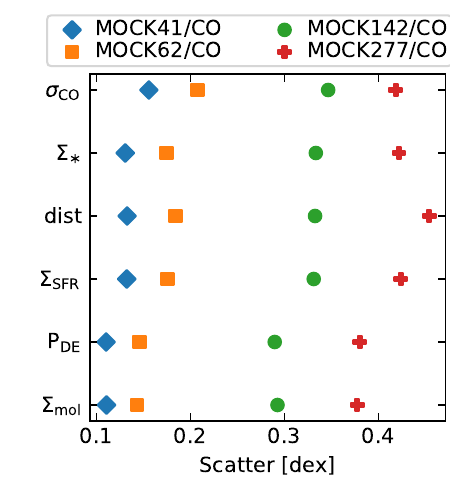}
    \caption{Average scatter of mock intensity (left panel) and mock-to-CO line ratios (left panel) as function of galactic physical properties tested for different mock data sets (Mock41-Mock277).}
    \label{fig:scatter-mockintensities}
\end{figure}

The scatter of a relation between line emission and physical parameters is equally important in assessing the strength of the relation as the slope. As an example, if the relation between line intensity and two galactic properties show similar slopes, increased scatter in one may suggest a secondary dependency on other factors.  
Unfortunately, a lower SNR coincides with an increased scatter, which challenges the interpretation of the scatter calculated using the line emission of lines of different average SNR. 
We test this impact of SNR on the calculation of the scatter with our mock data and galactic parameters. 
Figure~\ref{fig:scatter-mockintensities} shows the average scatter of mock line intensities as function of physical galactic properties. 
The scatter is the medium absolute deviation of the line intensity as function of galactic properties after subtracting a linear relation fitted to the binned intensities (Section~\ref{sec:Methods:Binning}). 
Per galactic quantity, the maximum deviation in scatter from different mock data sets ranges from 0.2\,dex (relations with \Pde, \Smol) to 0.5\,dex (\vdisp, \Sstar, dist).
The variations in scatter between the different galactic quantities is expected, as some quantities (\Smol, \Pde) depend on the \CO intensity, which is the basis for the mock data set.

In addition to line intensity, we test the scatter of relations between mock-to-\CO line ratios and galactic physical properties in Figure~\ref{fig:scatter-mockintensities}. 
For each galactic property, the average scatter increases significantly with decreasing average SNR (lowest SNR for mock277), as expected.
For each mock data set, variations in the mock-to-CO relation with different galactic properties are small:  
We find the least variations in scatter  across galactic properties for the highest SNR data set (mock41, scatter varies by $\sim0.04$, followed by $\sim0.06$, $\sim0.06$, $\sim0.08$ for mock62, mock142 and mock277 respectively.)
These variations in the average scatter are driven by the varying SNR and different sorting due to the galactic parameters.

\section{Testing different prescriptions of \Smol}
\label{App:alphaCO}

\begin{figure}
    \centering
    \includegraphics[width=0.4\textwidth]{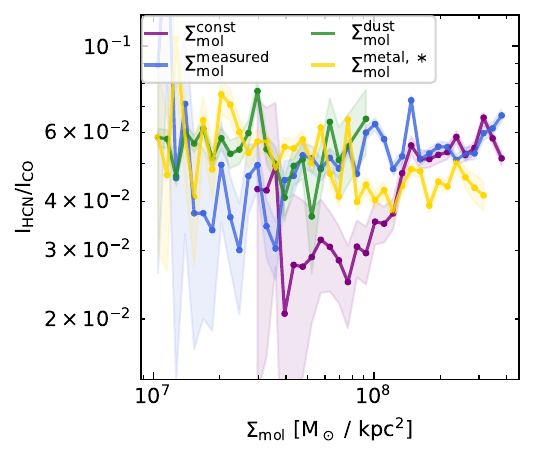}
    \caption{Average HCN/CO line ratio as function of \Smol, with \Smol being calculated with four different methods: a) a constant $\alpha_\mathrm{CO}$ conversion factor multiplied to \CO emission ($\Sigma^\mathrm{const}_\mathrm{mol}$) b) the measured $\alpha_\mathrm{CO}$ as described in Section\ref{sec:Data:CO}, $\Sigma^\mathrm{measured}_\mathrm{mol}$ c) a metallicity and stellar mass based $\alpha_\mathrm{CO}$, $\Sigma^{\mathrm{metal}, \ast}_\mathrm{mol}$, and an    
    independent dust-based estimate of molecular gas surface density from \citet{faustino_vieira_molecular_2024}, $\Sigma_\mathrm{mol}^{dust}$.}
    \label{fig:Fdense_vs_varyingSigmamol}
\end{figure}

We test the performance of the average HCN-to-\CO line ratio when comparing to different prescriptions of calculating \Smol in Figure~\ref{fig:Fdense_vs_varyingSigmamol}
The methods to calculate \Smol are described below. 

A common method to estimate the CO-to-H$_2$ conversion factors is via the metallicity and stellar mass surface density in the disk, which we refer to as $\Sigma^{\mathrm{metal}, \ast}_\mathrm{mol}$.
We first obtain the $\alpha_\mathrm{CO}$ distribution based on radial metallicity profiles from CHAOS \citep{croxall_chaos_2015}. These are converted into $\alpha_\mathrm{CO}$ values according to the prescription in \citet{bolatto_co--h2_2013}, which relies not only on the metallicity but also on kpc-surface density of stellar mass, atomic gas and molecular gas. 
The values were calculated iteratively as described in \citet{sun_molecular_2022}. 
Inside the SWAN FoV, $\alpha_\mathrm{CO}$ has a median of $1.99\,\mathrm{M}_\odot$\,(K\,km\,s$^{-1}$\,pc$^{-2}$)$^{-1}$, with 16$^\mathrm{th}$ and 84$^\mathrm{th}$ percentiles of $1.35$, $2.75\,\mathrm{M}_\odot$\,(K\,km\,s$^{-1}$\,pc$^{-2}$)$^{-1}$, respectively. It also has a strong radial gradient: the central 1\,kpc (diameter) median  $\alpha_\mathrm{CO}$ is $0.72\,\mathrm{M}_\odot$\,(K\,km\,s$^{-1}$\,pc$^{-2}$)$^{-1}$, the median $\alpha_\mathrm{CO}$ at  galactocentric distance  larger than 3\,kpc is $2.81\,\mathrm{M}_\odot$\,(K\,km\,s$^{-1}$\,pc$^{-2}$)$^{-1}$.

We add our first prescription, using a measured $\alpha_\mathrm{CO}$ that is applied to the integrated \CO intensities (see Section~\ref{sec:Data:CO}). We refer to this surface density as $\Sigma_\mathrm{mol}^\mathrm{measured}$. 

If the $\alpha_\mathrm{CO}$ value is not known precisely, a common method is to assume a constant based on Galactic measurements \citep{bolatto_co--h2_2013}. 
Commonly used in extragalactic targets is a value of $\alpha_\mathrm{CO} = 4.35 \mathrm{M}_\odot$\,(K\,km\,s$^{-1}$\,pc$^{-2}$)$^{-1}$. We calculate the gas surface density by applying this constant conversion factor to the \CO moment-0 map. We refer to this surface density as $\Sigma^\mathrm{const}_\mathrm{mol}$.

Lastly, we add measurements of gas surface density from \citet{faustino_vieira_molecular_2024} based on HST observations for comparison. 
We refer to this surface density as $\Sigma_\mathrm{mol}^{dust}$.

While there are difference in the average HCN/CO ratio as function of those different values of \Smol, we note that most of them are similar, comparably flat, or even result in negative trends ($\Sigma^{\mathrm{metal}, \ast}_\mathrm{mol}$).
The relation obtained using a constant CO-to-H$_2$ conversion factor reveals the steepest relation, but we note, that a constant value of $\alpha_\mathrm{CO}$ is not able to reflect the changes expected in environment, such as a generally increased $\alpha_\mathrm{CO}$ in galaxy centers \citep{schinnerer_molecular_2024}.

\section{The impact of the central environment on the average HCN/CO to \Smol relation}
\label{App:fdense_Sigmamol_radialtrend}

\begin{figure}
    \centering
    \includegraphics[width=0.4\textwidth]{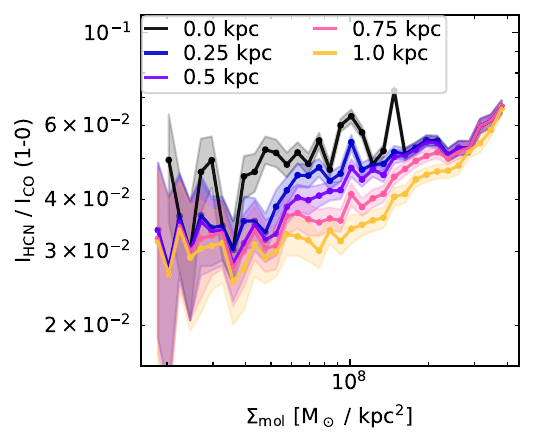}
    \caption{Average HCN/CO line ratio as function of \Smol when excluding pixels inside the central $r=0, 0.25, 0.5, 0.75$, and $1$\,kpc in radius. The ratio is binned as described in Section~\ref{sec:Methods:Binning} and we use the same values of \Smol, that we use in the main text.  }
    \label{fig:fdense_sigmamol_varyingcenter}
\end{figure}

We test the effect of the central pixels on the performance of the HCN/CO-to-\Smol relation by removing central apertures of increasing sizes in Figure~\ref{fig:fdense_sigmamol_varyingcenter}. 
Further, we fit the binned HCN/CO-to-\Smol relations and provide fit slopes in Table~\ref{tab:ApertureHCNCO}. We add Spearman correlation coefficients ($\rho_\mathrm{Sp}$) for the respective apertures in Table~\ref{tab:ApertureHCNCO}. We note that $\rho_\mathrm{Sp}$ is sensitive to the SNR of the line, therefore we expect brighter lines to have higher values of $\rho_\mathrm{Sp}$. 

While we can see an increase in the slopes of the HCN/CO-to-\Smol relation with increasing aperture that is removed from the analysis, the values stay overall small. At a radius of $\sim1\,$kpc the molecular ring is reached. 
Figure~\ref{fig:fdense_sigmamol_varyingcenter} shows that particularly pixels at intermediate values of \Smol are removed when excluding  central apertures of increasing sizes. 
The molecular outflow driven by the AGN is most prominent in a small region with the size of $\sim200\,$pc in the center \citep{querejeta_agn_2016}.  The actual impact of the AGN on the molecular gas, however, is under discussion \citep[][see also Section~\ref{sec:Discussion}]{querejeta_agn_2016,querejeta_dense_2019,Zhang2025} and some effects might reach out to radii of $\sim 600\,$pc north of the AGN. 
The slopes of the HCN/CO-to-\Smol relation however, show a continuous increase with increased size of the excluded aperture. 
The HCN/CO ratio as function of \Smol reaches a slope of up to $0.36\pm0.02$.
Spearman correlation coefficients do not change even when excluding the central 1\,kpc in radius. 
In all cases values of $\rho_\mathrm{Sp}$ are highest for the \hcop/CO ratio as function of \Smol, despite HCN being on average brighter than \hcop.

\begin{table}\label{tab:ApertureHCNCO}
\caption{Fitting CO line ratios (LR) as function of \Smol when excluding central apertures of varying sizes}
\begin{tabular}{ccccccc}
Cen. aperture [kpc] & LR & slope & slope err & $\rho_\mathrm{Sp}$ \\
\noalign{\smallskip}
\hline 
\noalign{\smallskip}
0.0 & \nnhp/CO & 0.51 & 0.05 & 0.18 \\
0.0 & HCN/CO & 0.07 & 0.03 & 0.29 \\
0.0 & HNC/CO & 0.21 & 0.03 & 0.29 \\
0.0 & \hcop/CO & 0.28 & 0.02 & 0.35 \\
\noalign{\smallskip}
\hline 
\noalign{\smallskip}
0.25 & \nnhp/CO & 0.56 & 0.05 & 0.18 \\
0.25 & HCN/CO & 0.2 & 0.02 & 0.30 \\
0.25 & HNC/CO & 0.33 & 0.02 & 0.301 \\
0.25 & \hcop/CO & 0.35 & 0.01 & 0.36 \\
\noalign{\smallskip}
\hline 
\noalign{\smallskip}
0.5 & \nnhp/CO & 0.62 & 0.05 & 0.18 \\
0.5 & HCN/CO & 0.26 & 0.02 & 0.31 \\
0.5 & HNC/CO & 0.38 & 0.02 & 0.32 \\
0.5 & \hcop/CO & 0.38 & 0.01 & 0.37 \\
\noalign{\smallskip}
\hline 
\noalign{\smallskip}
0.75 & \nnhp/CO & 0.69 & 0.05 & 0.16 \\
0.75 & HCN/CO & 0.33 & 0.02 & 0.27 \\
0.75 & HNC/CO & 0.46 & 0.02 & 0.29 \\
0.75 & \hcop/CO & 0.41 & 0.02 & 0.35 \\
\noalign{\smallskip}
\hline 
\noalign{\smallskip}
1.0 & \nnhp/CO & 0.68 & 0.06 & 0.15 \\
1.0 & HCN/CO & 0.36 & 0.02 & 0.20 \\
1.0 & HNC/CO & 0.51 & 0.02 & 0.25 \\
1.0 & \hcop/CO & 0.43 & 0.02 & 0.33 \\
\end{tabular}
\end{table}

\section{Environmental dependency of line emission and line ratios}
\label{App:Environments}

\begin{figure*}
    \centering
    \includegraphics[width = 0.95\textwidth]{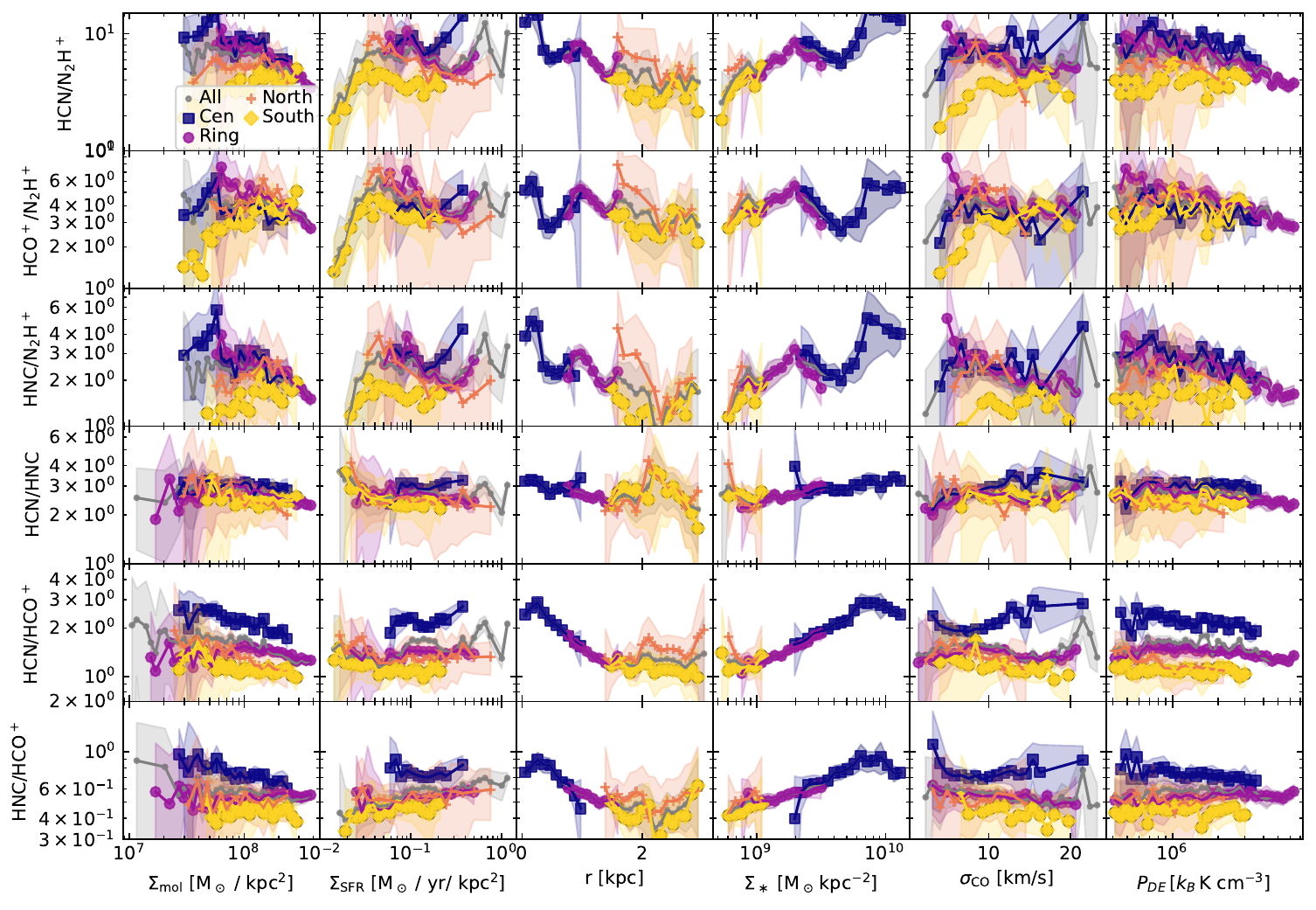}
    \includegraphics[width = 0.95\textwidth]{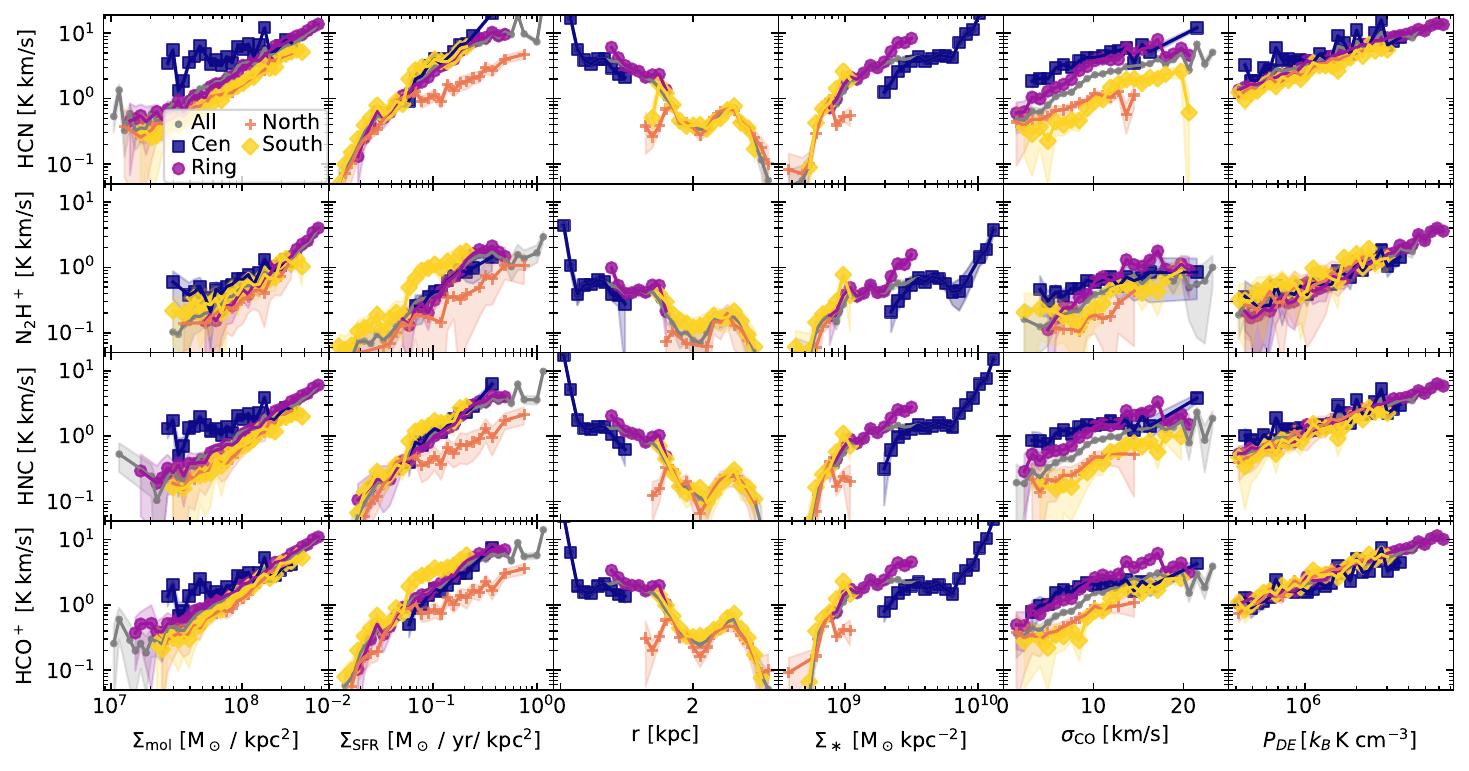}
    \caption{Average line ratios (top panels) and average line intensities (bottom panels) and as function of \Smol, \SSFR, $r$, \Sstar, \vdisp and \Pde for different environments in the disk (see Figure~\ref{fig:Gallery}. }
    \label{fig:Environments}
\end{figure*}

Figure~\ref{fig:Environments} shows the average line intensities and average line ratios for all dense gas tracers as function of all environmental parameters tested for (\Smol, \SSFR, \Sstar, $r$, \Pde, \vdisp) for different environments in the disk.
The binning is the same as for Figure~\ref{fig:average_intensity_sigmas} and Figure~\ref{fig:average_lineratio_n2hp}), but for pixels inside the corresponding environment. 
No masking is applied for average line intensities and line ratios (compare Section~\ref{sec:Methods:Binning}).
The center and ring environments are based on a kinematic analysis of \citet{colombo_pdbi_2014-1}, and we additionally separate the disk into northern and southern parts based on the galaxy center, to test for asymmetries between the spiral arms. All environments are shown in Figure~\ref{fig:Gallery}. 

For all molecules, the trend between line intensity and \Smol is offset to slightly higher line intensities in the center compared to the molecular ring  and both northern and southern disk, which agree well. 
This is in agreement with the inverse radial dependency of the line emission found in Section~\ref{sec:ResultsEnv:Intensity}.  
Similarly, there is an offset relation between average line intensity and \vdisp from one environment to the other, with the average line intensity being higher in center and ring compared to both spiral arms at the same values of \vdisp. Despite the offset, the shapes of the relations are very similar across all environments.   

For \SSFR, however, the opposite can be seen: Center, ring and southern disk agree on a single distribution of line intensity as function of \SSFR, but the northern disk depends more shallow on \SSFR. 
While the line ratios differ for the center environment as function of \SSFR, the line ratios of the northern and southern arm mostly agree. 
The line ratios with \nnhp in the denominator are generally slightly lower in the southern arm compared to the northern one, but the data points agree well within their uncertainties.

Average line ratios among the dense gas tracing lines HCN, HNC, \hcop and \nnhp generally show very similar behavior with galactic properties across the individual environment.  
The most remarkable difference seen is the relation betweeen line ratios with \nnhp and their dependency on \Smol. We explore those in more detail in the main text.

\begin{table*}
\centering
\caption{Fitting average line intensities as function of physical properties per environment}\label{tab:Fitparams_Int_vs_all_env}
\begin{tabular}{ccc|cccc}
\hline 
\hline 
\noalign{\smallskip}
 & \multicolumn{2}{c}{Full FoV} & Center & Ring & North & South \\
\noalign{\smallskip}
line & slope & offset [dex] & slope & slope & slope & slope  \\
\noalign{\smallskip}
\hline 
\noalign{\smallskip}
\multicolumn{6}{c}{Binned line intensities per \Smol }\\
\noalign{\smallskip}
\hline 
\noalign{\smallskip}
CO & 0.98$\pm$0.01 & -6.1$\pm$0.0 & 0.78$\pm$0.02 & 0.98$\pm$0.01 & 0.99$\pm$0.01 & 0.99$\pm$0.02 \\
HCN & 1.04$\pm$0.04 & -7.8$\pm$0.3 & 0.51$\pm$0.15 & 1.17$\pm$0.02 & 1.15$\pm$0.04 & 1.23$\pm$0.04 \\
HNC & 1.17$\pm$0.04 & -9.3$\pm$0.3 & 0.51$\pm$0.13 & 1.29$\pm$0.02 & 1.43$\pm$0.05 & 1.25$\pm$0.06 \\
\hcop & 1.26$\pm$0.02 & -9.8$\pm$0.2 & 0.63$\pm$0.12 & 1.26$\pm$0.02 & 1.44$\pm$0.04 & 1.29$\pm$0.05 \\
\nnhp & 1.49$\pm$0.05 & -12.3$\pm$0.4 & 0.73$\pm$0.11 & 1.78$\pm$0.04 & 1.27$\pm$0.08 & 1.01$\pm$0.08 \\
\noalign{\smallskip}
\hline 
\noalign{\smallskip}
\multicolumn{6}{c}{Binned line intensities per \Pde }\\
\noalign{\smallskip}
\hline 
\noalign{\smallskip}
CO & 0.63$\pm$0.01 & -2.0$\pm$0.1 & 0.72$\pm$0.02 & 0.7$\pm$0.01 & 0.7$\pm$0.01 & 0.69$\pm$0.02 \\
HCN & 0.83$\pm$0.03 & -4.4$\pm$0.2 & 0.64$\pm$0.13 & 0.85$\pm$0.02 & 0.82$\pm$0.02 & 0.88$\pm$0.02 \\
HNC & 0.89$\pm$0.02 & -5.3$\pm$0.1 & 0.59$\pm$0.12 & 0.92$\pm$0.02 & 1.01$\pm$0.03 & 0.91$\pm$0.04 \\
\hcop & 0.9$\pm$0.02 & -5.1$\pm$0.1 & 0.74$\pm$0.12 & 0.9$\pm$0.01 & 1.0$\pm$0.03 & 0.92$\pm$0.03 \\
\nnhp & 1.06$\pm$0.03 & -6.6$\pm$0.2 & 0.92$\pm$0.09 & 1.22$\pm$0.04 & 0.91$\pm$0.08 & 0.71$\pm$0.04 \\
\noalign{\smallskip}
\hline 
\noalign{\smallskip}
\multicolumn{6}{c}{Binned line intensities per \SSFR }\\
\noalign{\smallskip}
\hline 
\noalign{\smallskip}
CO & 0.7$\pm$0.03 & 2.3$\pm$0.0 & 0.6$\pm$0.1 & 0.82$\pm$0.04 & 0.6$\pm$0.05 & 1.05$\pm$0.09 \\
HCN & 0.99$\pm$0.04 & 1.4$\pm$0.0 & 1.28$\pm$0.12 & 1.08$\pm$0.04 & 0.71$\pm$0.06 & 1.28$\pm$0.1 \\
HNC & 0.95$\pm$0.04 & 0.9$\pm$0.0 & 1.21$\pm$0.1 & 1.12$\pm$0.05 & 0.67$\pm$0.06 & 1.29$\pm$0.11 \\
\hcop & 0.91$\pm$0.04 & 1.1$\pm$0.0 & 1.19$\pm$0.07 & 1.06$\pm$0.05 & 0.66$\pm$0.07 & 1.28$\pm$0.1 \\
\nnhp & 0.99$\pm$0.06 & 0.6$\pm$0.0 & 1.17$\pm$0.12 & 1.32$\pm$0.12 & 0.89$\pm$0.05 & 1.29$\pm$0.08 \\
\noalign{\smallskip}
\hline 
\noalign{\smallskip}
\multicolumn{6}{c}{Binned line intensities per dist }\\
\noalign{\smallskip}
\hline 
\noalign{\smallskip}
CO & -0.68$\pm$0.1 & 1.6$\pm$0.0 & -0.12$\pm$0.1 & -0.69$\pm$0.14 & -0.35$\pm$0.36 & -0.89$\pm$0.29 \\
HCN & -1.04$\pm$0.04 & 0.4$\pm$0.0 & -1.11$\pm$0.07 & -1.53$\pm$0.2 & -0.24$\pm$0.4 & -1.95$\pm$0.48 \\
HNC & -0.96$\pm$0.05 & -0.0$\pm$0.0 & -1.06$\pm$0.07 & -1.18$\pm$0.22 & -0.6$\pm$0.43 & -2.07$\pm$0.51 \\
\hcop & -0.84$\pm$0.06 & 0.2$\pm$0.0 & -0.99$\pm$0.09 & -0.84$\pm$0.21 & -0.61$\pm$0.41 & -1.64$\pm$0.48 \\
\nnhp & -0.77$\pm$0.07 & -0.4$\pm$0.0 & -0.78$\pm$0.1 & -0.62$\pm$0.41 & 1.05$\pm$0.7 & -1.41$\pm$0.4 \\
\noalign{\smallskip}
\hline 
\noalign{\smallskip}
\multicolumn{6}{c}{Binned line intensities per \Sstar }\\
\noalign{\smallskip}
\hline 
\noalign{\smallskip}
CO & 0.54$\pm$0.08 & -3.4$\pm$0.7 & 0.12$\pm$0.1 & 0.88$\pm$0.11 & 1.11$\pm$0.71 & 2.0$\pm$0.52 \\
HCN & 0.94$\pm$0.07 & -8.3$\pm$0.7 & 1.36$\pm$0.13 & 1.45$\pm$0.12 & 0.78$\pm$0.85 & 3.13$\pm$0.52 \\
HNC & 0.83$\pm$0.08 & -7.7$\pm$0.7 & 1.23$\pm$0.12 & 1.25$\pm$0.13 & 0.54$\pm$1.05 & 3.66$\pm$0.55 \\
\hcop & 0.68$\pm$0.09 & -6.1$\pm$0.8 & 1.07$\pm$0.14 & 1.08$\pm$0.13 & 0.68$\pm$0.98 & 3.1$\pm$0.55 \\
\nnhp & 0.68$\pm$0.07 & -6.6$\pm$0.7 & 0.79$\pm$0.15 & 0.99$\pm$0.2 & 2.43$\pm$1.05 & 3.12$\pm$0.59 \\
\noalign{\smallskip}
\hline 
\noalign{\smallskip}
\multicolumn{6}{c}{Binned line intensities per \vdisp }\\
\noalign{\smallskip}
\hline 
\noalign{\smallskip}
CO & 1.14$\pm$0.06 & 0.5$\pm$0.1 & 0.44$\pm$0.13 & 1.19$\pm$0.11 & 0.82$\pm$0.11 & 1.58$\pm$0.14 \\
HCN & 1.16$\pm$0.06 & -0.8$\pm$0.1 & 0.95$\pm$0.09 & 1.19$\pm$0.11 & 0.79$\pm$0.1 & 1.52$\pm$0.16 \\
HNC & 1.17$\pm$0.06 & -1.3$\pm$0.1 & 0.71$\pm$0.1 & 1.25$\pm$0.13 & 1.09$\pm$0.14 & 1.58$\pm$0.18 \\
\hcop & 1.32$\pm$0.07 & -1.2$\pm$0.1 & 0.62$\pm$0.1 & 1.32$\pm$0.12 & 0.96$\pm$0.13 & 1.72$\pm$0.19 \\
\nnhp & 1.4$\pm$0.11 & -1.8$\pm$0.1 & 0.67$\pm$0.12 & 1.75$\pm$0.24 & 1.34$\pm$0.27 & 0.95$\pm$0.19 \\
\noalign{\smallskip}
\hline 
\noalign{\smallskip}
\end{tabular}
\tablefoot{Fit parameters fitting a linear (in log space) to molecular line emission as function of (from top to bottom) \Smol, \Pde, \SSFR, dist, \Sstar, \vdisp and (from left to right) per environment. }
\end{table*}

\begin{table*}
\centering
\caption{Fitting average CO line ratios as function of galactic physical properties per environment}\label{tab:Fitparams_COLR_vs_all_env}
\begin{tabular}{ccc|cccc}
\hline 
\hline 
\noalign{\smallskip}
 & \multicolumn{2}{c}{Full FoV} & Center & Ring & North & South \\
\noalign{\smallskip}
line & slope & offset [dex] & slope & slope & slope & slope  \\
\noalign{\smallskip}
\hline 
\noalign{\smallskip}
\multicolumn{6}{c}{Binned CO line ratios per \Smol }\\
\noalign{\smallskip}
\hline 
\noalign{\smallskip}
HCN/CO & 0.07$\pm$0.03 & -1.8$\pm$0.3 & -0.27$\pm$0.13 & 0.2$\pm$0.02 & 0.17$\pm$0.04 & 0.27$\pm$0.03 \\
HNC/CO & 0.2$\pm$0.03 & -3.3$\pm$0.3 & -0.25$\pm$0.12 & 0.31$\pm$0.02 & 0.45$\pm$0.05 & 0.32$\pm$0.05 \\
\hcop/CO & 0.28$\pm$0.02 & -3.7$\pm$0.2 & -0.12$\pm$0.11 & 0.28$\pm$0.02 & 0.45$\pm$0.04 & 0.36$\pm$0.04 \\
\nnhp/CO & 0.51$\pm$0.05 & -6.2$\pm$0.4 & -0.04$\pm$0.1 & 0.8$\pm$0.05 & 0.31$\pm$0.09 & 0.02$\pm$0.08 \\

\noalign{\smallskip}
\hline 
\noalign{\smallskip}
\multicolumn{6}{c}{Binned CO line ratios per \Pde }\\
\noalign{\smallskip}
\hline 
\noalign{\smallskip}
HCN/CO  & 0.2$\pm$0.03 & -2.4$\pm$0.2 & 0.05$\pm$0.12 & 0.16$\pm$0.01 & 0.13$\pm$0.02 & 0.21$\pm$0.02 \\
HNC/CO & 0.25$\pm$0.03 & -3.2$\pm$0.2 & 0.02$\pm$0.11 & 0.24$\pm$0.02 & 0.33$\pm$0.03 & 0.29$\pm$0.03 \\
\hcop/CO & 0.25$\pm$0.02 & -3.0$\pm$0.1 & 0.11$\pm$0.11 & 0.21$\pm$0.01 & 0.32$\pm$0.03 & 0.27$\pm$0.03 \\
\nnhp/CO & 0.4$\pm$0.02 & -4.5$\pm$0.1 & 0.2$\pm$0.08 & 0.55$\pm$0.04 & 0.24$\pm$0.09 & 0.02$\pm$0.05 \\
\noalign{\smallskip}
\hline 
\noalign{\smallskip}
\multicolumn{6}{c}{Binned CO line ratios per \SSFR }\\
\noalign{\smallskip}
\hline 
\noalign{\smallskip}
HCN/CO  & 0.45$\pm$0.04 & -0.8$\pm$0.0 & 0.6$\pm$0.17 & 0.31$\pm$0.02 & 0.26$\pm$0.03 & 0.42$\pm$0.03 \\
HNC/CO & 0.46$\pm$0.03 & -1.2$\pm$0.0 & 0.68$\pm$0.15 & 0.36$\pm$0.02 & 0.25$\pm$0.04 & 0.51$\pm$0.05 \\
\hcop/CO & 0.38$\pm$0.02 & -1.1$\pm$0.0 & 0.7$\pm$0.11 & 0.29$\pm$0.02 & 0.23$\pm$0.04 & 0.45$\pm$0.03 \\
\nnhp/CO & 0.41$\pm$0.03 & -1.6$\pm$0.0 & 0.75$\pm$0.09 & 0.57$\pm$0.08 & 0.45$\pm$0.04 & 0.3$\pm$0.07 \\
\noalign{\smallskip}
\hline 
\noalign{\smallskip}
\multicolumn{6}{c}{Binned CO line ratios per per dist }\\
\noalign{\smallskip}
\hline 
\noalign{\smallskip}
HCN/CO  & -0.78$\pm$0.04 & -1.2$\pm$0.0 & -0.8$\pm$0.09 & -0.79$\pm$0.12 & 0.05$\pm$0.15 & -0.78$\pm$0.25 \\
HNC/CO & -0.7$\pm$0.04 & -1.7$\pm$0.0 & -0.77$\pm$0.07 & -0.49$\pm$0.13 & -0.38$\pm$0.26 & -0.73$\pm$0.32 \\
\hcop/CO & -0.49$\pm$0.05 & -1.4$\pm$0.0 & -0.67$\pm$0.09 & -0.16$\pm$0.12 & -0.39$\pm$0.19 & -0.51$\pm$0.23 \\
\nnhp/CO & -0.45$\pm$0.04 & -2.0$\pm$0.0 & -0.5$\pm$0.05 & 0.09$\pm$0.34 & 1.56$\pm$0.38 & -0.27$\pm$0.16 \\
\noalign{\smallskip}
\hline 
\noalign{\smallskip}
\multicolumn{6}{c}{Binned CO line ratios per \Sstar }\\
\noalign{\smallskip}
\hline 
\noalign{\smallskip}
HCN/CO  & 0.64$\pm$0.04 & -7.1$\pm$0.3 & 0.89$\pm$0.12 & 0.58$\pm$0.05 & 0.82$\pm$0.2 & 1.59$\pm$0.17 \\
HNC/CO & 0.58$\pm$0.05 & -7.0$\pm$0.5 & 0.99$\pm$0.12 & 0.38$\pm$0.06 & 1.47$\pm$0.31 & 1.92$\pm$0.31 \\
\hcop/CO & 0.33$\pm$0.05 & -4.5$\pm$0.5 & 0.63$\pm$0.14 & 0.2$\pm$0.06 & 1.08$\pm$0.35 & 1.52$\pm$0.19 \\
\nnhp/CO & 0.38$\pm$0.05 & -5.5$\pm$0.4 & 0.73$\pm$0.11 & 0.14$\pm$0.16 & 0.09$\pm$1.05 & 0.71$\pm$0.59 \\
\noalign{\smallskip}
\hline 
\noalign{\smallskip}
\multicolumn{6}{c}{Binned CO line ratios per \vdisp }\\
\noalign{\smallskip}
\hline 
\noalign{\smallskip}
HCN/CO  & 0.05$\pm$0.07 & -1.4$\pm$0.1 & 0.46$\pm$0.16 & -0.03$\pm$0.04 & -0.05$\pm$0.07 & -0.02$\pm$0.07 \\
HNC/CO & 0.06$\pm$0.06 & -1.8$\pm$0.1 & 0.29$\pm$0.15 & 0.04$\pm$0.06 & 0.22$\pm$0.12 & -0.09$\pm$0.1 \\
\hcop/CO & 0.2$\pm$0.03 & -1.7$\pm$0.0 & 0.21$\pm$0.11 & 0.11$\pm$0.05 & 0.14$\pm$0.08 & 0.13$\pm$0.1 \\
\nnhp/CO & 0.28$\pm$0.07 & -2.4$\pm$0.1 & 0.27$\pm$0.14 & 0.51$\pm$0.14 & 0.51$\pm$0.38 & -0.51$\pm$0.12 \\
\noalign{\smallskip}
\hline 
\noalign{\smallskip}
\end{tabular}
\tablefoot{Same as Table~\ref{tab:Fitparams_Int_vs_all_env}, but for CO line ratios}
\end{table*}

\section{Pixel-by-pixel distribution of line intensities and line ratios as function of galaxy properties}
\label{app:pixelbased}

\begin{figure*}
    \centering
    \includegraphics[width = 1\textwidth]{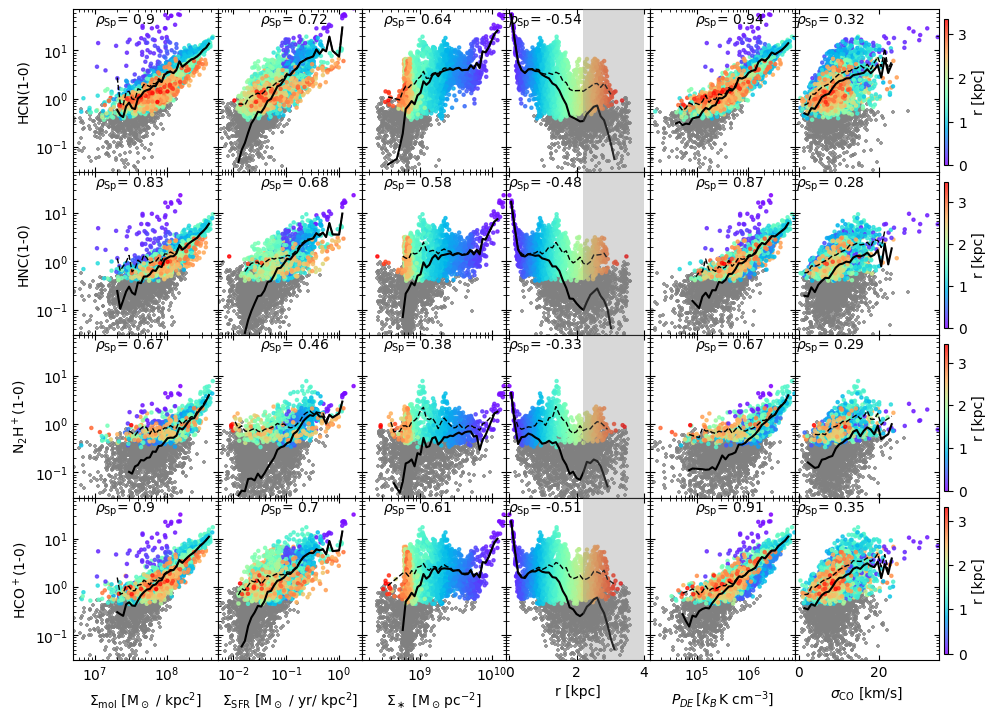 }
    \caption{Pixel-by-pixel integrated line intensities for HCN, HNC, \nnhp, \hcop(1-0) as function of \Smol, \SSFR, \Sstar, galactocentric radius, \Pde and \vdisp for both detected emission (> 3$\sigma$, colored by galactocentric distance) as well as all data (grey).
    Pixels with detected emission are colored by galactocentric radius. 
    We show the binned average for all data (black solid line) and significant data (black dashed line).  
    We add Spearman correlation coefficients using all data in the top left corner of each panel, but note that those coefficients strongly depend on the SNR of the line and a decrease in coefficients from brighter to fainter lines is expected (i.e. from HCN to \nnhp). All p-values are below 5\%. 
    }
    \label{fig:LineIntensity_pixelpixel}
\end{figure*}

We show the pixel-by-pixel distribution of line intensities of HCN, HNC, \nnhp and \hcop as function of galactic properties (\Smol, \SSFR, \Sstar, r, \Pde, \vdisp) in Figure~\ref{fig:LineIntensity_pixelpixel}. 
The data is colored by galactocentric radius, so that the impact of the central pixels can be clearly seen in the intensity-to-\Smol relation: All of the tested lines have bright emission in the center, that follows a distribution offset from the rest of the data points. 

Figure~\ref{fig:LineIntensity_pixelpixel} also shows the binned intensity trends utilizing all data points and only data points where emission is significantly detected. 
As discussed in Section~\ref{sec:Methods:Binning} and Appendix~\ref{sec:Mockdata:binning}, excluding data points from the analysis can significantly bias the results. 

We add Spearman correlation coefficients $\rho_\mathrm{Sp}$ using all data (no masking) in Figure~\ref{fig:LineIntensity_pixelpixel}. 
These coefficients are strongly dependent on the SNR of the data and therefore we expect to find a decrease in coefficient value from brighter to fainter lines, or in our case, from HCN to  \nnhp. 
For each line we find $\rho_\mathrm{Sp}$ highest for line intensity as function of \Pde, followed by \Smol, \SSFR, \Sstar, r and \vdisp.

\begin{figure*}
    \centering
    \includegraphics[width = 1\textwidth]{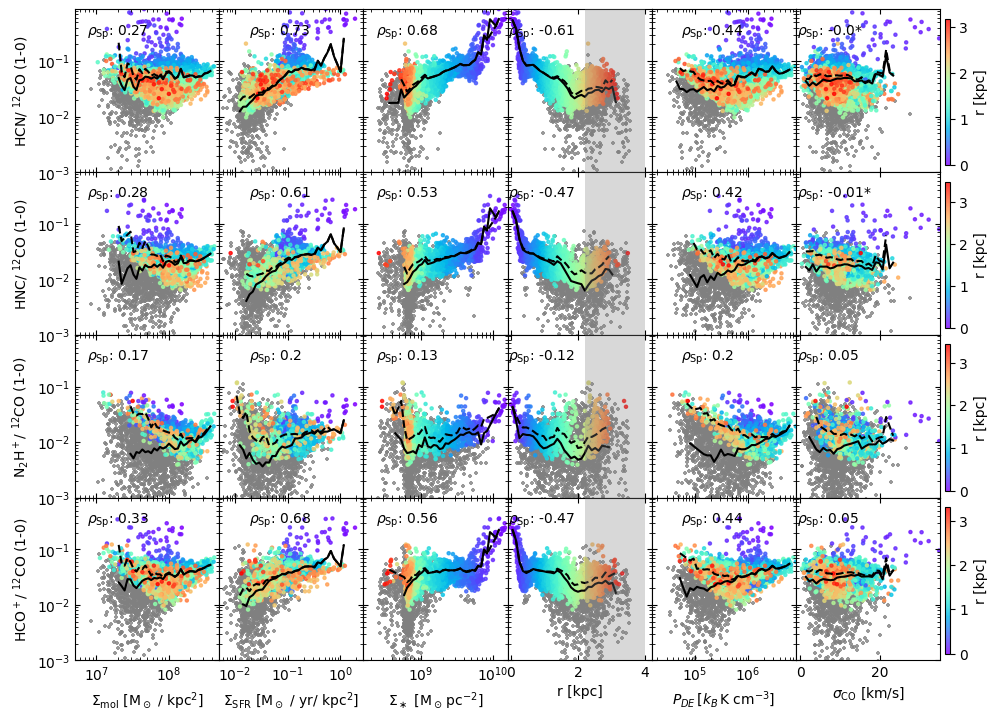}
    \caption{Same as Figure~\ref{fig:LineIntensity_pixelpixel} but for \CO line ratios. We mark Spearman correlation coefficients for which the corresponding p-value is below 5\% with an asterisk. 
    }
    \label{fig:DGfractions_pixelpixel}
\end{figure*}

\begin{figure*}
    \centering
    \includegraphics[width = 1\textwidth]{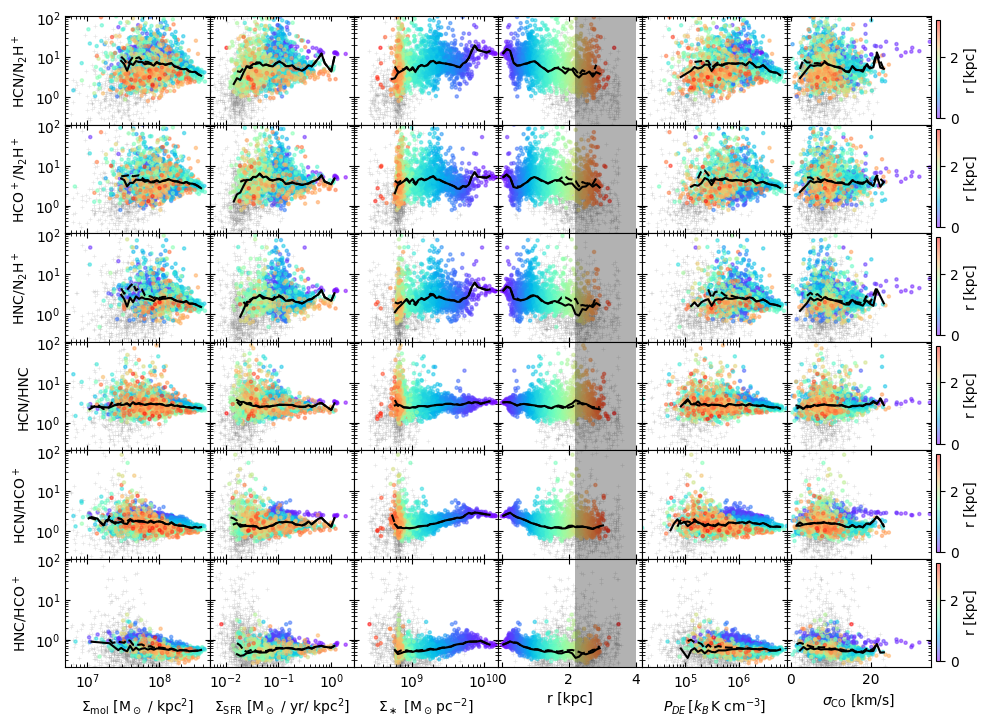}
    \caption{Same as Figure~\ref{fig:LineIntensity_pixelpixel} but for line ratios between dense gas tracing lines HCN, HNC, \hcop and \nnhp. 
    }
    \label{fig:DGlineratios_pixelbypixel}
\end{figure*}

Similarly, we show the pixel-by-pixel distribution of line intensities of HCN, HNC, \nnhp and \hcop divided by \CO emission as function of galactic properties (\Smol, \SSFR, \Sstar, r, \Pde, \vdisp) in Figure~\ref{fig:DGfractions_pixelpixel}. 
Lastly, we also show the pixel-by-pixel distribution of line ratios among the dense gas tracing lines in Figure~\ref{fig:DGlineratios_pixelbypixel}.

\end{appendix}
\end{document}